\documentclass[11pt]{article}
\usepackage[T1]{fontenc}
\usepackage[latin9]{inputenc}
\usepackage{geometry}
\usepackage{float}
\usepackage{amsmath}
\usepackage{amssymb}
\usepackage{stackrel}
\usepackage{graphicx}
\usepackage{esint}

\makeatletter

\providecommand{\tabularnewline}{\\}

\makeatother

\usepackage[english]{babel}
\begin{document}

\title{Model independence of the measurement of the $\mathrm{e}^{+}\mathrm{e^{-}\rightarrow\mathrm{ZH}}$
cross section using $\mathrm{\mathrm{\mathrm{Z\rightarrow\mu^{+}\mu^{-}}}}$
and $\mathrm{Z\rightarrow\mathrm{e^{+}\mathrm{e^{-}}}}$ at the ILC}

\author{J.Yan, K.Fujii, J. Tian}

\maketitle
\begin{center}
\textit{High Energy Accelerator Research Organization (KEK), Tsukuba
305-0801, Japan}
\par\end{center}
\begin{abstract}
The model independent measurement of the absolute cross section ($\sigma_{\mathrm{ZH}}$)
of the Higgsstrahlung process $\mathrm{e^{+}e^{-}\rightarrow ZH}$
is an unique measurement at the ILC indispensable for measuring the
Higgs couplings and their deviations from the Standard Model in order
to identify new physics models. The performance in measuring $\sigma_{\mathrm{ZH}}$
using events in which the Higgs boson recoils against a Z boson which
decays into a pair of muons or electrons has been demonstrated based
on full simulation of the ILD detector for three center of mass energies
$\sqrt{s}$ = 250, 350, and 500 GeV, and two beam polarizations $\left(P\mathrm{e^{-}},P\mathrm{e^{+}}\right)$
=($-$80\%, +30\%) and (+80\%, $-$30\%). This paper demonstrates
in detail that the analysis which achieved these results are model
independent to the sub-percent level. Data selection methods are designed
to optimize the precisions of $\sigma_{\mathrm{ZH}}$ and at the same
time minimize the bias on the measured $\sigma_{\mathrm{ZH}}$ due
to discrepancy in signal efficiencies among Higgs decay modes. Under
conservative assumptions which take into account unknown Higgs decay
modes, the relative bias on $\sigma_{\mathrm{ZH}}$ is shown to be
smaller than 0.2\% for all center-of-mass energies, which is five
times below even the smallest $\sigma_{\mathrm{ZH}}$ statistical
uncertainties expected from the leptonic recoil measurements in a
full 20 years ILC physics program. 
\end{abstract}

\section{INTRODUCTION}

It is one of the most important missions of high energy particle physics
to uncover the physics behind electroweak symmetry breaking (EWSB).
The discovery of the Standard Model (SM)-like Higgs boson at the Large
Hadron Collider (LHC) in 2012 \cite{ATLAS,CMS} proved the basic idea
of the SM that the vacuum filled with the Higgs condensate broke the
electroweak symmetry. The SM assumes one doublet of complex scalar
fields for the Higgs sector. However, apart from the fact that it
is the simplest, there is no reason to prefer the Higgs sector in
the SM over any other model that is consistent with experiments. Moreover,
the SM does not explain why the Higgs field became condensed in vacuum.
To answer this question, we need physics beyond the SM (\textquotedblleft BSM\textquotedblright )
which necessarily alters the properties of the Higgs boson. Each new
physics model predicts its own size and pattern of the deviations
of Higgs boson properties from their SM predictions. In order to discriminate
these new physics models, we need to measure with high precision as
many types of couplings as possible and as model independently as
possible. Because the deviations predicted by most new physics models
are typically no larger than a few percent, the coupling measurements
must achieve a precision of 1\% or better for a statistically significant
measurement. This level of sensitivity is available only in the clean
experimental environment of lepton colliders. 

The International Linear Collider (ILC) \cite{TDR} is a proposed
$\mathrm{e^{+}e^{-}}$ collider covering center-of-mass energy range
of 200 to 500 GeV, with expandability to 1 TeV. Among the most important
aspects of its physics program \cite{ILCPhysics} are the measurements
of Higgs couplings with unprecedented precision so as to find their
deviations from the SM and match their deviation pattern with predictions
of various new physics models.

Most of the Higgs boson measurements at the LHC are measurements of
cross section times branching ratio (BR). This is also true at the
ILC with one important exception, the measurement of the absolute
size of an inclusive Higgs production cross section by applying the
recoil technique to the Higgsstrahlung process $\mathrm{e^{+}e^{-}\rightarrow ZH}$.
The recoil technique involves measuring only the momenta of the decay
products of the Z boson which recoils against the Higgs boson, and
hence in principle is independent of the Higgs decay mode. The measurement
of this cross section $\sigma_{\mathrm{ZH}}$ is indispensable for
extracting the branching ratios, the Higgs total width, and couplings
from cross section times branching ratio measurements. The recoil
technique, which is only possible at a lepton collider owing to the
well-known initial state, is applicable even if the Higgs boson decays
invisibly and hence allows us to determine $\sigma_{\mathrm{ZH}}$
in a completely model independent way, as will be shown in this paper.
Especially high precision measurements of $\sigma_{\mathrm{ZH}}$
and $M_{\mathrm{H}}$ are possible by applying the recoil technique
to Higgsstrahlung events where the Z boson decays to a pair of electrons
or muons, which profits from excellent tracking momentum resolution
and relatively low background levels. Furthermore, in this channel
model independence for the measurement of $\sigma_{\mathrm{ZH}}$
can be demonstrated in practice.

A study reported in \cite{Recoil} evaluates the performance of measuring
$\sigma_{\mathrm{ZH}}$ and the Higgs boson mass $M_{\mathrm{H}}$
using Higgsstrahlung events with leptonic Z boson decays $\mathrm{e^{+}e^{-}\rightarrow ZH\rightarrow\mathrm{\mathit{\mathrm{l^{+}l^{-}}}H}}$
($\mathrm{l}$ = e or $\mu$) for three center-of-mass energies (250,
350, and 500 GeV), as well as two beam polarizations $\left(P\mathrm{e}^{-},P\mathrm{e}^{+}\right)$
=($-$80\%, +30\%) and (+80\%, $-$30\%), which will be denoted as
$\mathrm{e_{L}^{-}}\mathrm{e_{R}^{+}}$ and $\mathrm{e_{R}^{-}}\mathrm{e_{L}^{+}}$,
respectively. The results in \cite{Recoil} will be scaled to the
``H20'' program \cite{ILCRun}, which designates that during a 20
year period, a total of 2000, 200, and 4000 $\mathrm{fb^{-1}}$ will
be accumulated at $\sqrt{s}$= 250, 350, and 500 GeV, respectively.
This paper reports a study which demonstrates that the measurement
of $\sigma_{\mathrm{ZH}}$ in \cite{Recoil} is model independent
to a level well below the expected statistical precision from the
full ILC physics program. %
\footnote{An analysis using hadronic decays of the Z boson at a center-of-mass
energy of 350 GeV has been presented in \cite{key-8hadronic recoil},
in which $\sigma_{\mathrm{ZH}}$ was measured with a Higgs decay mode
efficiency dependence of the order of 15\%.%
} The methods of signal selection and background rejection studied
here are those used for producing the results in \cite{Recoil}. 

This paper is structured as follows: Section \ref{sec:The-Higgs-Recoil}
explains the recoil measurement; Section \ref{sec:Analysis} introduces
the simulation tools, the ILC detector concept, and the signal and
physics background processes; Section \ref{sec:Event-Selection} presents
the methods of data selection optimized for this analysis; Section
\ref{sec:Extraction} describes the efforts to minimize Higgs decay
mode bias and evaluates the bias on the measured $\sigma_{\mathrm{ZH}}$;
Finally Section \ref{sec:SUMMARY-AND-CONCLUSIONS} summarizes the
analysis and concludes the paper.

\section{HIGGS BOSON MEASUREMENTS USING THE RECOIL TECHNIQUE \label{sec:The-Higgs-Recoil}}

\begin{figure}[H]
\begin{centering}
\includegraphics[scale=0.68]{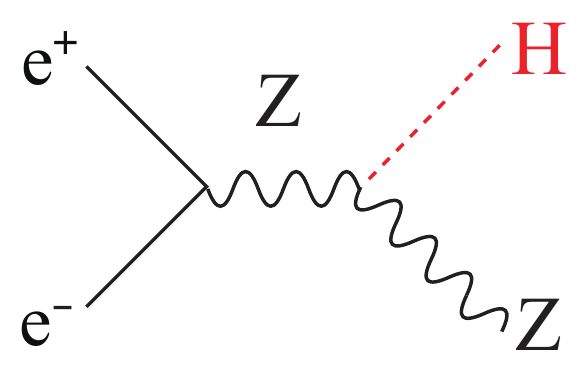} \quad{}\includegraphics[scale=0.53]{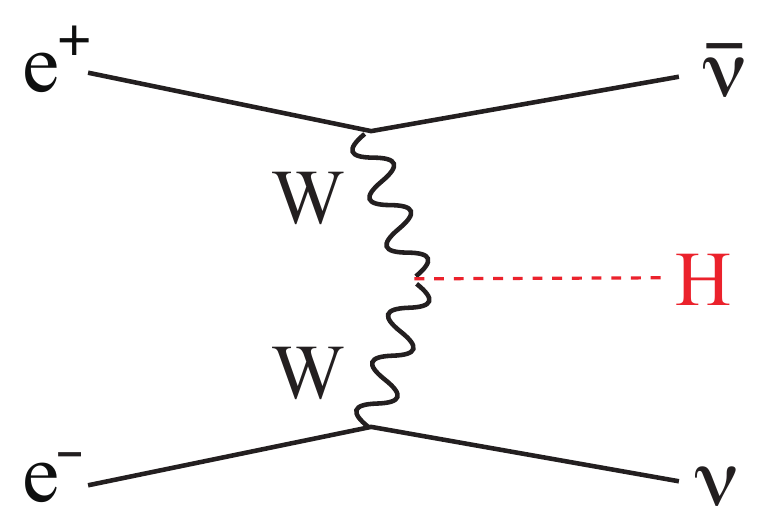}\quad{}
\includegraphics[scale=0.53]{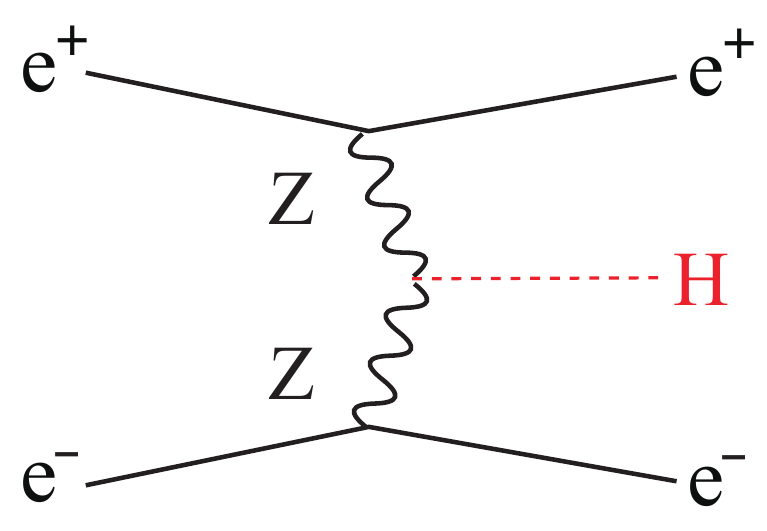}
\par\end{centering}

\protect\caption{The lowest order Feynman diagrams of the three major Higgs production
processes at the ILC: (left) Higgsstrahlung process $\mathrm{e^{+}e^{-}\rightarrow ZH}$
, (center) WW fusion process $\mathrm{e^{+}e^{-}\rightarrow\nu\overline{\nu}H}$
, and (right) ZZ fusion process $\mathrm{e^{+}e^{-}\rightarrow e^{+}e^{-}H}$.
\label{Feynmann}}
\end{figure}

\begin{figure}[H]
\begin{centering}
\includegraphics[scale=0.32]{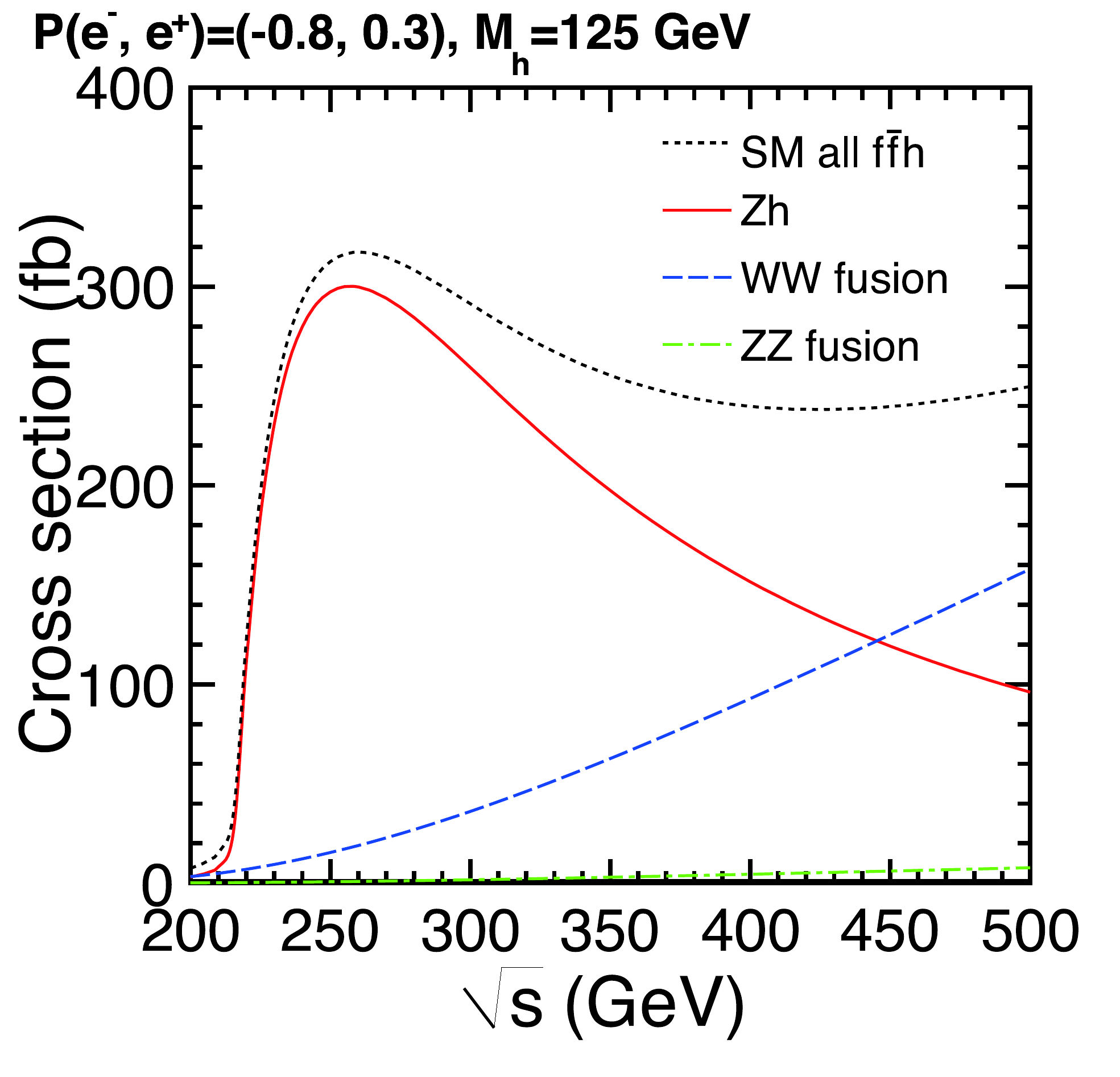}
\par\end{centering}

\protect\caption{The Higgs production cross section as a function of $\sqrt{s}$ assuming
$M_{\mathrm{H}}$=125 GeV for the following Higgs production processes:
Higgsstrahlung (solid), WW fusion (dashed), and ZZ fusion (dotted).
(Figure taken from \cite{TDR}.) \label{crossSec}}
\end{figure}

The major Higgs production processes at the ILC are Higgsstrahlung
and WW fusion, whose lowest order Feynman diagrams are illustrated
in Figure \ref{Feynmann}, along with the ZZ fusion process which
has a significantly smaller cross section than the other two processes
at ILC center-of-mass energies. Figure \ref{crossSec} shows the production
cross sections as a function of $\sqrt{s}$, assuming a Higgs boson
mass of 125 GeV. 

The Higgsstrahlung process with a Z boson decaying into a pair of
electrons or muons: $\mathrm{e^{+}e^{-}\rightarrow ZH\rightarrow\mathit{\mathrm{l^{+}l^{-}}}H}$
( $\mathrm{l}$ = e or $\mu$) will be hereafter referred to as $\mathrm{e^{+}e^{-}H}$
and $\mathrm{\mu{}^{+}\mathit{\mu}^{-}H}$, respectively. The leptonic
recoil technique is based on the Z boson identification by the invariant
mass of the dilepton system being consistent with the Z boson mass,
and the reconstruction of the mass of the rest of the final-state
system recoiling against the Z boson ($M_{\mathrm{rec}}$), corresponding
to the Higgs boson mass, which is calculated as

\begin{equation}
M_{\mathrm{rec}}=\left(\sqrt{s}-E_{\mathrm{l^{+}l^{-}}}\right)^{2}-\left|\overrightarrow{p}_{\mathrm{l^{+}l^{-}}}\right|^{2}\:,\label{eq:kinematics1}
\end{equation}
where $E_{\mathrm{l^{+}l^{-}}}\equiv E_{\mathrm{l^{+}}}+E_{\mathrm{l^{-}}}$
and $\overrightarrow{p}_{\mathrm{l^{+}l^{-}}}\equiv\overrightarrow{p}_{\mathrm{l}^{+}}+\overrightarrow{p}_{\mathrm{l}^{-}}$
are the energy and momentum of the lepton pair from Z boson decay.
The $M_{\mathrm{rec}}$ calculated using Equation \ref{eq:kinematics1}
is expected to form a peak corresponding to Higgs boson production.
From the location of the $M_{\mathrm{rec}}$ peak and the area beneath
it the Higgs boson mass and the signal yield can be extracted. The
production cross section ($\sigma_{\mathrm{ZH}}$) can be obtained
as : 

\begin{equation}
\sigma_{\mathrm{ZH}}=\frac{N_{S}}{BR\mathrm{\left(\mathrm{Z\rightarrow\mathrm{l^{+}l^{-}}}\right)\varepsilon}_{S}L}\:,\label{eq:xsec}
\end{equation}
where $N_{S}$ is the number of selected signal events, $\varepsilon_{\mathrm{S}}$
is the efficiency of signal event selection, and $L$ is the total
integrated luminosity. In principle, $\varepsilon_{\mathrm{S}}$ and
hence $\sigma_{\mathrm{ZH}}$ are independent of how the Higgs boson
decays, since only the leptons from the Z decay need to be measured
in the recoil technique. In practice, however, this is not completely
guaranteed since there is a possibility of confusion between the leptons
from the Z boson decay and those from the Higgs boson decay. Thus
this paper aims to demonstrate that the signal efficiency is indeed
independent of assumptions regarding Higgs boson decay, based on the
Higgs recoil analysis given in detail in in \cite{Recoil}.

\section{ANALYSIS FRAMEWORK, DETECTOR SIMULATION, AND EVENT GENERATION \label{sec:Analysis}}

\subsection{Analysis framework}

This study used the simulation and reconstruction tools contained
in the software package ILCSoft v01-16 \cite{ILCSoft}. All parameters
of the incoming beams are simulated with the GUINEA-PIG package \cite{Guinea Pig}
and the beam spectrum, including beamstrahlung and ISR, are explicitly
taken into consideration based on the parameters in the TDR. The beam
crossing angle of 14 mrad in the current ILC design is taken into
account. The $\mathrm{\mu{}^{+}\mathit{\mu}^{-}H}$, $\mathrm{e^{+}e^{-}H}$,
and SM background samples (see Section \ref{sub:Signal-and-Background}
for details) are generated using the WHIZARD 1.95 \cite{whizard}
event generator. The input mass of the Higgs boson is 125 GeV, and
its SM decay branching ratios are assumed \cite{key-4Higgs handbook}.
The model for the parton shower and hadronization is taken from PYTHIA
6.4 \cite{Pythia}. The generated events are passed through the ILD
\cite{ILD} simulation performed with the MOKKA \cite{Mokka} software
package based on GEANT4 \cite{Geant4}. Event reconstruction is performed
using the Marlin \cite{Marlin} framework. The PandoraPFA \cite{PFA}
algorithm is used for calorimeter clustering and the analysis of track
and calorimeter information based on the particle flow approach.

\subsection{The ILD concept\label{sub:The-ILD-Detector}}

The International Large Detector (ILD) concept \cite{ILD} is one
of the two detectors being designed for the ILC. It features a hybrid
tracking system with excellent momentum resolution. The jet energy
resolution is expected to be better than 3\% for jets with energies
$\geq$ 100 GeV, thanks to its highly granular calorimeters optimized
for Particle Flow reconstruction \cite{PFA}. This section describes
the ILD sub-detectors important for this study. 

The vertex detector (VTX), consisting of three double layers of extremely
fine Si pixel sensors with the innermost radius at 15 mm, measures
particle tracks with a typical spatial resolution of 2.8 $\mathrm{\mu m}$.
The hybrid tracking system consists of a time projection chamber (TPC)
which provides up to 224 points per track, excellent spatial resolution
of better than 100 $\mathrm{\mu m}$, and $dE/dx$ - based particle
identification, as well as Si-strip sensors placed in the barrel region
both inside and outside the TPC and in the endcap region outside the
TPC in order to further improve track momentum resolution. The tracking
system measures charged particle momenta to a precision of $\frac{\delta p_{t}}{p_{t}^{2}}=2\times10^{-5}$$\mathrm{\; GeV^{-1}}$.
Outside of the tracking system sits the ECAL, a Si-W sampling electromagnetic
calorimeter with an inner radius of 1.8 m, finely segmented $5\times5$
$\mathrm{mm^{2}}$ transverse cell size and 30 longitudinal layers
equivalent to 24 radiation lengths. The HCAL, a steel-scintillator
type hadronic calorimeter which surrounds the ECAL, has an outer radius
of 3.4 m, $3\times3$ $\mathrm{cm^{2}}$ transverse tiles, and 48
longitudinal layers corresponding to 5.9 interaction lengths. Radiation
hard calorimeters for monitoring the luminosity and quality of the
colliding beams are installed in the forward region. The tracking
system and calorimeters are placed inside a superconducting solenoid
which provides a magnetic field of 3.5 T. An iron yoke outside the
solenoid coil returns the magnetic flux, and is instrumented with
scintillator-based muon detectors.

\subsection{Signal and background processes\label{sub:Signal-and-Background}}

\begin{figure}[H]
\centering{}\includegraphics[scale=0.8]{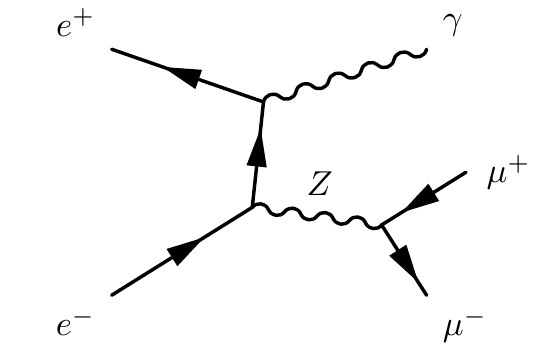}\quad{} \includegraphics[scale=0.8]{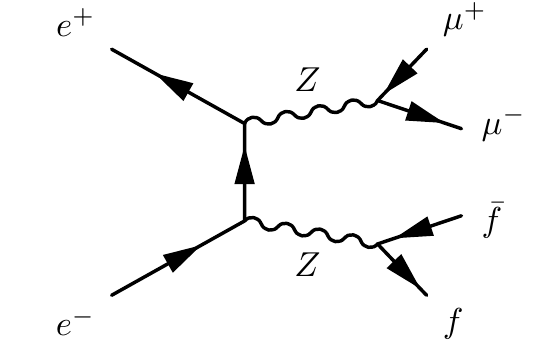}
\quad{}\includegraphics[scale=0.8]{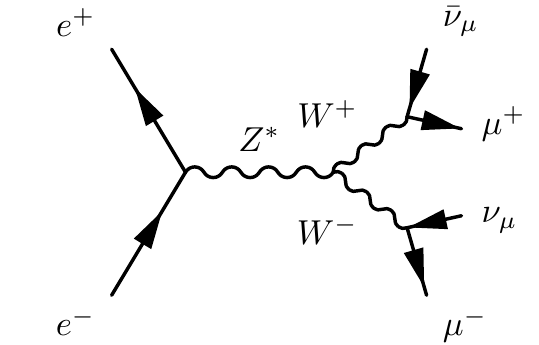} \protect\caption{The Feynman diagrams contributing to the major background processes
for the Higgs recoil analysis in the $\mathrm{\mu{}^{+}\mathit{\mu}^{-}H}$
channel: 2f\_l background with $\mu\mu$ in the final state and an
ISR photon (left), 4f\_sl background with ZZ as intermediate state
(center), 4f\_l background with WW as intermediate state (right).
\label{fig:The-FeynmanBG}}
\end{figure}

The Higgsstrahlung signal is selected by identifying a pair of prompt,
isolated, and oppositely charged muons or electrons with well-measurable
momentum whose invariant mass $M_{\mathrm{l^{+}l^{-}}}$ ($\mathrm{l}$=$\mathrm{e}$
or $\mu$) is close to the Z boson mass ($M_{\mathrm{Z}}$). The $\mathrm{\mu{}^{+}\mathit{\mu}^{-}H}$
and $\mathrm{e^{+}e^{-}H}$ channels are analyzed independently and
then statistically combined. Figure \ref{fig:The-FeynmanBG} shows
the Feynman diagrams of the dominant 4-fermion and 2-fermion processes.
Table \ref{tab:Processes-and-cross} gives the cross sections of signal
and major background processes assuming $M_{\mathrm{H}}$=125 GeV.
For each process, all SM tree-level diagrams are included by WHIZARD.
These processes are grouped as follows from the perspective of finding
leptons in the final state:
\begin{itemize}
\item $\mathrm{l^{+}l^{-}}\mathrm{H}$ ($\mathrm{l}$=$\mathrm{e}$ or $\mu$)
: The Higgsstrahlung signal process with Z decaying to $\mathrm{l^{+}l^{-}}$.
The $\mathrm{e^{+}e^{-}H}$ channel contains an admixture of the ZZ
fusion process, which is removed at the early stages of the analysis.
\item 2-fermion leptonic (2f\_l): final states consisting of a charged lepton
pair or a neutrino pair. The intermediate states are Z or $\gamma^{*}$. 
\item 4-fermion leptonic (4f\_l): final states of 4 leptons consisting of
mainly processes through ZZ and WW intermediate states. Those events
containing a pair of electrons or muons are a background of the $\mathrm{\mu{}^{+}\mathit{\mu}^{-}H}$
and $\mathrm{e^{+}e^{-}H}$ channels, respectively. 
\item 4-fermion semileptonic (4f\_sl): final states of a pair of charged
leptons and a pair of quarks, consisting of mainly processes through
ZZ and WW intermediate states. In the former case, one Z boson decays
to a pair of charged leptons or neutrinos, and the other to quarks.
In the latter case, one W boson decays to a charged lepton and a neutrino
of the same flavor and the other to quarks.
\item 4(2)-fermion hadronic (4(2)f\_h): final states of 4 (2) quarks. Since
the probability of finding isolated leptons is very small for these
final states, these events are removed almost completely at the lepton
identification stage (see Section \ref{sub:Selection-of-Best}).
\end{itemize}
The analysis in this paper and \cite{Recoil} are conducted for the
center-of-mass energies 250, 350, and 500 GeV, and two beam polarization
$\mathrm{e_{L}^{-}e_{R}^{+}}$ and $\mathrm{e_{R}^{-}e_{L}^{+}}$.
From Table \ref{tab:Processes-and-cross}, it can be seen that the
signal cross sections for $\mathrm{e_{R}^{-}e_{L}^{+}}$ is smaller
by a factor of 1.5 with respect to $\mathrm{e_{L}^{-}e_{R}^{+}}$,
whereas the total background is suppressed by a factor of 2 and some
individual background processes are suppressed by a factor of up to
10. The methods and performance of signal selection and background
rejection are presented in Section \ref{sec:Event-Selection}.

The Monte Carlo (MC) samples are generated for the cases in which
$\left(P\mathrm{e^{-}},P\mathrm{e^{+}}\right)$ =($-$100\%, +100\%)
and (+100\%, $-$100\%). The standard samples used in \cite{Recoil}
are generated for signal and background processes with the statistics
as shown in Table \ref{tab:Processes-and-cross}, and the events are
and normalized to the assumed integrated luminosities, cross sections,
and polarizations. Another type of signal sample is generated with
high statistics of more than 40k for each major SM Higgs decay mode,
mainly for the purpose of the model independence study in this paper. 

\begin{table}[H]
\begin{centering}
\begin{tabular}{|c|c|c|c|c|}
\hline 
$\sqrt{s}$ = 250 GeV & \multicolumn{1}{c}{cross} & section & \multicolumn{1}{c}{$N_{\mathrm{Gen}}$} & \tabularnewline
\hline 
polarization & left & right & left & right\tabularnewline
\hline 
$\mathrm{\mu{}^{+}\mathit{\mu}^{-}H}$  & 10.4 fb & 7.03 fb & 17.1k & 11.0k\tabularnewline
\hline 
$\mathrm{e^{+}e^{-}H}$ & 10.9 fb & 7.38 fb & 17.6k & 11.2k\tabularnewline
\hline 
2f\_l & 38.2 pb & 35.0 pb & 2.63M & 2.13M\tabularnewline
\hline 
2f\_h & 78.1 pb & 46.2 pb & 1.75M & 1.43M\tabularnewline
\hline 
4f\_l & 5.66 pb & 1.47 pb & 2.25M & 0.35M\tabularnewline
\hline 
4f\_sl & 18.4 pb & 2.06 pb & 4.43M & 0.36M\tabularnewline
\hline 
4f\_h & 16.8 pb & 1.57 pb & 2.50M & 0.24M\tabularnewline
\hline 
total background & 157.1 pb & 86.3 pb & 13.6M & 4.51M\tabularnewline
\hline 
\end{tabular}
\par\end{centering}

\bigskip{}

\begin{centering}
\begin{tabular}{|c|c|c|c|c|}
\hline 
$\sqrt{s}$ = 350 GeV & \multicolumn{1}{c}{cross} & section & \multicolumn{1}{c}{$N_{\mathrm{Gen}}$} & \tabularnewline
\hline 
polarization & left & right & left & right\tabularnewline
\hline 
$\mathrm{\mu{}^{+}\mathit{\mu}^{-}H}$  & 6.87 fb & 4.63 fb & 11.3k & 8.0k\tabularnewline
\hline 
$\mathrm{e^{+}e^{-}H}$ & 10.24 fb & 6.68 fb & 17.9k & 9.0k\tabularnewline
\hline 
2f\_l & 33.5 pb & 31.5 pb & 2.71M & 1.94M\tabularnewline
\hline 
2f\_h & 38.6 pb & 23.0 pb & 1.60M & 0.89M\tabularnewline
\hline 
4f\_l & 4.90 pb & 1.48 pb & 3.07M & 0.48M\tabularnewline
\hline 
4f\_sl & 14.5 pb & 1.70 pb & 4.77M & 0.37M\tabularnewline
\hline 
4f\_h & 12.6 pb & 1.11 pb & 2.49M & 0.22M\tabularnewline
\hline 
total background & 104.1 pb & 58.7 pb & 14.6M & 3.89M\tabularnewline
\hline 
\end{tabular}
\par\end{centering}

\bigskip{}

\begin{centering}
\begin{tabular}{|c|c|c|c|c|}
\hline 
$\sqrt{s}$ = 500 GeV & \multicolumn{1}{c}{cross} & section & \multicolumn{1}{c}{$N_{\mathrm{Gen}}$} & \tabularnewline
\hline 
polarization & left & right & left & right\tabularnewline
\hline 
$\mathrm{\mu{}^{+}\mathit{\mu}^{-}H}$  & 3.45 fb & 2.33 fb & 6.0k & 4.0k\tabularnewline
\hline 
$\mathrm{e^{+}e^{-}H}$ & 11.3 fb & 7.11 fb & 15.0k & 7.5k\tabularnewline
\hline 
2f\_l & 6.77 pb & 5.96 pb & 0.42M & 0.36M\tabularnewline
\hline 
2f\_h & 19.6 pb & 11.7 pb & 1.51M & 0.84M\tabularnewline
\hline 
4f\_l & 10.6 pb & 7.48 pb & 0.60M & 0.34M\tabularnewline
\hline 
4f\_sl & 13.2 pb & 2.94 pb & 0.97M & 99.9k\tabularnewline
\hline 
4f\_h & 8.65 pb & 0.74 pb & 0.69M & 18.0k\tabularnewline
\hline 
total background & 58.9 pb & 28.8 pb & 4.18M & 1.65M\tabularnewline
\hline 
\end{tabular}
\par\end{centering}

\protect\caption{Cross sections and number of generated MC events ($N_{\mathrm{Gen}}$)
of signal and major background processes at each center-of-mass energy
and 100\% left and right beam polarizations, as calculated by the
WHIZARD generator. Here, ``left'' and ``right'' polarization correspond
to the cases where $\left(P\mathrm{e^{-}},P\mathrm{e^{+}}\right)$
=($-$100\%, +100\%) and (+100\%, $-$100\%), respectively.\label{tab:Processes-and-cross}}
\end{table}

\section{ANALYSIS \label{sec:Event-Selection}}

First, the signal events are selected by identifying a pair of leptons
($\mathrm{\mathrm{e^{+}\mathrm{e^{-}}}}$ or $\mu^{+}\mu^{-}$) produced
in the decay of the Z boson against which the Higgs recoils. Then
the recovery of FSR/bremsstrahlung photons are performed. Finally
background events are rejected through a series of cuts on several
kinematic variables.

\subsection{Selection of best lepton pair\label{sub:Selection-of-Best}}

\subsubsection{Isolated lepton finder \label{sub:Isolated-Lepton-Finder}}

Table \ref{LepID} summarizes the criteria for selecting an isolated
lepton. Here, $p_{\mathrm{track}}$ is the measured track momentum,
$E_{\mathrm{ECAL}}$ is the energy deposit in the ECAL, $E_{\mathrm{CAL,tot}}$
is the energy deposit in both ECAL and HCAL, $E_{\mathrm{yoke}}$
is the energy deposit inside the muon detector, and $d_{0}$ and $z_{0}$
are the transverse and longitudinal impact parameters. These criteria
are described as follows:
\begin{enumerate}
\item An electron deposits nearly all its energy in the ECAL while a muon
passes the ECAL and HCAL as a minimal ionizing particle. Therefore
$E_{\mathrm{ECAL}}$, $E_{\mathrm{CAL,tot}}$, and $p_{\mathrm{track}}$
are compared for each final state particle.
\item The leptons from $\tau$ decay or b/c quark jets are suppressed by
requirements on $d_{0}$ and $z_{0}$ with respect to their measurement
uncertainties. 
\item In order to avoid selecting leptons in hadronic jets, the leptons
are required to have sufficient $p_{\mathrm{track}}$, and to satisfy
an isolation requirement based on a multi-variate double cone method,
as described in \cite{Double cone}. 
\end{enumerate}
\begin{table}[H]
\begin{centering}
\begin{tabular}{|c|c|c|}
\hline 
 & $\mu$ ID & $\mathrm{e}$ ID\tabularnewline
\hline 
\hline 
momentum and  & $p_{\mathrm{track}}>5\quad\mathrm{GeV}$ & $p_{\mathrm{track}}>5\quad\mathrm{GeV}$\tabularnewline
\cline{2-3} 
energy deposit & $E_{\mathrm{CAL,tot}}/p_{\mathrm{track}}<0.3$ & $0.5<E_{\mathrm{CAL,tot}}/p_{\mathrm{track}}<1.3$\tabularnewline
\cline{2-3} 
 & $E_{\mathrm{yoke}}<1.2\quad\mathrm{GeV}$ & $E_{\mathrm{ECAL}}/E_{\mathrm{CAL,tot}}>0.9$\tabularnewline
\hline 
impact parameter & $\left|d_{0}/\delta d_{0}\right|<5$ & $\left|d_{0}/\delta d_{0}\right|<50$\tabularnewline
\cline{2-3} 
 & $\left|z_{0}/\delta z_{0}\right|<5$ & $\left|z_{0}/\delta z_{0}\right|<5$\tabularnewline
\hline 
 & \multicolumn{1}{c}{isolation criteria} & isolation criteria\tabularnewline
\hline 
\end{tabular}
\par\end{centering}

\protect\caption{The criteria for the identification of isolated leptons ($\mu$ and
$e$). \label{LepID}}
\end{table}

\subsubsection{Selection of the best lepton pair\label{sub:pair}}

For each event, two isolated leptons of the same flavor and opposite
charges are selected as the candidate pair for analysis. In this stage,
it is essential to distinguish a pair of leptons produced in the decay
of the Z boson recoiling against the Higgs boson (``correct pair'')
from those produced in the Higgs boson decay (``wrong pair''). This
is important for achieving precise $M_{\mathrm{H}}$ measurements
and for preventing Higgs decay mode dependence. For the Higgsstrahlung
process, the invariant mass $M_{\mathrm{l^{+}l^{-}}}$ ($\mathrm{l}$
= $\mathrm{e}$ or $\mathrm{\mu}$) of the dilepton system and recoil
mass $M_{\mathrm{rec}}$ should be close to the Z boson mass $M_{\mathrm{Z}}$=91.187
GeV \cite{PDG} and the Higgs boson mass $M_{\mathrm{H}}$=125 GeV
(in this study), respectively. The decay modes which contain an extra
source of leptons, such as the $\mathrm{H\rightarrow ZZ^{*}}$ and
$\mathrm{H\rightarrow WW^{*}}$ modes, have a higher ratio of ``wrong
pairs''. 

\begin{figure}[H]
\centering{}\includegraphics[scale=0.37]{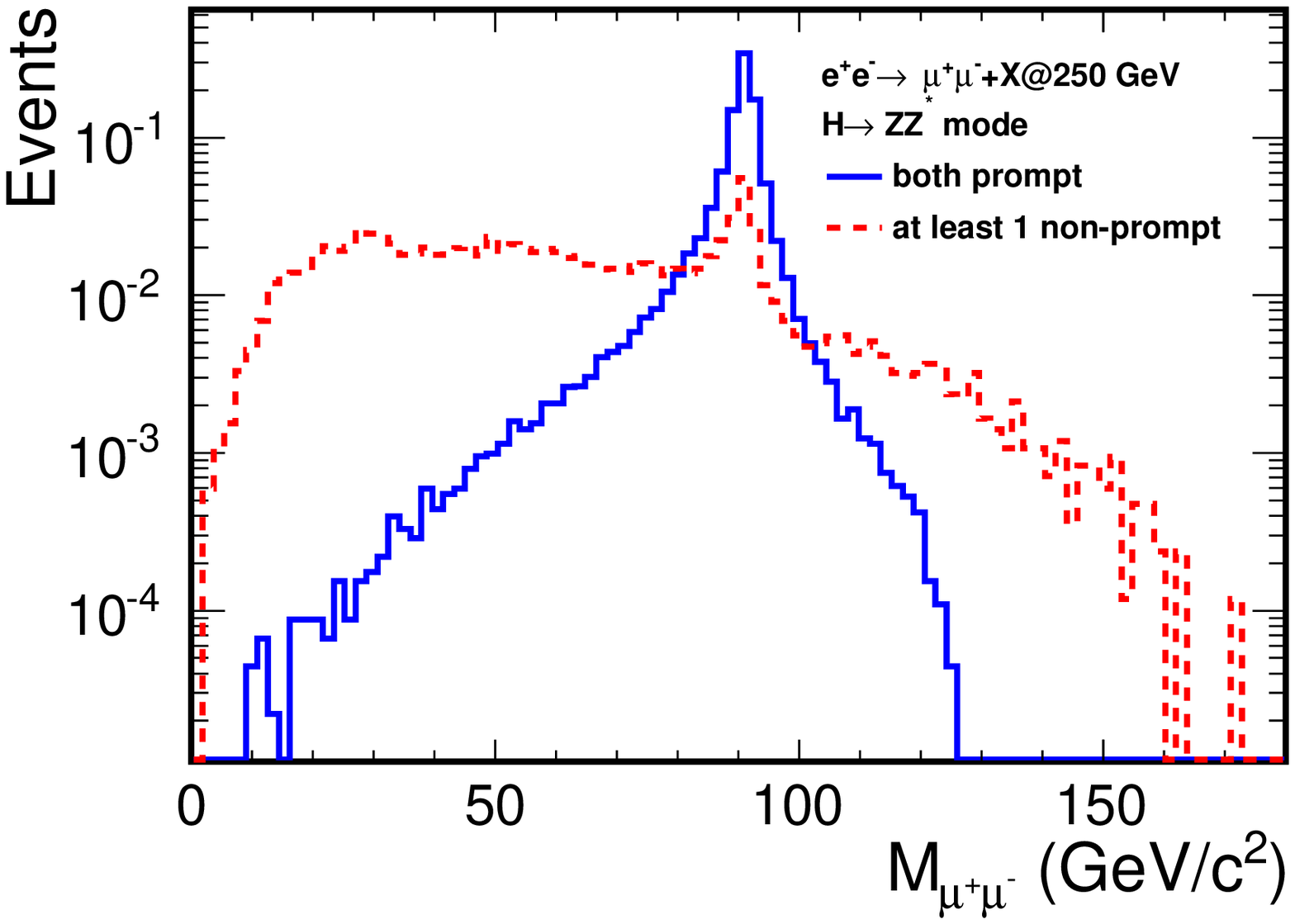} \includegraphics[scale=0.37]{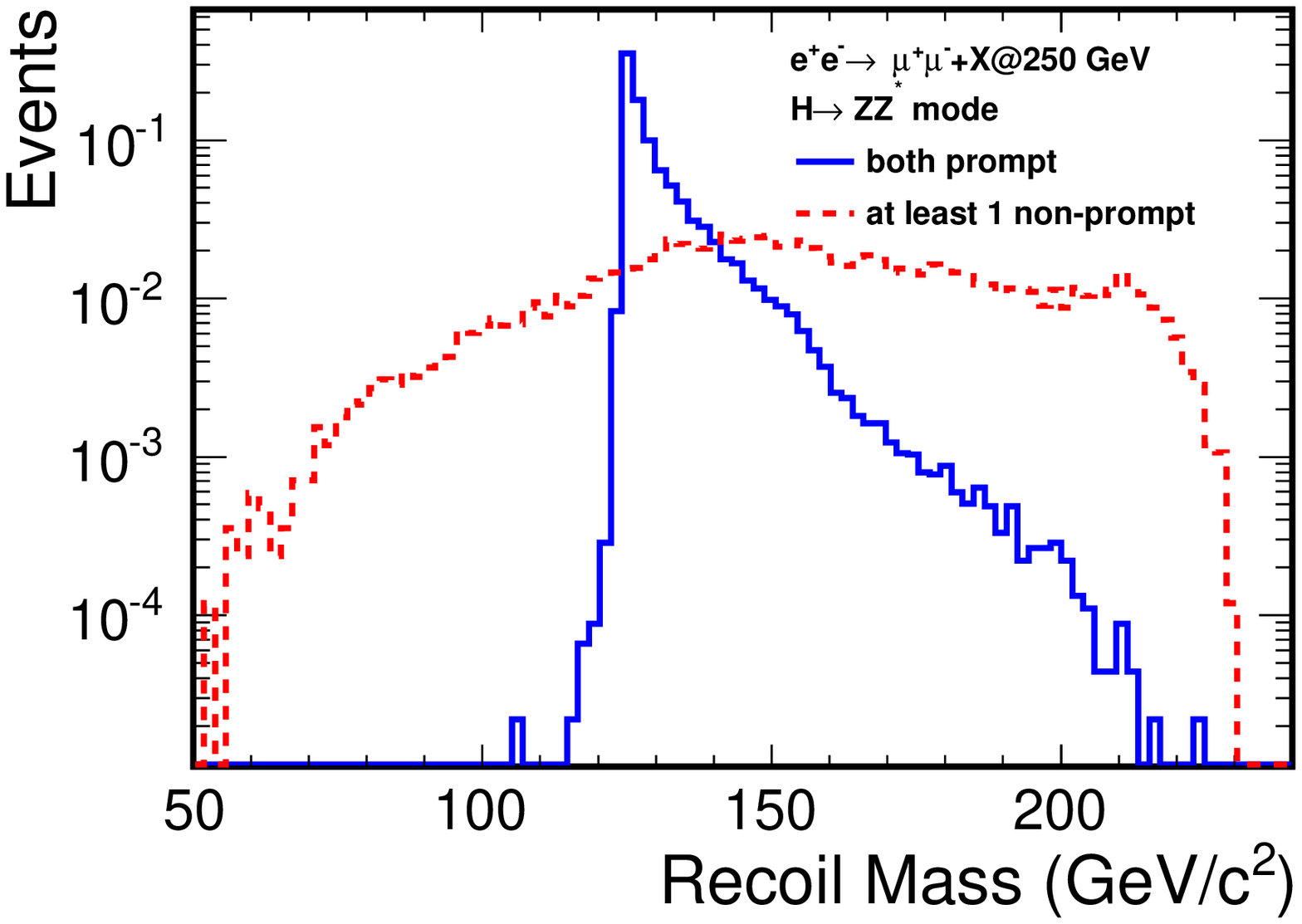}\protect\caption{Comparison of the distributions of $M_{\mathrm{l^{+}l^{-}}}$ (left)
and $M_{\mathrm{rec}}$ (right) between \textquotedblleft correct\textquotedblright{}
and \textquotedblleft wrong\textquotedblright{} lepton pairs. This
is an example of the $\mathrm{H\rightarrow ZZ^{*}}$ decay mode in
the $\mathrm{\mu{}^{+}\mathit{\mu}^{-}H}$ channel at $\sqrt{s}$
= 250 GeV.\label{MassCompare}}
\end{figure}

The best lepton pair candidate is selected based on the following
criteria. First, the requirement $\left|M_{\mathrm{l^{+}l^{-}}}-M_{\mathrm{Z}}\right|<40(60)\:\mathrm{GeV}$
is implemented for $\mu\left(\mathrm{e}\right)$. In the case where
both leptons originate from a single Z boson produced in Higgs boson
decay, $M_{\mathrm{rec}}$ tends to deviate from $M_{\mathrm{H}}$
even if $M_{\mathrm{l^{+}l^{-}}}$ is close to $M_{\mathrm{Z}}$.
Therefore the next step is to select, taking into account both $M_{\mathrm{l^{+}l^{-}}}$
and $M_{\mathrm{rec}}$, the pair which minimizes the following $\chi^{2}$
function:

\begin{equation}
\chi^{2}\left(M_{\mathrm{l^{+}l^{-}}},M_{\mathrm{rec}}\right)=\frac{\left(M_{\mathrm{l^{+}l^{-}}}-M_{\mathrm{Z}}\right)^{2}}{\sigma_{M_{\mathrm{l^{+}l^{-}}}}^{2}}+\frac{\left(M_{\mathrm{rec}}-M_{\mathrm{H}}\right)^{2}}{\sigma_{M_{\mathrm{rec}}}^{2}}\:,\label{eq:chi2}
\end{equation}
$\sigma_{M_{\mathrm{l^{+}l^{-}}}}$ and $\sigma_{M_{\mathrm{rec}}}$
are determined by a Gaussian fit to the distributions of $M_{\mathrm{l^{+}l^{-}}}$
and $M_{\mathrm{rec}}$ for each channel. Using the $\mathrm{H\rightarrow ZZ^{*}}$
mode in the $\mathrm{\mu{}^{+}\mathit{\mu}^{-}H}$ channel at $\sqrt{s}$=250
GeV as an example, Figure \ref{MassCompare} compares the distributions
of $M_{\mathrm{l^{+}l^{-}}}$ and $M_{\mathrm{rec}}$ between \textquotedblleft correct\textquotedblright{}
(solide line) and \textquotedblleft wrong\textquotedblright{} (dotted
line) pairs, defined as those in which at least one lepton is from
Higgs boson decay. Here, the ``correct'' and ``wrong'' pairs are
separated using the MC truth information of the pairs selected by
the above-mentioned pairing algorithm. One can see, only in the case
of the ``correct pairs'', a clean $M_{\mathrm{l^{+}l^{-}}}$ peak
at $M_{\mathrm{Z}}$ signaling Z boson production, and a clean $M_{\mathrm{rec}}$
peak corresponding to the Higgs boson production. At $\sqrt{s}$ =
250 GeV, the efficiency of the dilepton finder described above in
finding a pair of isolated leptons is about 94\% and about 89\% for
the $\mathrm{\mu{}^{+}\mathit{\mu}^{-}H}$ and $\mathrm{e^{+}e^{-}H}$
channels, respectively. Meanwhile ``wrong pairs'' as well as the
backgrounds in Section \ref{sub:Signal-and-Background} are significantly
suppressed. 

The shape of the $M_{\mathrm{rec}}$ distribution is affected by radiative
and resolution effects. The radiative effects comprise of beamstrahlung,
Initial State Radiation (ISR), Final State Radiation (FSR) and bremsstrahlung.
Because events are moved from the peak region of the $M_{\mathrm{rec}}$
distribution to the tail, the measurement precision is degraded. On
the other hand, resolution effects determine the peak width of the
distribution and thus the measurement uncertainties. The dominant
resolution effects are the beam energy spread induced by the accelerator
and the uncertainty of the detector response, dominated by the track
momentum resolution. Compared to these, the SM Higgs decay width of
about 4 MeV is negligible. While ISR and FSR are unsuppressible physical
effects, beamstrahlung, bremsstrahlung, and resolution effects can
be mitigated by optimization in the design of accelerator and detector.

\subsection{Recovery of Bremsstrahlung and FSR Photons \label{sub:Recovery-of-Bremsstrahlung}}

\begin{figure}[H]
\begin{centering}
\includegraphics[scale=0.35]{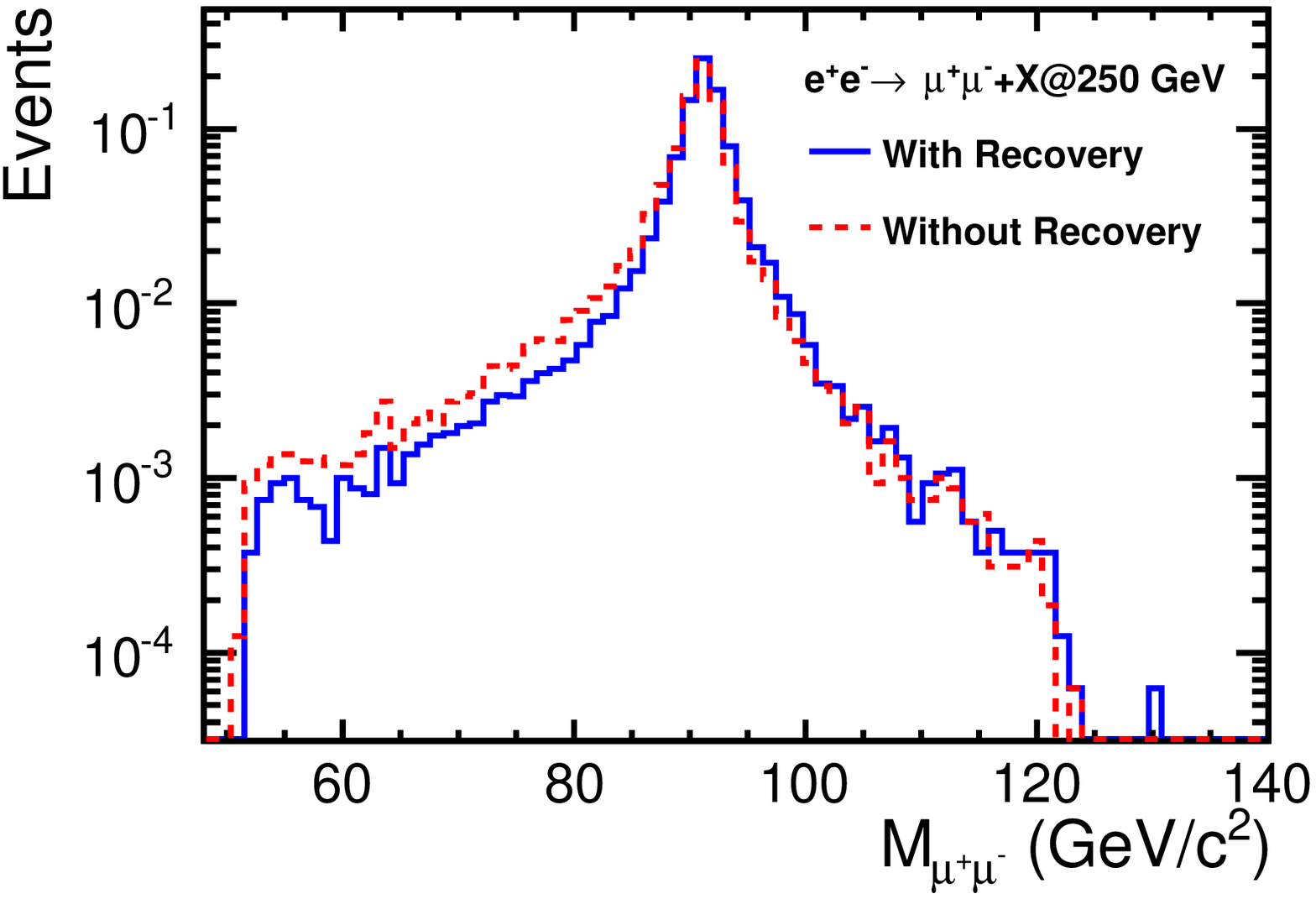} \includegraphics[scale=0.35]{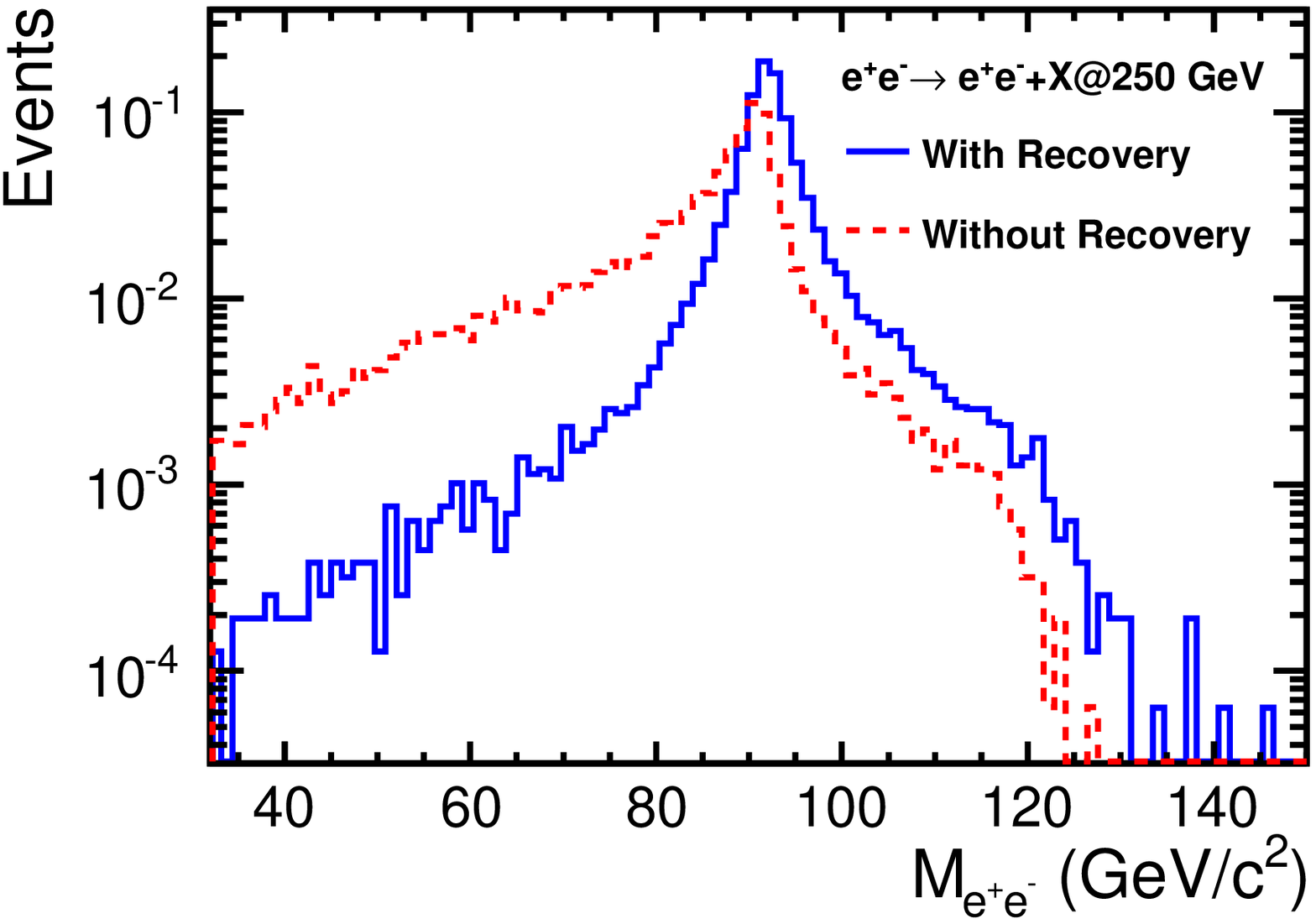}
\par\end{centering}

\centering{}\includegraphics[scale=0.35]{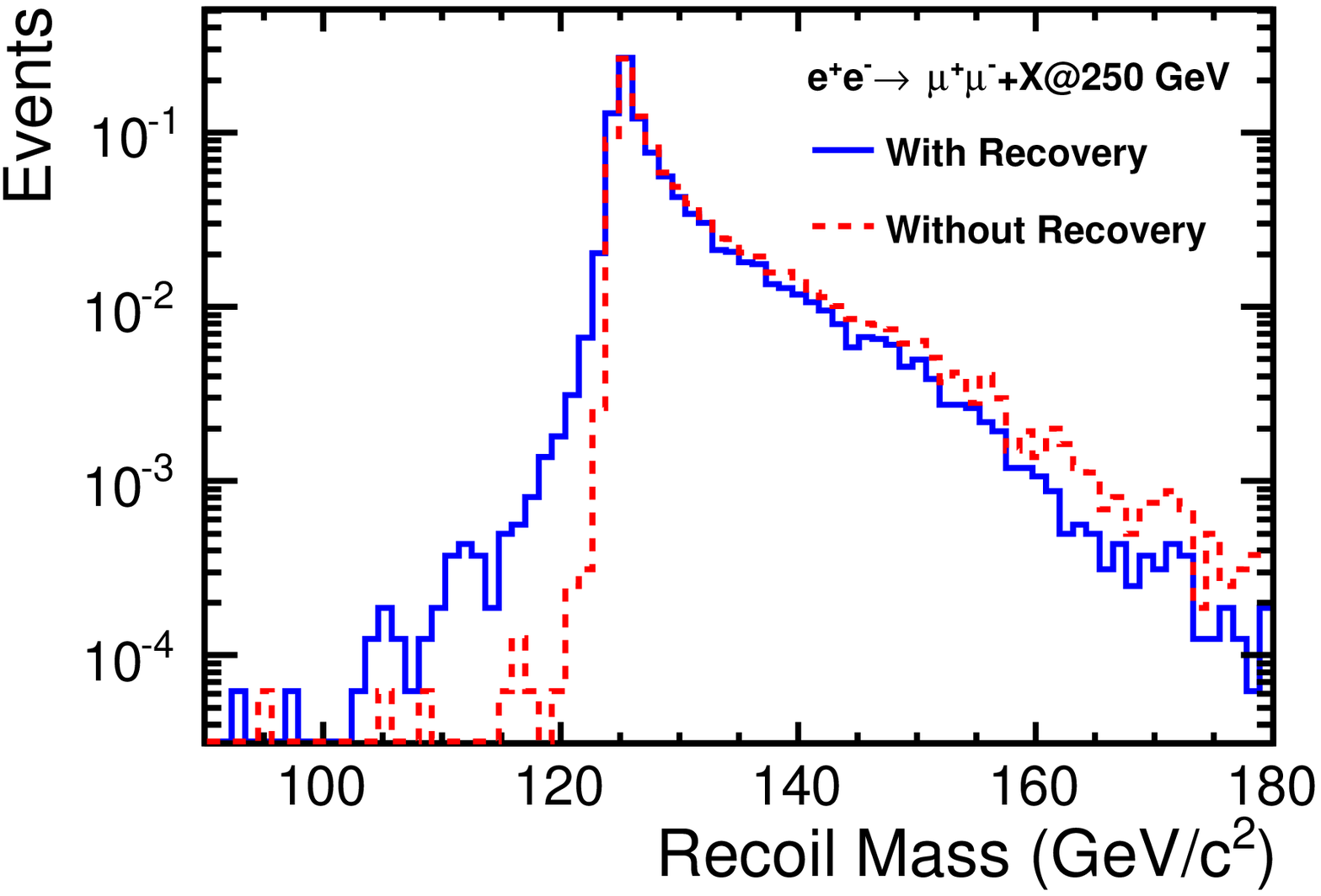} \includegraphics[scale=0.35]{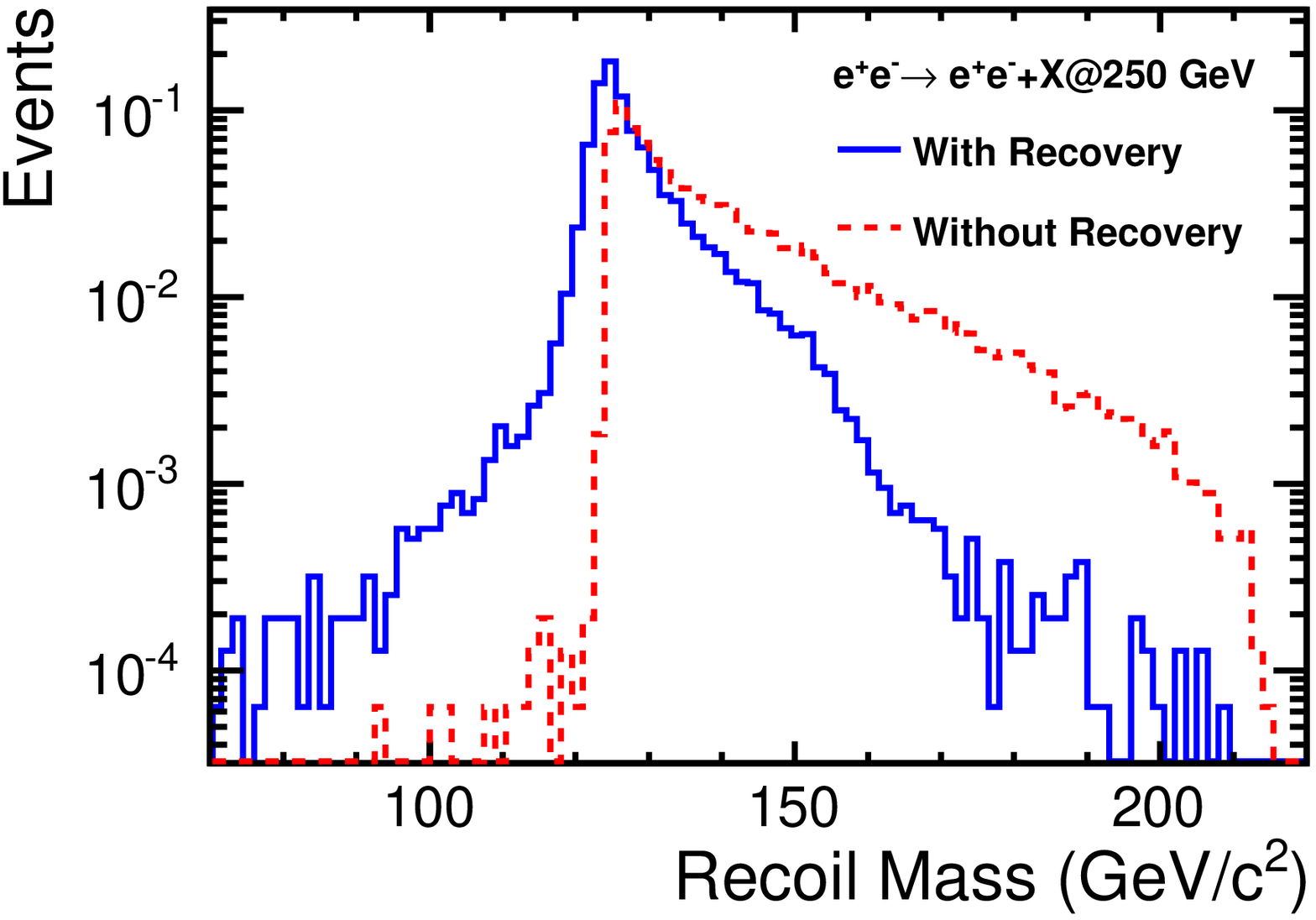}\protect\caption{Comparison of the $M_{\mathrm{l^{+}l^{-}}}$ (top) and $M_{\mathrm{rec}}$
(bottom) spectra between the cases with (blue) and without (red) bremsstrahlung/FSR
recovery for $\sqrt{s}$ = 250 GeV. The left and right shows the $\mathrm{\mu{}^{+}\mathit{\mu}^{-}H}$
and $\mathrm{e^{+}e^{-}H}$ channels, respectively. \label{Recovery}}
\end{figure}

The bremsstrahlung and FSR of the final state leptons degrade measurement
precision of $\sigma_{\mathrm{ZH}}$ and $M_{\mathrm{H}}$, particularly
for the $\mathrm{e^{+}e^{-}H}$ channel, which has a broader peak
and longer tail to lower values than the $\mathrm{\mu{}^{+}\mathit{\mu}^{-}H}$
channel. The recovery of bremsstrahlung and FSR photons is implemented
for both $\mathrm{\mu{}^{+}\mathit{\mu}^{-}H}$ and $\mathrm{e^{+}e^{-}H}$
channels. A bremsstrahlung/FSR photon is identified using its polar
angle with respect to the final state lepton; if the cosine of the
polar angle exceeds 0.99, the photon four momentum is combined with
that of the lepton. Figure \ref{Recovery} compares the reconstructed
$M_{\mathrm{l^{+}l^{-}}}$ and $M_{\mathrm{rec}}$ spectra before
(dotted line) and after (solid line) bremsstrahlung/FSR recovery for
$\sqrt{s}$=250 GeV. It can be seen that the recovery process pushes
the events at the lower end of the $M_{\mathrm{l^{+}l^{-}}}$ spectrum
(corresponding to the tail in the higher region of the $M_{\mathrm{rec}}$
spectrum) back to the peak. In the case of the $\mathrm{e^{+}e^{-}H}$
channel, the precision of $\sigma_{\mathrm{ZH}}$ could become degraded
by more than 50\%, and $M_{\mathrm{H}}$ by more than 20\% without
the recovery process. The change in the $\mathrm{\mu{}^{+}\mathit{\mu}^{-}H}$
channel is negligible.

\subsection{Background rejection\label{sub:Background-Rejection}}

\begin{figure}[H]
\begin{centering}
\includegraphics[scale=0.35]{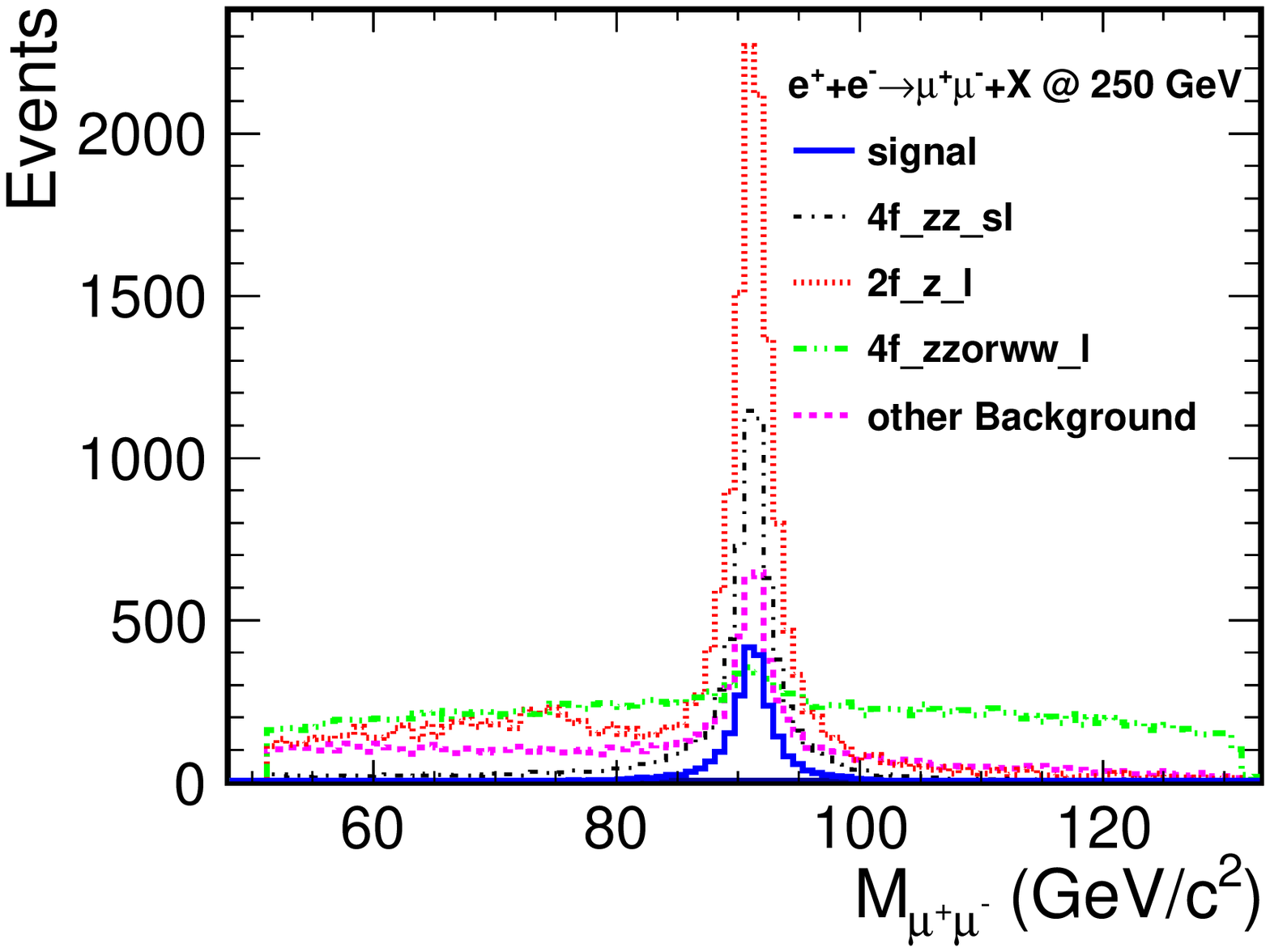}\includegraphics[scale=0.35]{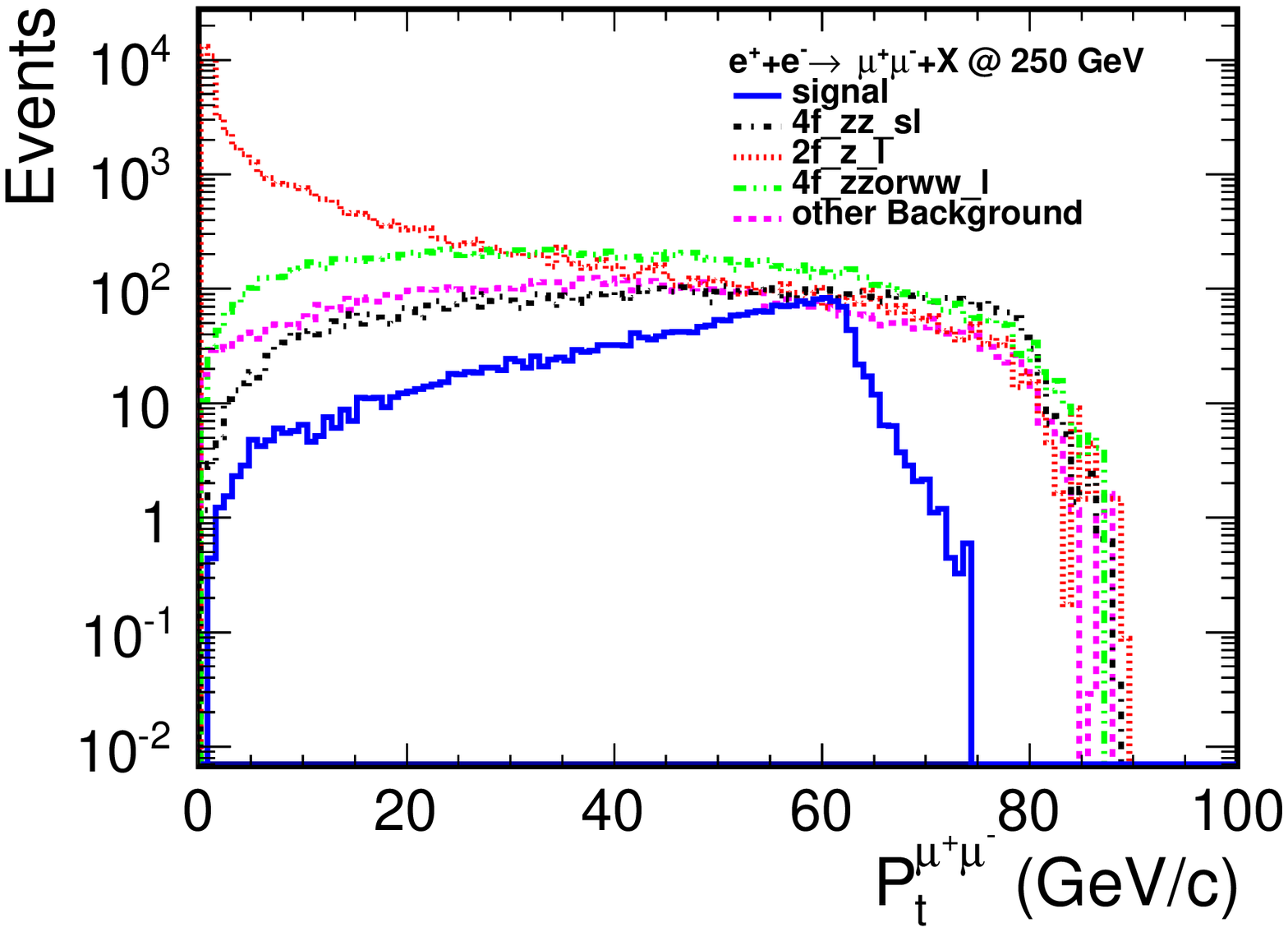}
\par\end{centering}

\begin{centering}
\includegraphics[scale=0.35]{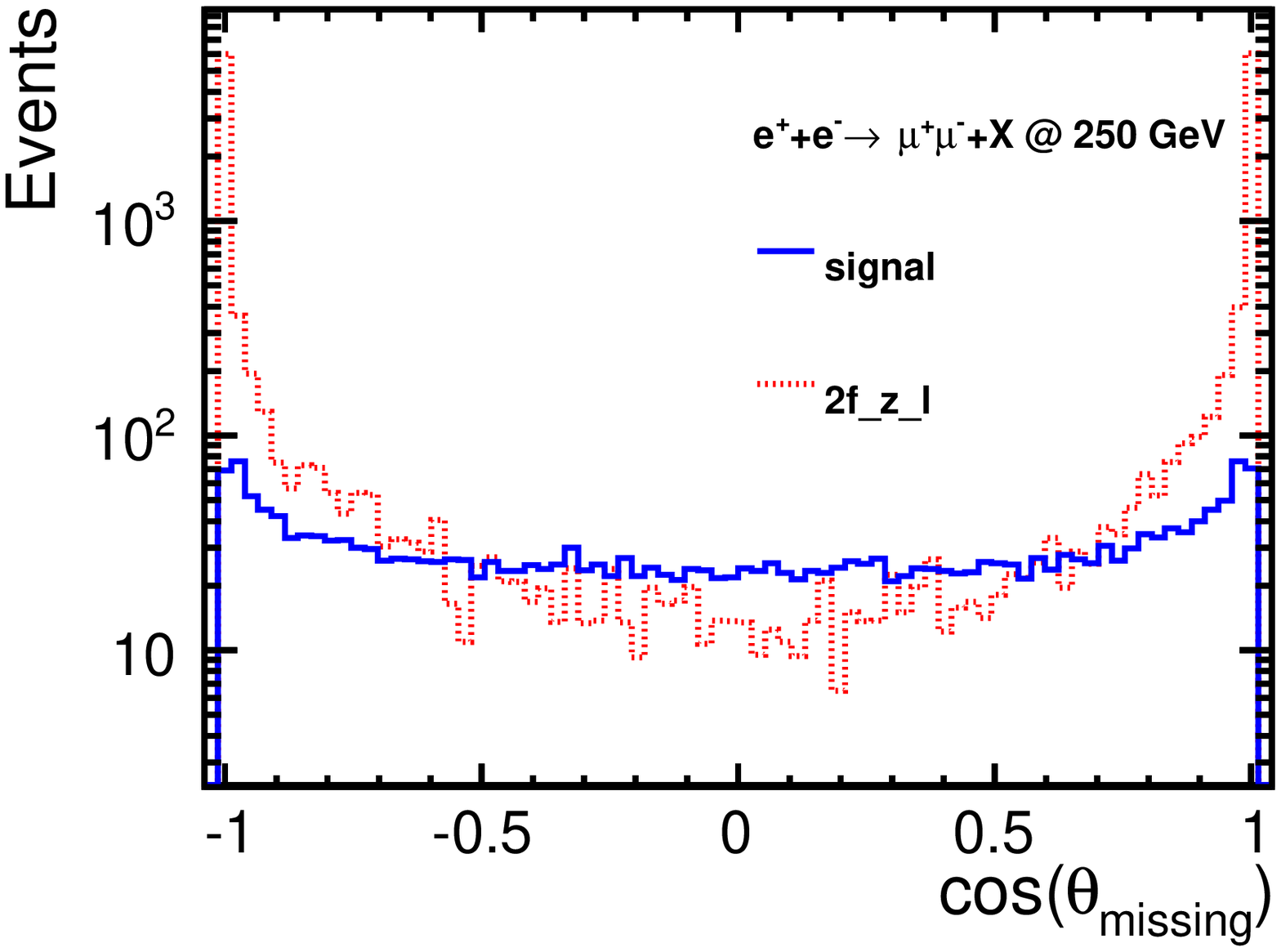}\includegraphics[scale=0.35]{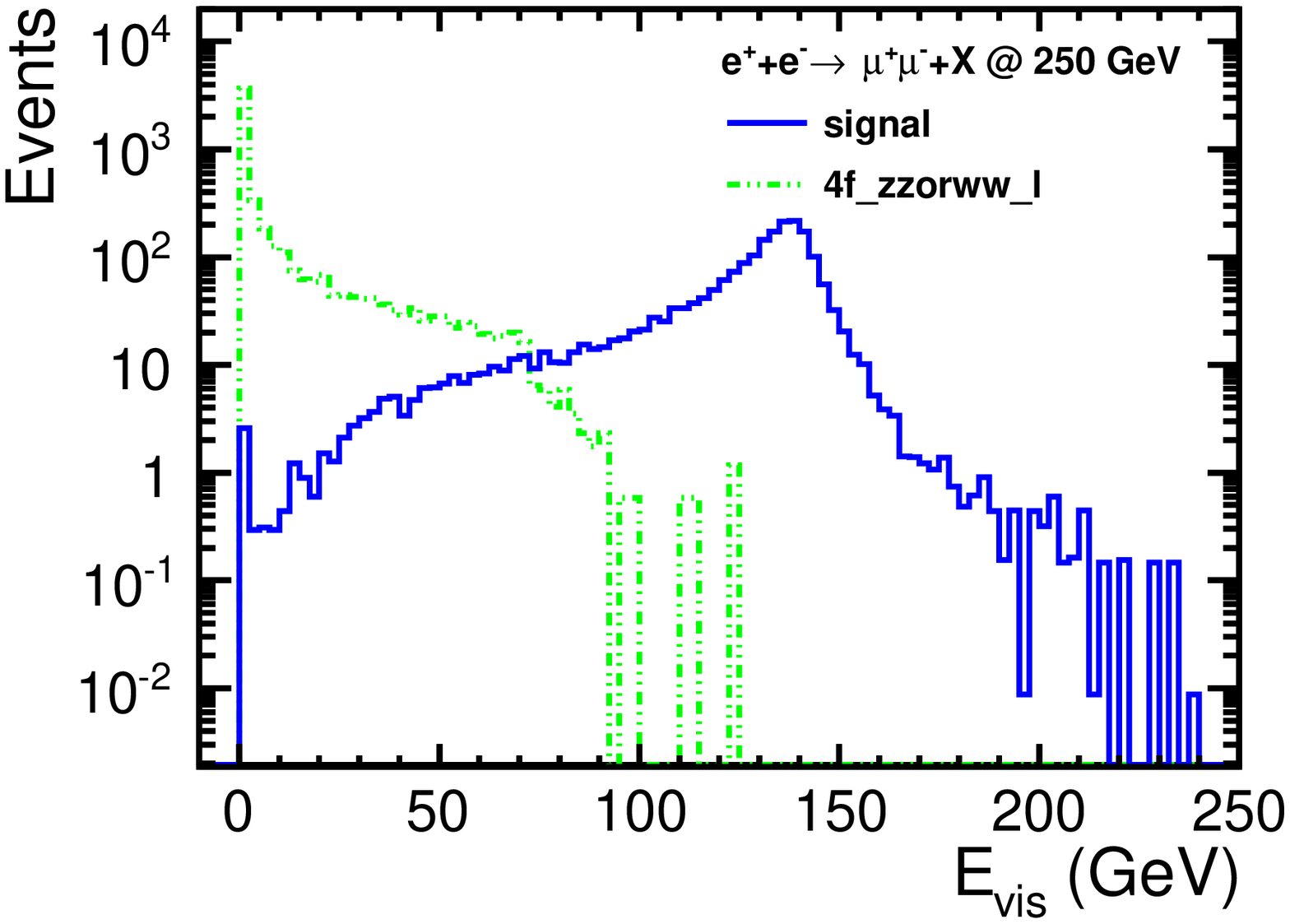}
\par\end{centering}

\centering{}\protect\caption{(top left) The $M_{\mu^{+}\mu^{-}}$ distributions of signal and the
major background processes, after a loose precut on $M_{\mathrm{rec}}$.
(top right) The $p_{\mathrm{\mathrm{T}}}^{\mu^{+}\mu^{-}}$ distributions
of signal and the major background processes, after a loose precut
on $M_{\mathrm{rec}}$ and a cut on $M_{\mu^{+}\mu^{-}}$. (bottom
left) The $\mathrm{cos\left(\theta_{missing}\right)}$ distributions
of signal and 2-fermion background, after a loose precut on $M_{\mathrm{rec}}$
and cuts have been applied on $M_{\mu^{+}\mu^{-}}$ and $p_{\mathrm{\mathrm{T}}}^{\mu^{+}\mu^{-}}$.
(bottom right) The distributions of $E_{\mathrm{vis}}$ (after excluding
the dilepton energy) of the signal and the 4f\_zzorww\_l processes,
after a loose precut on $M_{\mathrm{rec}}$ and cuts have been applied
on $M_{\mu^{+}\mu^{-}}$, $p_{\mathrm{\mathrm{T}}}^{\mu^{+}\mu^{-}}$.,
$\mathrm{cos\left(\theta_{missing}\right)}$, and the BDT response
of the MVA analysis. \label{fig:-Minv}}
\end{figure}

\begin{figure}[H]
\centering{}\includegraphics[scale=0.4]{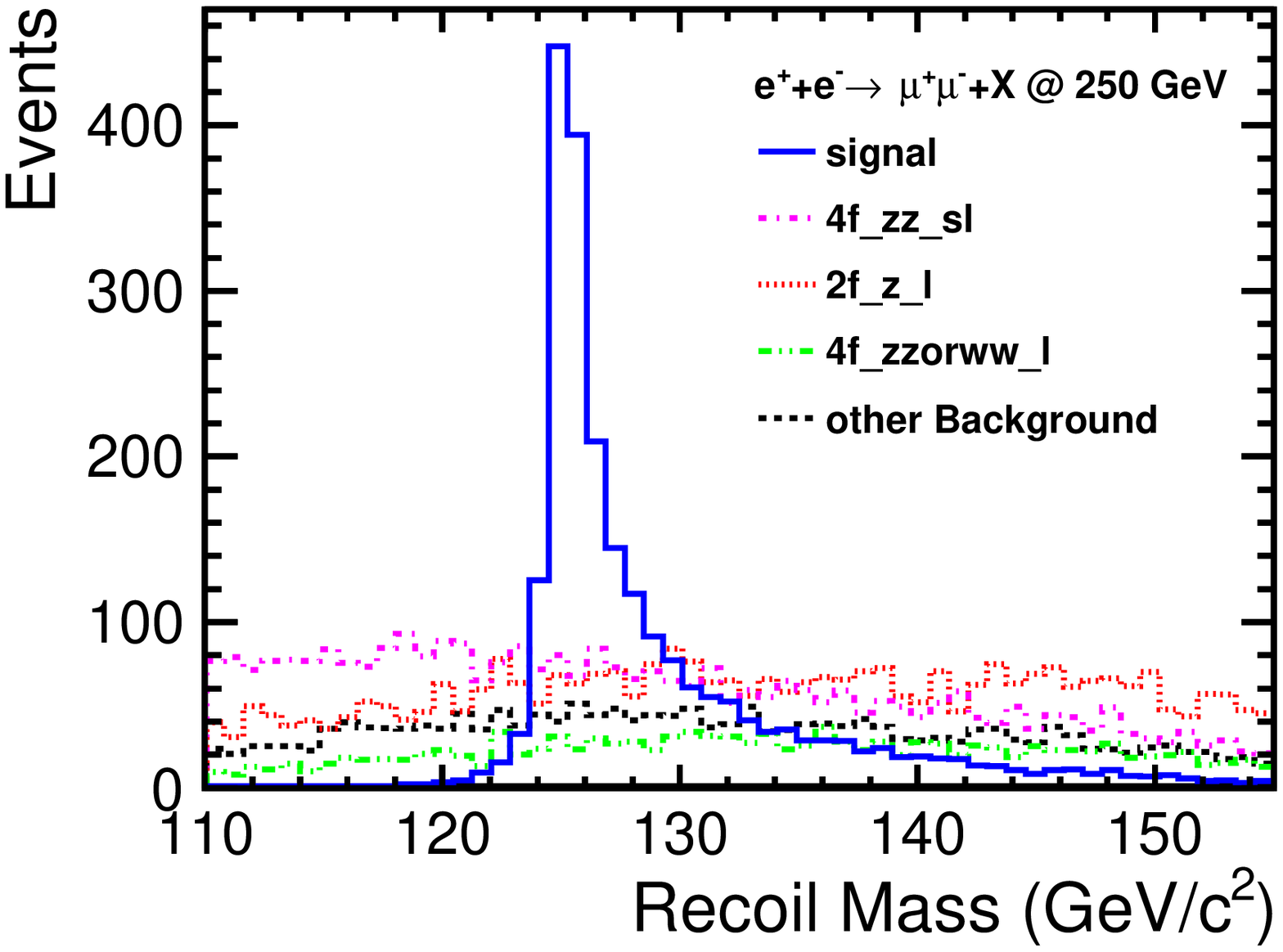} \includegraphics[scale=0.4]{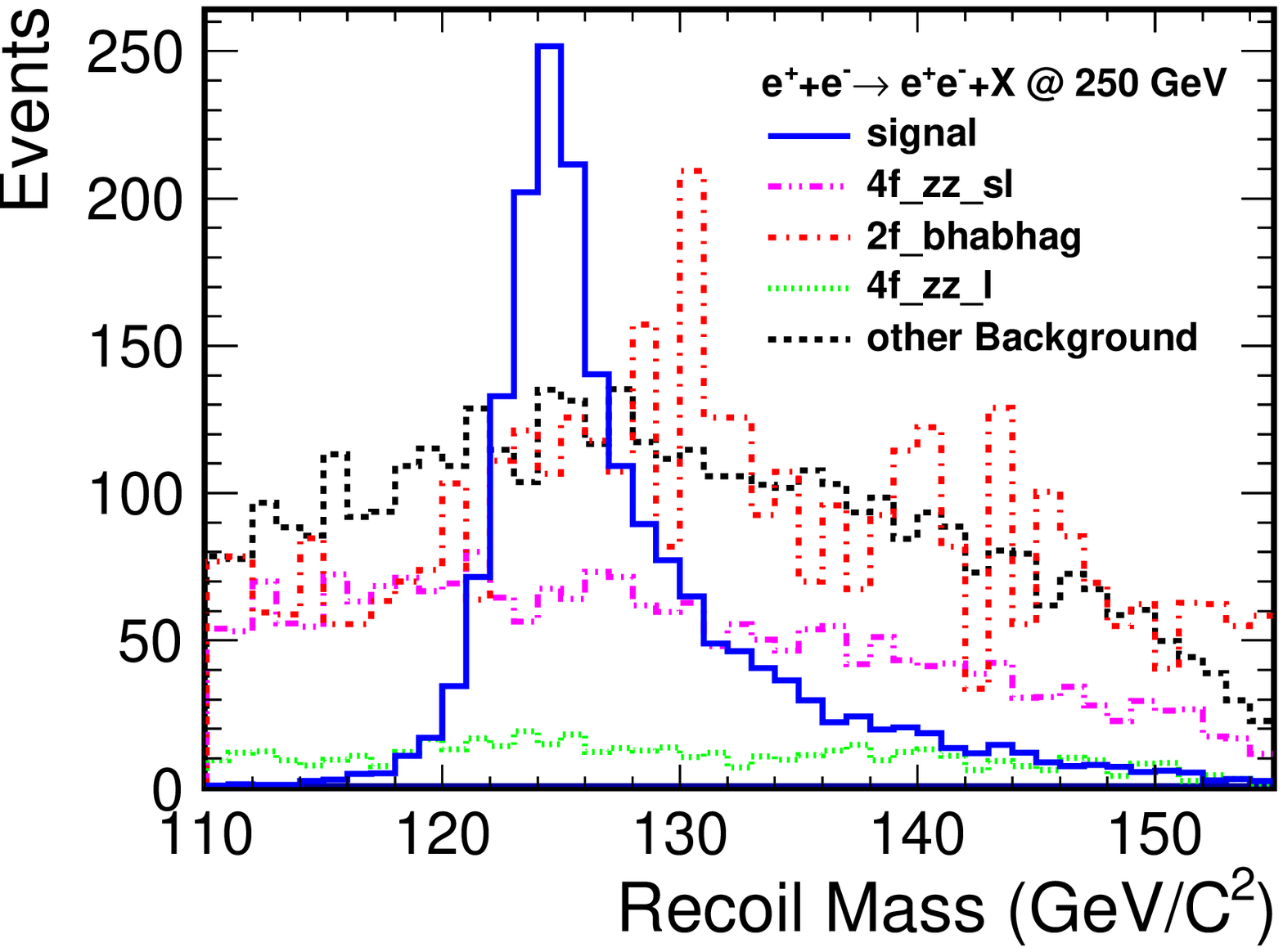}\protect\caption{The histograms of the recoil mass of the signal and the major residual
background processes left in a wide window around the signal $M_{\mathrm{rec}}$
peak, shown here for the $\mathrm{\mu{}^{+}\mathit{\mu}^{-}H}$ (left)
and $\mathrm{e{}^{+}\mathrm{\mathrm{\mathit{\mathrm{e}}^{-}}}H}$
(right) channels at $\sqrt{s}$=250 GeV, after all cuts described
in the main text have been applied. \label{fig:stk} }
\end{figure}

After the signal selection process, background events are rejected
by applying cuts on various kinematic properties. While the cut values
are adjusted for each center-of-mass energy, the overall strategies
are similar. Unless specified otherwise, the plots in this section
are made using the standard samples of the $\mathrm{\mu{}^{+}\mathit{\mu}^{-}H}$
channel and $\mathrm{\mathrm{e_{L}^{-}}\mathit{\mathrm{e_{R}^{+}}}}$
at $\sqrt{s}$=250 GeV, and are normalized to the assumed integrated
luminosities, cross sections, and polarizations (see Section \ref{sub:Signal-and-Background}).
In these plots, 4f\_zz\_l(sl) represents background with ZZ intermediate
states and two pairs of $\mu\mu$ / $\tau\tau$ (a pair of $\mu\mu$
/ $\tau\tau$ and a pair of quarks), 2f\_z\_l and 2f\_bhabhag represents
background with final states of $\mu\mu$ / $\tau\tau$ and $\mathrm{ee}$,
respectively, and 4f\_zzorww\_l represents background with $\mu\mu\nu\nu$
or $\tau\tau\nu\nu$ as the final state. First, a loose precut on
$M_{\mathrm{rec}}$ is applied as $M_{\mathrm{rec}}\in${[}100, 300{]}
GeV. Then the following cuts are applied in this order:
\begin{itemize}
\item since the invariant mass $M_{\mathrm{l^{+}l^{-}}}$ ($\mathrm{l}$
= $\mathrm{e}$ or $\mu$) of the dilepton system should be close
to the Z boson mass for the Higgsstrahlung process, a criterion is
imposed as $M_{\mathrm{l^{+}l^{-}}}\in${[}73, 120{]} GeV. The top
left plot in Figure \ref{fig:-Minv} compares the $M_{\mathrm{l^{+}l^{-}}}$
of signal and major background processes. 
\item for the signal, the transverse dilepton momentum $p_{\mathrm{T}}^{\mathrm{l^{+}l^{-}}}$
should peak at a certain value determined by kinematics. In contrast,
the $p_{\mathrm{T}}^{\mathrm{l^{+}l^{-}}}$ of 2-fermion background
peaks towards small values. This motivates the cut $p_{\mathrm{T}}^{\mathrm{l^{+}l^{-}}}$>
10 GeV. In addition, an upper limit on $p_{\mathrm{T}}^{\mathrm{l^{+}l^{-}}}$
is imposed to suppress background processes whose $p_{\mathrm{T}}^{\mathrm{l^{+}l^{-}}}$
extend to large values. The top right plot in Figure \ref{fig:-Minv}
compares the $p_{\mathrm{T}}^{\mathrm{l^{+}l^{-}}}$ of the signal
and major background processes. 
\item $\theta_{\mathrm{missing}}$, the polar angle of the missing momentum,
discriminates against events which are unbalanced in longitudinal
momentum, in particular those 2-fermion events in which ISR emitted
approximately collinear with the incoming beams escapes detection
in the beam pipe. The bottom left plot in Figure \ref{fig:-Minv}
shows the distribution of $\mathrm{cos\left(\theta_{missing}\right)}$
between the signal and major background processes. A cut is made at
$\mathrm{\left|cos\left(\theta_{missing}\right)\right|}<0.98$, which
cuts 2f\_l background by approximately two thirds.
\item multi-variate cut: While the $p_{\mathrm{T}}^{\mathrm{l^{+}l^{-}}}$
and $\mathrm{cos\left(\theta_{missing}\right)}$ cuts are effective
for removing 2-fermion background, the signatures of 4-fermion backgrounds
are harder to distinguish from the Higgsstrahlung signal, especially
in the case of one of the dominant background processes $\mathrm{\mathrm{\mathrm{\mathrm{e^{+}\mathrm{e^{-}}}\rightarrow ZZ\rightarrow\mathit{\mathrm{llqq}}}}}$
($l$=$\mathrm{e}$ or $\mu$). Nevertheless, further rejection of
residual background events is achieved by a multi-variate (MVA) cut
based on the Boosted Decision Tree (BDT) method \cite{TMVA} using
a combination of the variables $M_{\mathrm{l^{+}l^{-}}}$, $\mathrm{cos\left(\theta_{Z}\right)}$,
$\mathrm{cos\left(\theta_{lep}\right)}$, $\mathrm{cos\left(\theta_{track,1}\right)}$
and $\mathrm{cos\left(\theta_{track,2}\right)}$. Here, $\theta_{\mathrm{Z}}$
is the polar angle of the Z boson, $\theta_{\mathrm{lep}}$ is the
angle between the leptons, and $\theta_{\mathrm{track,1,2}}$ is the
polar angle of each lepton track. The BDT response is calculated using
weights obtained from training samples consisting of simulated signal
and background events. The MVA cut is optimized for each channel to
maximize $\sigma_{\mathrm{ZH}}$ precision, and is very effective
for increasing signal significance. For example, in the case of the
$\mathrm{\mu{}^{+}\mathit{\mu}^{-}H}$ channel at $\sqrt{s}$=250
GeV, the number of background events is reduced by more than 35\%
by the MVA cut, whereas the loss of signal events is only about 5\%. 
\item recoil mass cut: $\sigma_{\mathrm{ZH}}$ and $M_{\mathrm{H}}$ are
obtained by fitting the $M_{\mathrm{rec}}$ spectrum within a wide
window around the signal $M_{\mathrm{rec}}$ peak. This is designated
to be $M_{\mathrm{rec}}\in${[}110, 155{]} GeV for $\sqrt{s}$=250
GeV, {[}100, 200{]} GeV for $\sqrt{s}$=350 GeV, and {[}100, 250{]}
GeV for $\sqrt{s}$= 500 GeV.
\item visible energy cut: $E_{\mathrm{vis}}$, defined as the visible energy
excluding that from the isolated lepton pair, is required to be above
a certain value (10 GeV for $\sqrt{s}$=250 and 350 GeV and 25 GeV
for $\sqrt{s}$=500 GeV) in order to suppress one of the dominant
residual backgrounds which has $\mathrm{ll\nu\nu}$ ($\mathrm{l}$
= $\mathrm{e}$ or $\mu$) in the final state. The bottom right plot
in Figure \ref{fig:-Minv} compares the distributions of $E_{\mathrm{vis}}$
between signal and $\mathrm{ll\nu\nu}$ background. For example, in
the case of the $\mathrm{\mu{}^{+}\mathit{\mu}^{-}H}$ channel at
$\sqrt{s}$=250 GeV, the $\mathrm{ll\nu\nu}$ background occupies
about 30\% and 10\% of all residual backgrounds without and with the
$E_{\mathrm{vis}}$ cut, respectively. This reduces background events
by 30-50\% and further improves the precision on $\sigma_{\mathrm{ZH}}$
and $M_{\mathrm{H}}$ by 10-15\% in the case of the $\mathrm{\mathrm{e_{L}^{-}}\mathit{\mathrm{e_{R}^{+}}}}$
polarization\cite{Mode dependence}, where the contribution of $\mathrm{ll\nu\nu}$
background with WW intermediate states is significant. Although the
$E_{\mathrm{vis}}$ cut also excludes signal events in which the Higgs
boson decays invisibly, Higgs decay model independence is maintained
by combining the results obtained from this analysis with a dedicated
analysis for invisible Higgs decays \cite{invisible,invisibleHad}.
This is explained by the fact that the $\mathrm{ZH}$ cross section
for the SM Higgs boson can be expressed as $\sigma_{\mathrm{ZH}}=\sigma_{\mathrm{ZH,vis}}+\sigma_{\mathrm{\mathrm{ZH,inv}is}}$,
where $\sigma_{\mathrm{ZH,vis}}$ and $\sigma_{\mathrm{ZH,invis}}$
, which are the cross sections of the visible and invisible decay
events, respectively, can both be measured individually and model
independently.
\end{itemize}
For the case of the $\mathrm{\mu{}^{+}\mathit{\mu}^{-}H}$ channel
at $\sqrt{s}$= 250 GeV, Table \ref{tab:cut eff} shows the number
of remaining signal and background, signal efficiency and significance
after each cut. Similar outcomes are seen for $\sqrt{s}$=350 and
500 GeV since similar data selection methods are used. Figure \ref{fig:stk}
shows distributions of the $M_{\mathrm{rec}}$ of signal and major
residual background processes for $\sqrt{s}$=250 GeV. The major residual
backgrounds are 4f\_sl and 2f\_l defined in Section \ref{sub:Signal-and-Background}. 

\begin{table}[H]
\begin{centering}
\begin{tabular}{|c|c|c|c|c|c|c|c|}
\hline 
$\int\mathcal{L}dt$  & $\mathrm{\mu{}^{+}\mathit{\mu}^{-}H}$  & signal & signal &  &  &  & total\tabularnewline
 250 $\mathrm{fb^{-1}}$ & $\mathrm{e_{L}^{-}e_{R}^{+}}$ & efficiency & significance & 2f\_l & 4f\_l & 4f\_sl & background\tabularnewline
\hline 
\hline 
no cut & 2603 & 100\% & 0.42 & 9.54$\times10^{6}$ & 3.15$\times10^{6}$ & 4.98$\times10^{6}$ & 1.98$\times10^{7}$\tabularnewline
\hline 
Lepton ID+Precut & 2439 & 93.70\% & 7.46 & 61675 & 34451 & 8218 & 104344\tabularnewline
\hline 
$M_{\mathrm{l^{+}l^{-}}}\in${[}73, 120{]} GeV & 2382 & 91.51\% & 8.09 & 54352 & 22543 & 7446 & 84341\tabularnewline
\hline 
$p_{\mathrm{T}}^{\mathrm{l^{+}l^{-}}}\in${[}10, 70{]} GeV & 2335 & 89.70\% & 11.17 & 15429 & 19648 & 6245 & 41322\tabularnewline
\hline 
$\mathrm{\left|cos\mathit{\theta_{\mathrm{missing}}}\right|}$< 0.98 & 2335 & 89.70\% & 12.71 & 5594 & 19539 & 6245 & 31378\tabularnewline
\hline 
MVA & 2310 & 88.74\% & 15.03 & 4195 & 12530 & 4586 & 21311\tabularnewline
\hline 
$M_{\mathrm{rec}}\in${[}110, 155{]} GeV & 2296 & 88.21\% & 16.37 & 3522 & 10423 & 3433 & 17378\tabularnewline
\hline 
$E_{\mathrm{vis}}$ > 10 GeV & 2293 & 88.09\% & 20.94 & 3261 & 2999 & 3433 & 9694\tabularnewline
\hline 
\end{tabular}
\par\end{centering}

\protect\caption{The number of events left after each cut for the $\mathrm{\mu{}^{+}\mathit{\mu}^{-}H}$
channel and $\mathrm{e_{L}^{-}e_{R}^{+}}$ at $\sqrt{s}$=250 GeV.
Also given are the efficiency and signal significance (defined as
$\frac{N_{S}}{\sqrt{N_{S}+N_{B}}}$ ,where $N_{S(B)}$ is the number
of signal (background)) for the Higgsstrahlung signal. Precut represents
the loose cut $M_{\mathrm{rec}}\in${[}100, 300{]} GeV.\label{tab:cut eff}}
\end{table}

\section{DEMONSTRATION OF HIGGS DECAY MODE INDEPENDENCE \label{sec:Extraction}}

In the recoil method, $\sigma_{\mathrm{ZH}}$ is measured without
any explicit assumption regarding Higgs decay modes. This section
demonstrates that the $\sigma_{\mathrm{ZH}}$ measured using the methods
described in \cite{Recoil} based on the data selection in Section
\ref{sec:Event-Selection} does not depend on the underlying model
which determines the Higgs decay modes and their branching ratios.
As can be understood from Equation \ref{eq:xsec}, the key question
here is whether the $\sigma_{\mathrm{ZH}}$ extracted using the measured
number of signal events ($N_{\mathrm{S}}$) and the signal selection
efficiency ($\varepsilon_{\mathrm{S}}$) from the Monte Carlo samples
would be biased when the Higgs boson decays differently from that
assumed in the samples.

\subsection{General strategies towards model independence}

First we introduce the general strategies towards a model independent
$\sigma_{\mathrm{ZH}}$ measurement. The direct observable $N_{\mathrm{S}}$
can be parameterised as 

\begin{equation}
N_{\mathrm{S}}=\underset{i}{\varSigma}N_{i}=\underset{i}{\varSigma}\sigma_{\mathrm{ZH}}R_{l}LB_{i}\varepsilon_{i}\,,\label{eq:BR1}
\end{equation}
where the summation goes through all Higgs decay modes. $N_{\mathrm{i}}$,
$B_{\mathrm{i}}$, and $\varepsilon_{\mathrm{i}}$ are the the number
of signal events, branching ratio and selection efficiency of Higgs
decay mode $i$, respectively. $L$ is the integrated luminosity,
and $R_{l}$ is the branching ratio of $\mathrm{Z\rightarrow l^{+}l^{-}}$.
If the signal efficiency equals to the same $\varepsilon$ for all
decay modes, Equation \ref{eq:BR1} becomes

\begin{equation}
N_{\mathrm{S}}=\sigma_{\mathrm{ZH}}R_{l}L\varepsilon\underset{i}{\varSigma}B_{i}\,.\label{eq:BR1-1}
\end{equation}
Since $\underset{i}{\varSigma}B_{i}=1$ stands in any case, $\sigma_{\mathrm{ZH}}$
can be extracted without assumptions on decay modes or branching ratios
as 

\begin{equation}
\sigma_{\mathrm{ZH}}=\frac{N_{\mathrm{S}}}{R_{l}L\varepsilon}\,,\label{eq:BR2-2}
\end{equation}
This is the ideal case which guarantees model independence. On the
other hand, if there exist discrepancies between the signal efficiencies
of each mode, $\sigma_{\mathrm{ZH}}$ has to be extracted as

\begin{equation}
\sigma_{\mathrm{ZH}}=\frac{N_{\mathrm{S}}}{R_{l}L\underset{i}{\varSigma}B_{i}\varepsilon_{i}}\equiv\frac{N_{\mathrm{S}}}{R_{l}L\overline{\varepsilon}}\,,\label{eq:BR2}
\end{equation}
where $\overline{\varepsilon}=\underset{i}{\varSigma}B_{i}\varepsilon_{i}$
is the expected efficiency for all decay modes. In this case, the
bias on $\sigma_{\mathrm{ZH}}$ depends on the determination of $\overline{\varepsilon}$.
This is discussed as follows in terms of three possible scenarios
of our knowledge of Higgs decay at the time of $\sigma_{\mathrm{ZH}}$
measurement.
\begin{itemize}
\item scenario A: all Higgs decay modes and the corresponding $B_{i}$ for
each mode are known. In this rather unlikely case, $\overline{\varepsilon}$
can be determined simply by summing up over all modes, leaving no
question of model independence.
\item scenario B: $B_{i}$ is completely unknown for every mode. We would
examine the discrepancy in $\epsilon_{i}$ by investigating as many
modes as possible, and retrieve the maximum and minimum of $\epsilon_{i}$
as $\varepsilon_{\mathrm{min}}\leq\epsilon_{i}\leq\varepsilon_{\mathrm{max}}$,
from which $\overline{\varepsilon}$ can be constrained as $\varepsilon_{\mathrm{min}}\underset{i}{\Sigma}B_{i}\leq\overline{\varepsilon}\leq\varepsilon_{\mathrm{max}}\underset{i}{\Sigma}B_{i}$.
Given that $\underset{i}{\varSigma}B_{i}=1$, this can be rewritten
as $\varepsilon_{\mathrm{min}}\leq\overline{\varepsilon}\leq\varepsilon_{\mathrm{max}}$.
Then from Equation \ref{eq:BR2}, $\sigma_{\mathrm{ZH}}$ can be constrained
as
\end{itemize}
\begin{equation}
\frac{N_{\mathrm{S}}}{R_{l}L\varepsilon_{\mathrm{max}}}\leq\sigma_{\mathrm{ZH}}\leq\frac{N_{\mathrm{S}}}{R_{l}L\varepsilon_{\mathrm{min}}}\,,\label{eq:BR2-1}
\end{equation}

which indicates that the possible relative bias on $\sigma_{\mathrm{ZH}}$
can be estimated as $\frac{\varepsilon_{\mathrm{\mathrm{max}}}-\varepsilon_{\mathrm{min}}}{\varepsilon_{\mathrm{\mathrm{max}}}+\varepsilon_{\mathrm{min}}}$.
This scenario is based on a considerably conservative assumption.
\begin{itemize}
\item scenario C: $B_{i}$ is known for some of the decay modes. Here, it
is assumed that the decay modes $i$ = 1 to $n$ with a total branching
ratio of $B_{0}=\stackrel[i=1]{n}{\Sigma}B_{i}$ are known, and that
the modes from $i$ = $n+1$ with a total branching ratio of $B_{x}=\underset{i=n+1}{\Sigma}B_{i}$
are unknown. In this case, we would know the efficiency of the known
modes as $\varepsilon_{0}=\frac{\stackrel[i=1]{n}{\Sigma}B_{i}\varepsilon_{i}}{B_{0}}$.
Meanwhile the efficiency for each unknown mode can be expressed as
$\varepsilon_{i}=\varepsilon_{0}+\delta\varepsilon_{i}$, where $\delta\varepsilon_{i}$
is the deviation in efficiency for each unknown mode $i$ from $\varepsilon_{0}$.
We can then write $\overline{\varepsilon}$ as
\end{itemize}
\begin{equation}
\overline{\varepsilon}=\stackrel[i=1]{n}{\Sigma}B_{i}\varepsilon_{i}+\underset{i=n+1}{\Sigma}B_{i}\varepsilon_{i}=B_{0}\varepsilon_{0}+B_{x}\varepsilon_{0}+\underset{i=n+1}{\Sigma}B_{i}\delta\varepsilon_{i}=\varepsilon_{0}+\underset{i=n+1}{\Sigma}B_{i}\delta\varepsilon_{i}\,.
\end{equation}

The relative bias for $\overline{\varepsilon}$ and hence for $\sigma_{\mathrm{ZH}}$
is a combination of the contribution from the unknown modes and the
known modes. The contribution from the unknown modes is derived as

\begin{equation}
\frac{\Delta\sigma_{\mathrm{ZH}}}{\sigma_{\mathrm{ZH}}}=\frac{\Delta\overline{\varepsilon}}{\overline{\epsilon}}<\underset{i=n+1}{\Sigma}B_{i}\frac{\delta\varepsilon_{\mathrm{max}}}{\varepsilon_{0}}=B_{x}\frac{\delta\varepsilon_{\mathrm{max}}}{\varepsilon_{0}}\,,\label{eq:BR3}
\end{equation}
where $\delta\varepsilon_{\mathrm{max}}$ is the maximum of $\left|\delta\varepsilon_{i}\right|$
for the unknown modes. As for the known modes, because $\overline{\varepsilon}=\stackrel[i=1]{n}{\Sigma}B_{i}\varepsilon_{i}=\stackrel[i=1]{n}{\Sigma}B_{i}\left(\varepsilon_{0}+\delta\varepsilon_{i}\right)$,
where $\delta\varepsilon_{i}\equiv\varepsilon_{i}-\varepsilon_{0}$
is the deviation in efficiency for each known mode, the uncertainty
due to a fluctuation in their branching ratios ($\Delta B_{i}$) can
be expressed as $\Delta\overline{\varepsilon}=\stackrel[i=1]{n}{\Sigma}\Delta B_{i}\varepsilon_{0}+\stackrel[i=1]{n}{\Sigma}\Delta B_{i}\delta\varepsilon_{i}=\stackrel[i=1]{n}{\Sigma}\Delta B_{i}\delta\varepsilon_{i}$.
Therefore the contribution from the known modes is derived as

\begin{equation}
\frac{\Delta\sigma_{\mathrm{ZH}}}{\sigma_{\mathrm{ZH}}}=\frac{\Delta\overline{\varepsilon}}{\overline{\epsilon}}=\sqrt{\stackrel[i=1]{n}{\Sigma}\Delta B_{i}^{2}\left(\frac{\varepsilon_{i}}{\varepsilon_{0}}-1\right)^{2}}\,.\label{eq:BR4}
\end{equation}

Scenario C is the most realistic as we will certainly have branching
ratio measurements from both the LHC and the ILC itself for a wide
range of Higgs decay modes. 

From the above formulation, it is apparent that the key to maintaining
model independence is to minimize the discrepancies in signal efficiency
between decay modes. The data selection methods in Section \ref{sec:Event-Selection}
are designed to satisfy this purpose while still achieving high precision
of $\sigma_{\mathrm{ZH}}$ and $M_{\mathrm{H}}$. To cover a large
number of Higgs decay modes and monitor their efficiencies, high statistics
signal samples ($\sim$ 40k events) are produced for each major SM
decay modes ($\mathrm{H\rightarrow\mathrm{bb}}$, cc, gg, $\tau\tau$,
$\mathrm{WW^{*}}$, $\mathrm{ZZ^{*}}$, $\gamma\gamma$, $\gamma\mathrm{Z}$),
and for each beam polarisation and center-of-mass energy, so that
the relative statistical error of each efficiency is below 0.2\% in
the end for any channel.

\subsection{Analysis strategies}

\subsubsection{Algorithms for lepton pairing\label{sub:LeptonPair}}

\begin{figure}[H]
\begin{centering}
\includegraphics[scale=0.4]{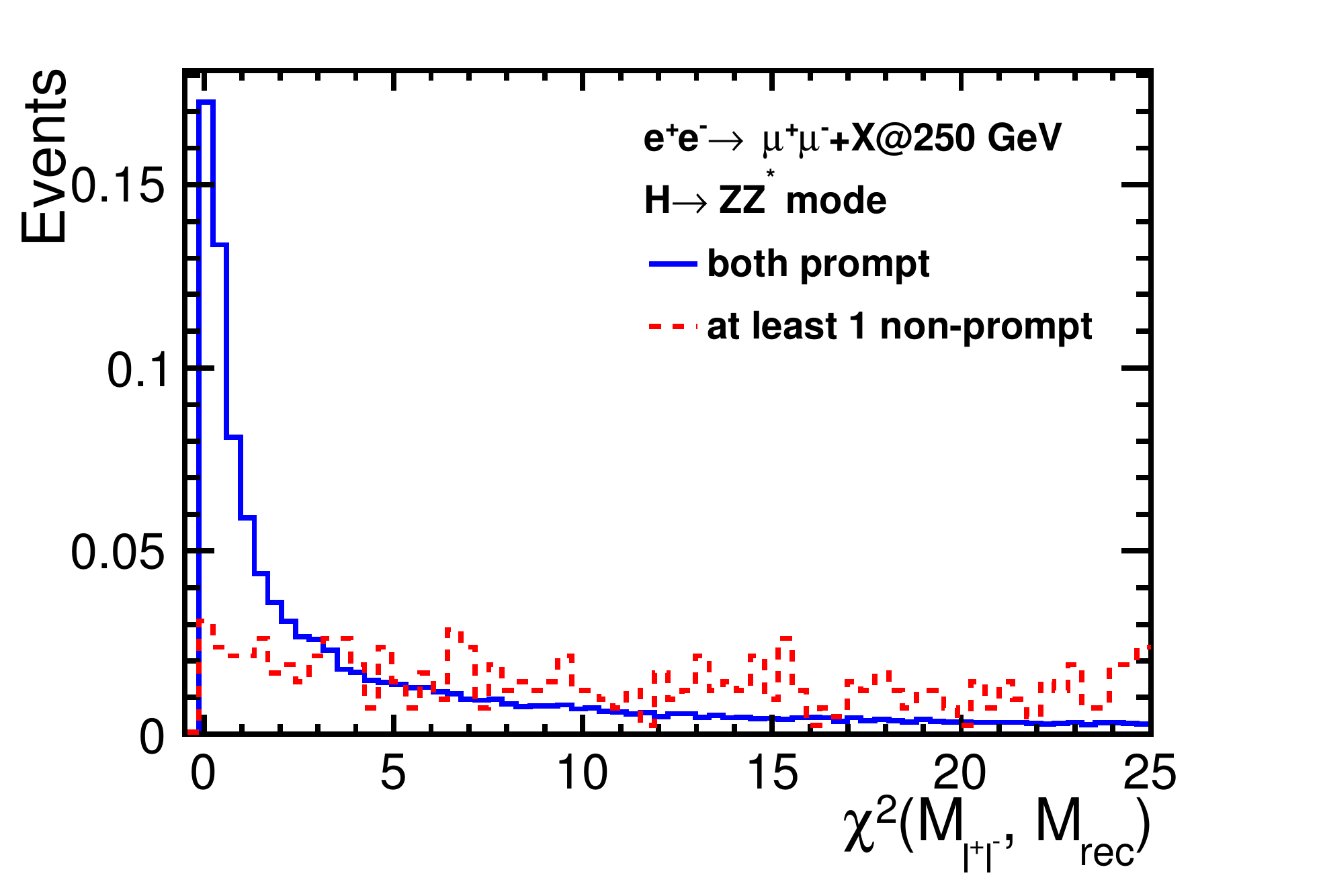}
\par\end{centering}

\protect\caption{The distribution of $\chi^{2}\left(M_{\mathrm{l^{+}l^{-}}},M_{\mathrm{r\mathrm{ec}}}\right)$
of the \textquotedblleft correct\textquotedblright{} and \textquotedblleft wrong\textquotedblright{}
lepton pairs in the $\mathrm{H\rightarrow ZZ^{*}}$ mode, shown for
the $\mathrm{\mu{}^{+}\mathit{\mu}^{-}H}$ channel and $\mathrm{e_{L}^{-}e_{R}^{+}}$
at $\sqrt{s}$ = 250 GeV. \label{fig:ResidualChi}}
\end{figure}

\begin{figure}[H]
\begin{centering}
\includegraphics[scale=0.3]{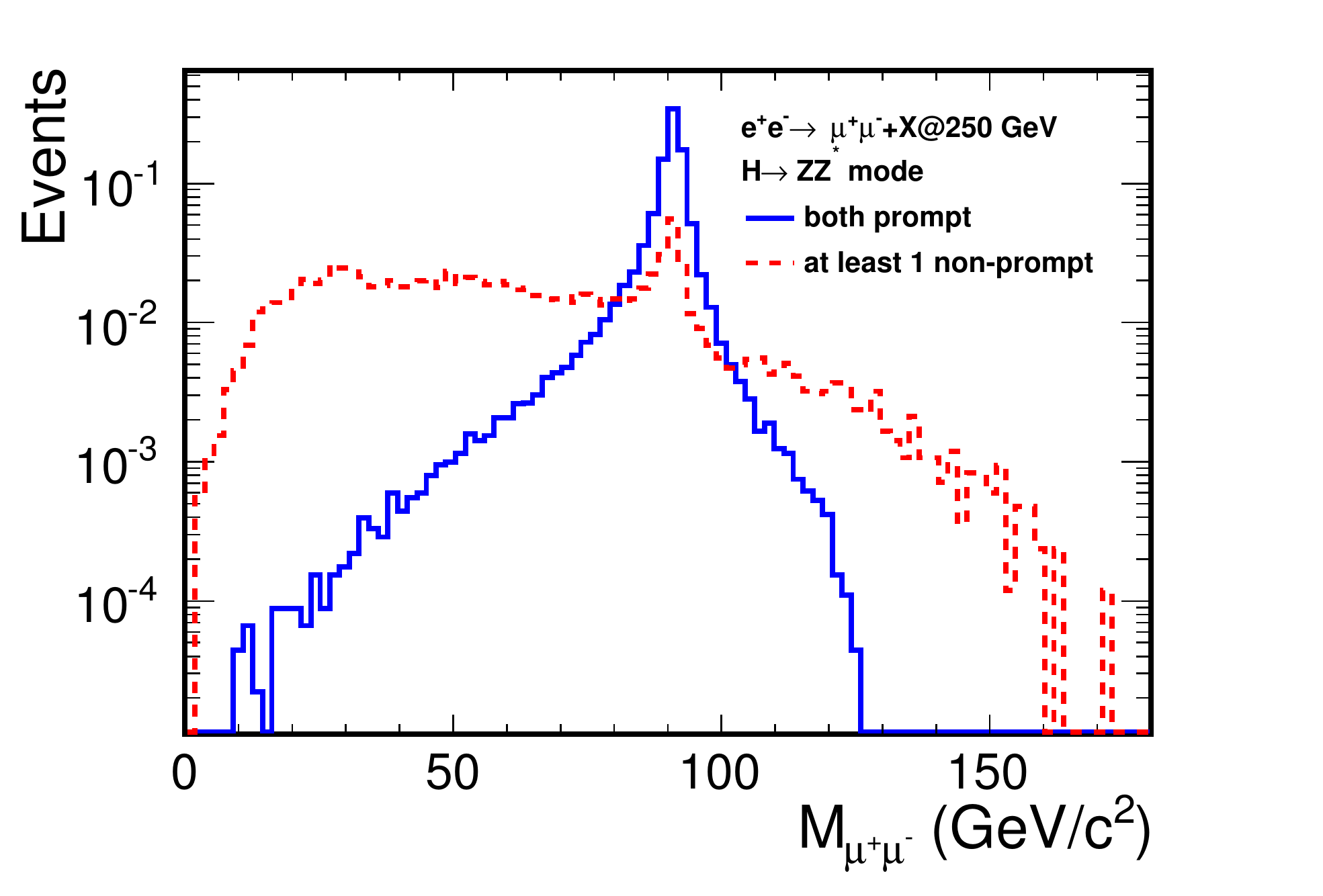}\includegraphics[scale=0.3]{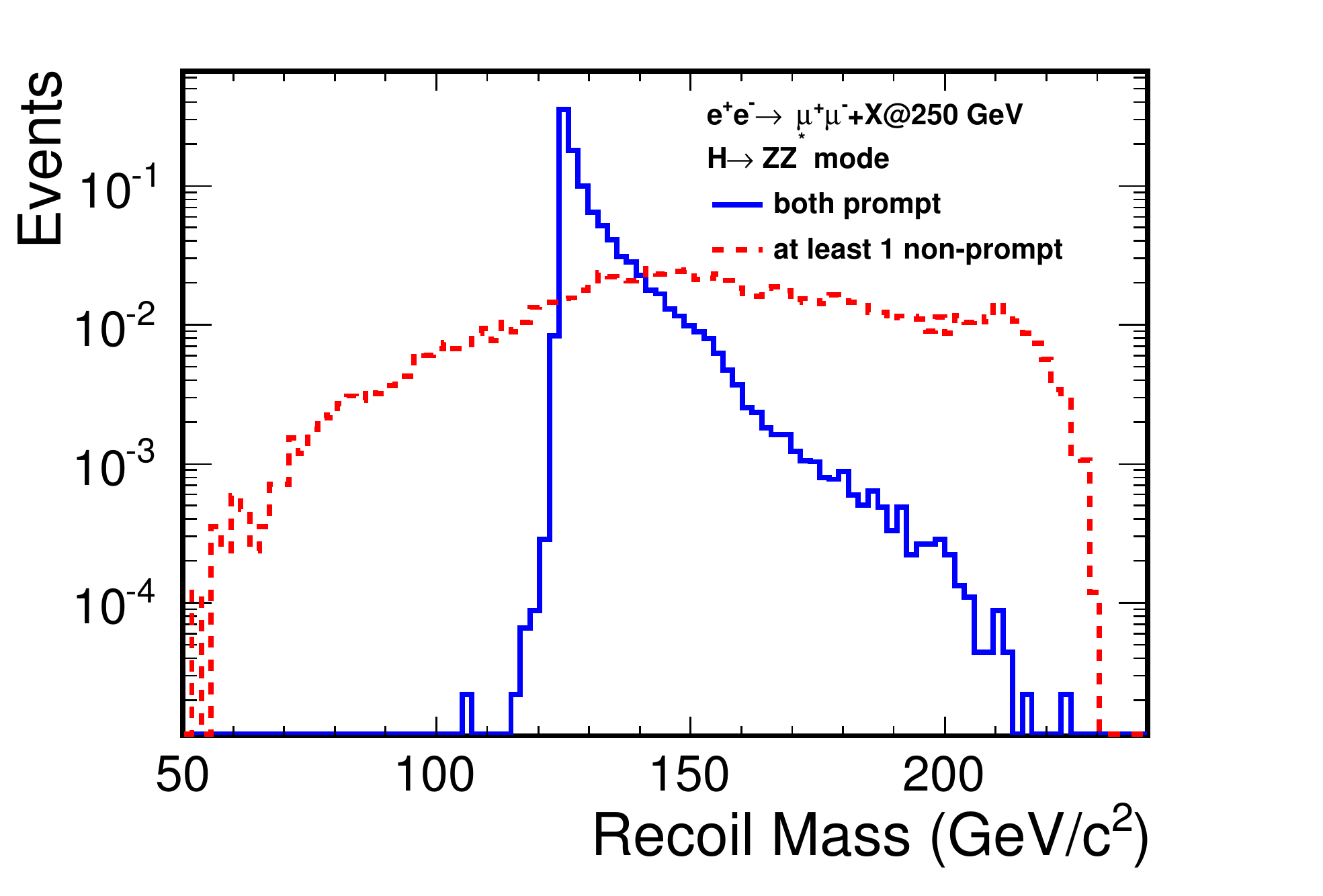}\includegraphics[scale=0.3]{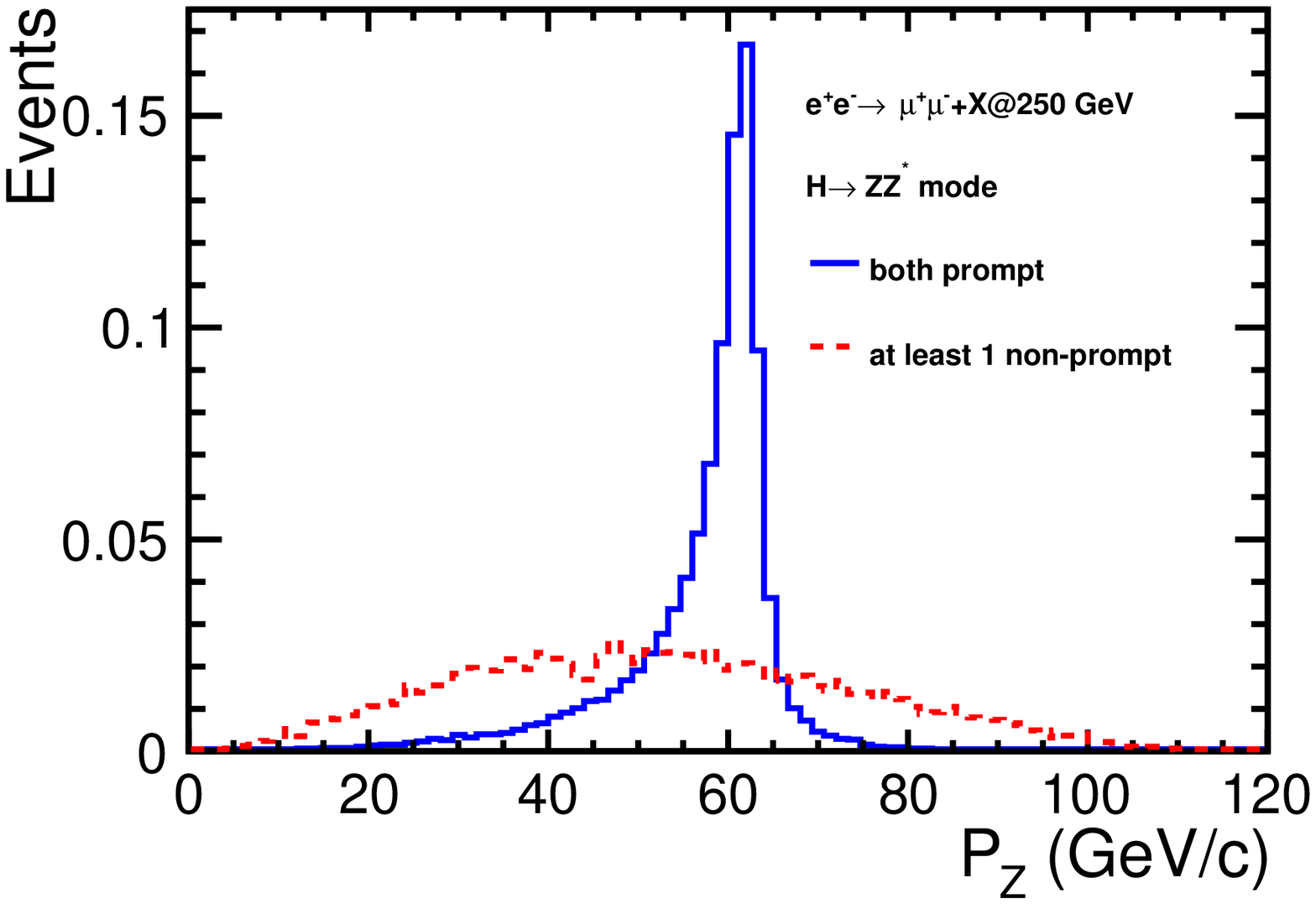} 
\par\end{centering}

\begin{centering}
\includegraphics[scale=0.3]{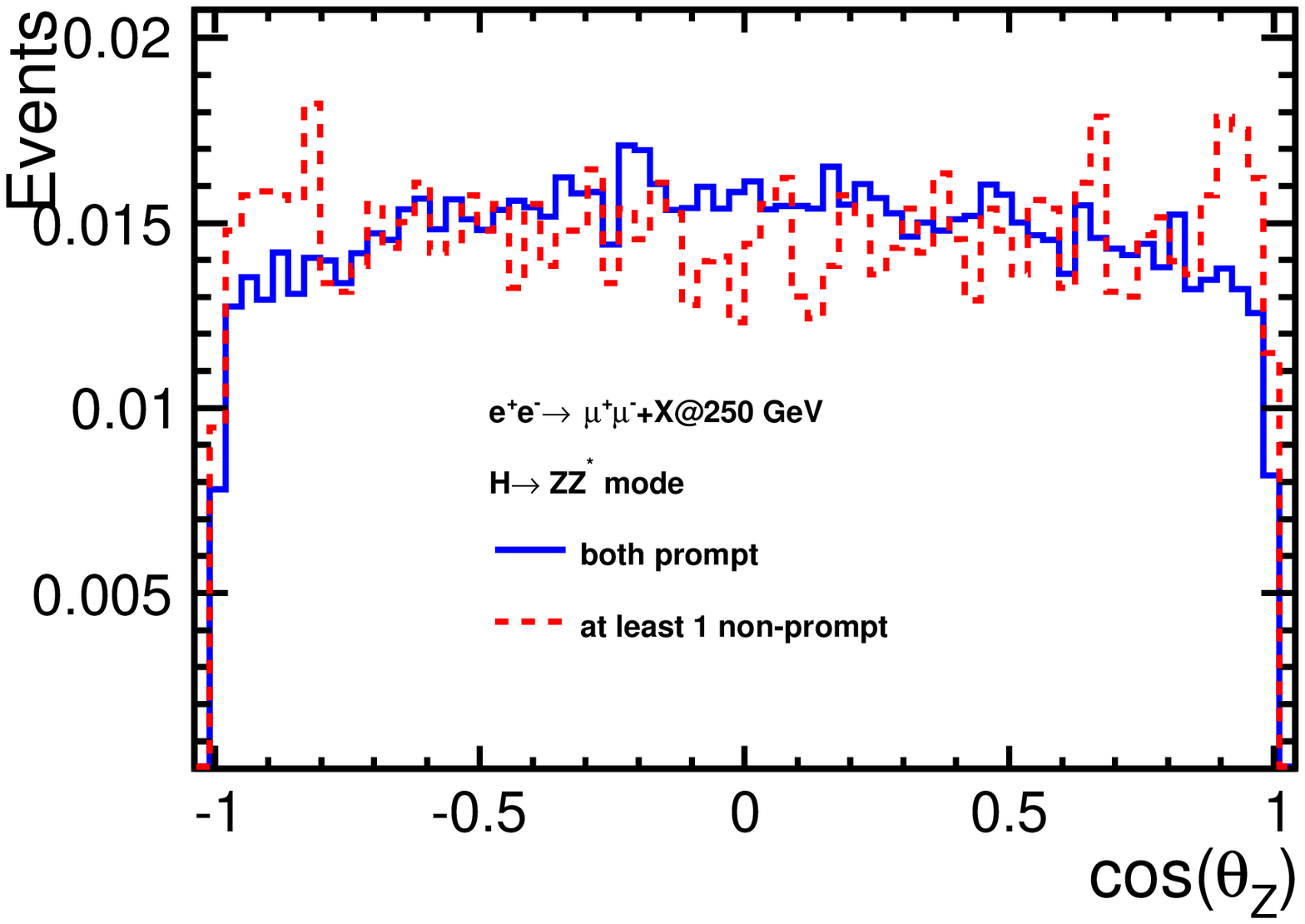}\includegraphics[scale=0.3]{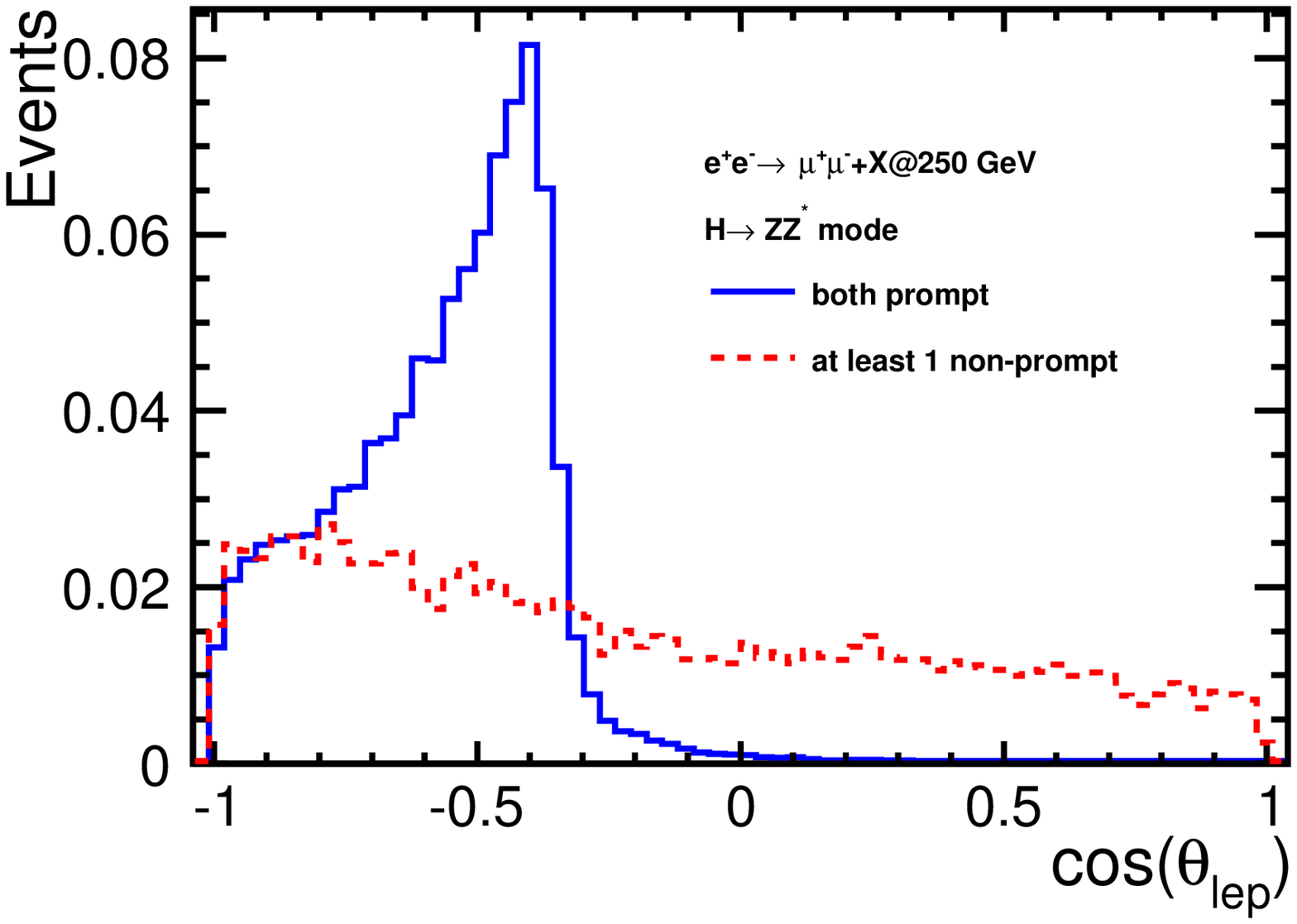}
\par\end{centering}

\centering{}\protect\caption{The distribution of the variables $M_{\mathrm{l^{+}l^{-}}}$, $M_{\mathrm{rec}}$,
$p_{\mathrm{\mathrm{Z}}}$ , $\cos\theta_{\mathrm{Z}}$, and $\cos\theta_{\mathrm{lep}}$
used for the training in the MVA lepton pairing method, shown here
for the \textquotedblleft correct pairs\textquotedblright{} and \textquotedblleft wrong
pairs\textquotedblright{} in the case of the $\mathrm{\mu{}^{+}\mathit{\mu}^{-}H}$
channel and and $\mathrm{e_{L}^{-}e_{R}^{+}}$ at $\sqrt{s}$ = 250
GeV. \label{TMVACompare}}
\end{figure}

The efforts to minimize bias start from the very beginning of the
data selection process. The isolated lepton selection mentioned in
Section \ref{sub:Isolated-Lepton-Finder} is tuned to take into account
the fact that each decay mode has a different density of particles
surrounding the leptons from Z boson decay. Section \ref{sub:pair}
mentioned that the decay modes which contain an extra source of leptons
receive the effect from ``lepton pairing mistake'', defined as the
case in which at least one of the leptons in the selected dilepton
pair is from Higgs boson decay. The analysis in \cite{Recoil} pairs
leptons using a method which minimizes a function $\chi^{2}\left(M_{\mathrm{l^{+}l^{-}}},M_{\mathrm{r\mathrm{ec}}}\right)$
(Equation \ref{eq:chi2}). Figure \ref{fig:ResidualChi} shows the
distribution of $\chi^{2}\left(M_{\mathrm{l^{+}l^{-}}},M_{\mathrm{r\mathrm{ec}}}\right)$
at $\sqrt{s}$=250 GeV. In this section, this ``$\chi^{2}$ method''
will be compared to two other types of lepton pairing algorithm. One
is the ``$M_{\mathrm{l^{+}l^{-}}}$ method'', which selects the
pair of leptons with $M_{\mathrm{l^{+}l^{-}}}$ closest to $M_{\mathrm{Z}}$
as the signal dilepton. The shortcoming of the $M_{\mathrm{l^{+}l^{-}}}$
method is that when both leptons are from the same Z boson originating
from Higgs decay, their $M_{\mathrm{l^{+}l^{-}}}$ would still be
close to $M_{\mathrm{Z}}$, whereas the corresponding $M_{\mathrm{rec}}$
tend to be deviated from $M_{\mathrm{H}}$. Another one is the ``MVA
method'', which selects a pair of leptons that maximizes a MVA response
formed from $M_{\mathrm{Z}}$ , $M_{\mathrm{rec}}$, $\cos\theta_{\mathrm{Z}}$,
$\cos\theta_{\mathrm{lep}}$, and $p_{\mathrm{\mathrm{Z}}}$. The
MVA evaluation is done using the MLP method and the weights are trained
using the $\mathrm{H\rightarrow ZZ^{*}}$ mode sample which has the
highest probability of wrong pairing. Figure \ref{TMVACompare} compares
the distribution of the MVA variables between correct and wrong pairs.
Regarding $M_{\mathrm{Z}}$ and $M_{\mathrm{rec}}$, $\cos\theta_{\mathrm{lep}}$,
and $p_{\mathrm{\mathrm{Z}}}$, those of the correct pairs peak around
the value calculated from kinematics whereas those of the wrong pairs
have a wider distribution. Regarding $\cos\theta_{\mathrm{Z}}$, correct
pairs have a more isotropic distribution than wrong pairs. 

Prior to pairing the leptons, a pre-cut on $M_{\mathrm{l^{+}l^{-}}}$
is implemented as $\left|M_{\mathrm{l^{+}l^{-}}}-M_{\mathrm{Z}}\right|<40(60)\:\mathrm{GeV}$
for $\mu\left(\mathrm{e}\right)$. For each of the three methods,
Table \ref{tab:CateZmm250} shows the pairing performance for the
$\mathrm{H\rightarrow ZZ^{*}}$ mode. The ratios are defined with
respect to the number of generated events. The MC statistical uncertainty
is about 0.1\%. It can be observed that there is no significant difference
between the $\chi^{2}$ method, which was eventually used in analysis,
and the MVA method, while both are better than the $M_{\mathrm{l^{+}l^{-}}}$
method. The pairing performance at $\sqrt{s}$=250 GeV using the $\chi^{2}$
method is shown for all major SM Higgs decay modes in Appendix \ref{sec:Performance-of-Lepton}. 

\begin{table}[H]
\begin{centering}
\begin{tabular}{|c|c|}
\hline 
C0 & number of generated events\tabularnewline
\hline 
C1 & number of selected $\mu$ (e) for $\mathrm{\mu{}^{+}\mathit{\mu}^{-}H}$
($\mathrm{\mathrm{e}{}^{+}\mathit{\mathrm{e}}^{-}H}$) channel\tabularnewline
\hline 
C2 & correct pairs\tabularnewline
\hline 
C3 & 1 prompt and 1 non-prompt lepton selected, with 2 prompt leptons found\tabularnewline
\hline 
C4 & 2 non-prompt leptons selected, with 2 prompt leptons found\tabularnewline
\hline 
C5 & only 1 prompt lepton found\tabularnewline
\hline 
C6 & no prompt leptons found\tabularnewline
\hline 
\end{tabular}\protect\caption{The categorization of the lepton pairing performance for the recoil
analysis. \label{tab:Cate}}

\par\end{centering}

\end{table}

\begin{table}[H]
\begin{centering}
\begin{tabular}{|c|c|c|c|c|c|c|}
\hline 
$\sqrt{s}$=250 GeV & \multicolumn{1}{c}{} & \multicolumn{1}{c}{$\mathrm{\mu{}^{+}\mathit{\mu}^{-}H}$} &  & \multicolumn{1}{c}{} & \multicolumn{1}{c}{$\mathrm{\mathrm{e}{}^{+}\mathit{\mathrm{e}}^{-}H}$ } & \tabularnewline
\cline{2-7} 
$\mathrm{H\rightarrow ZZ^{*}}$ & $\chi^{2}$  & MVA & $M_{\mathrm{l^{+}l^{-}}}$  & $\chi^{2}$  & MVA & $M_{\mathrm{l^{+}l^{-}}}$ \tabularnewline
\hline 
C0 & 100\% & 100\% & 100\% & 100\% & 100\% & 100\%\tabularnewline
\hline 
C1 & 94.15\% & 94.15\% & 94.15\% & 87.08\% & 87.08\% & 87.08\%\tabularnewline
\hline 
C2 & 93.17\% & 93.18\% & 92.44\% & 85.13\% & 85.09\% & 84.78\%\tabularnewline
\hline 
C3 & 0.728\% & 0.715\% & 1.46\% & 1.363\% & 1.412\% & 1.714\%\tabularnewline
\hline 
C4 & 0.342\% & 0.421\% & 1.13\% & 0.548\% & 0.795\% & 1.017\%\tabularnewline
\hline 
C5 & 0.250\% & 0.250\% & 0.250\% & 0.572\% & 0.572\% & 0.572\%\tabularnewline
\hline 
C6 & 0.002\% & 0.002\% & 0.002\% & 0.008\% & 0.008\% & 0.008\%\tabularnewline
\hline 
\end{tabular}
\par\end{centering}

\centering{}\protect\caption{The lepton pairing performance of the $\mathrm{H\rightarrow ZZ^{*}}$
mode and and $\mathrm{e_{L}^{-}e_{R}^{+}}$ at $\sqrt{s}$=250 GeV.
The categories C1-C6 are defined in Table \ref{tab:Cate}. The statistical
uncertainties are below 0.1\%. \label{tab:CateZmm250}}
\end{table}

\begin{figure}[H]
\begin{centering}
\includegraphics[scale=0.3]{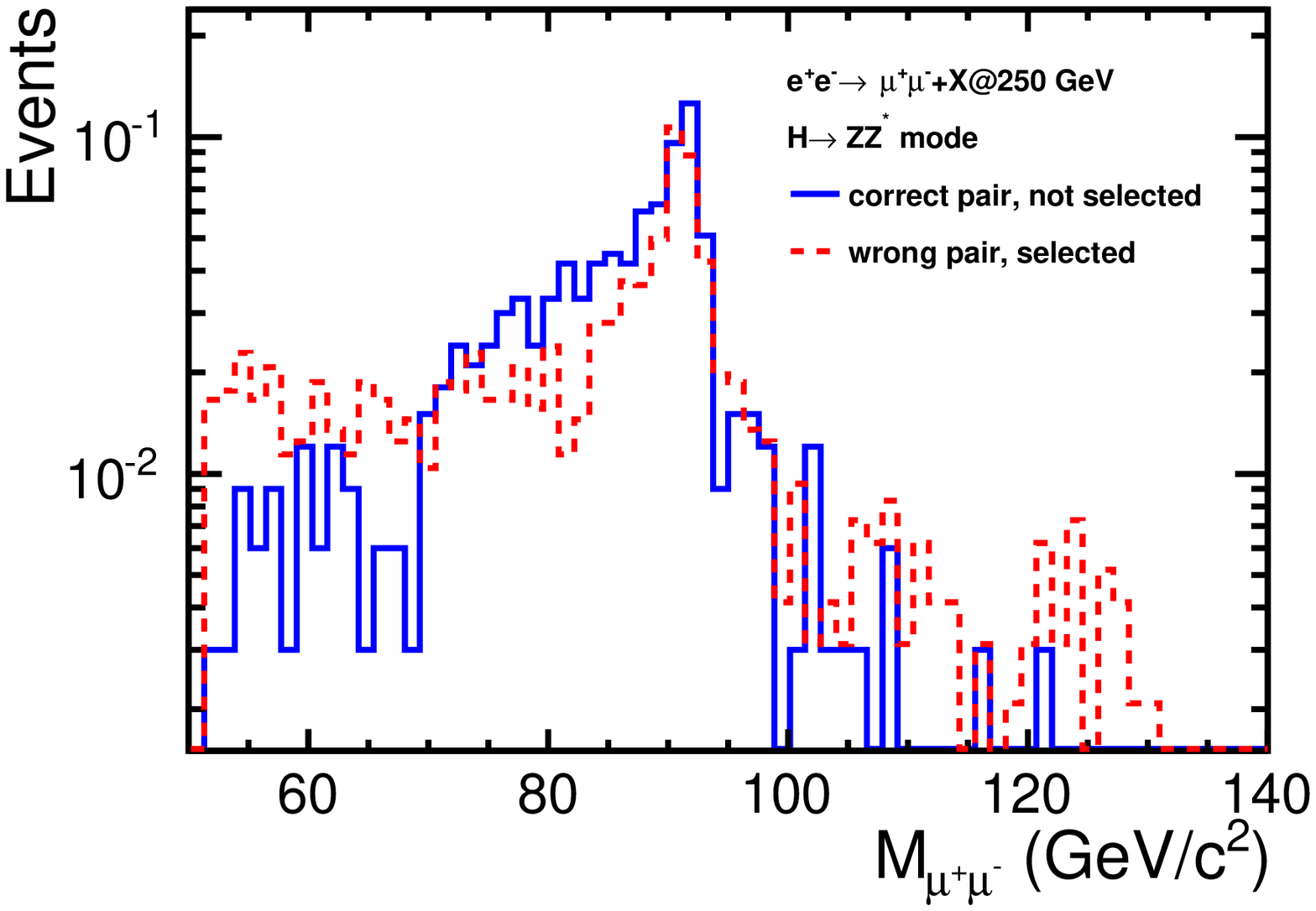}\includegraphics[scale=0.3]{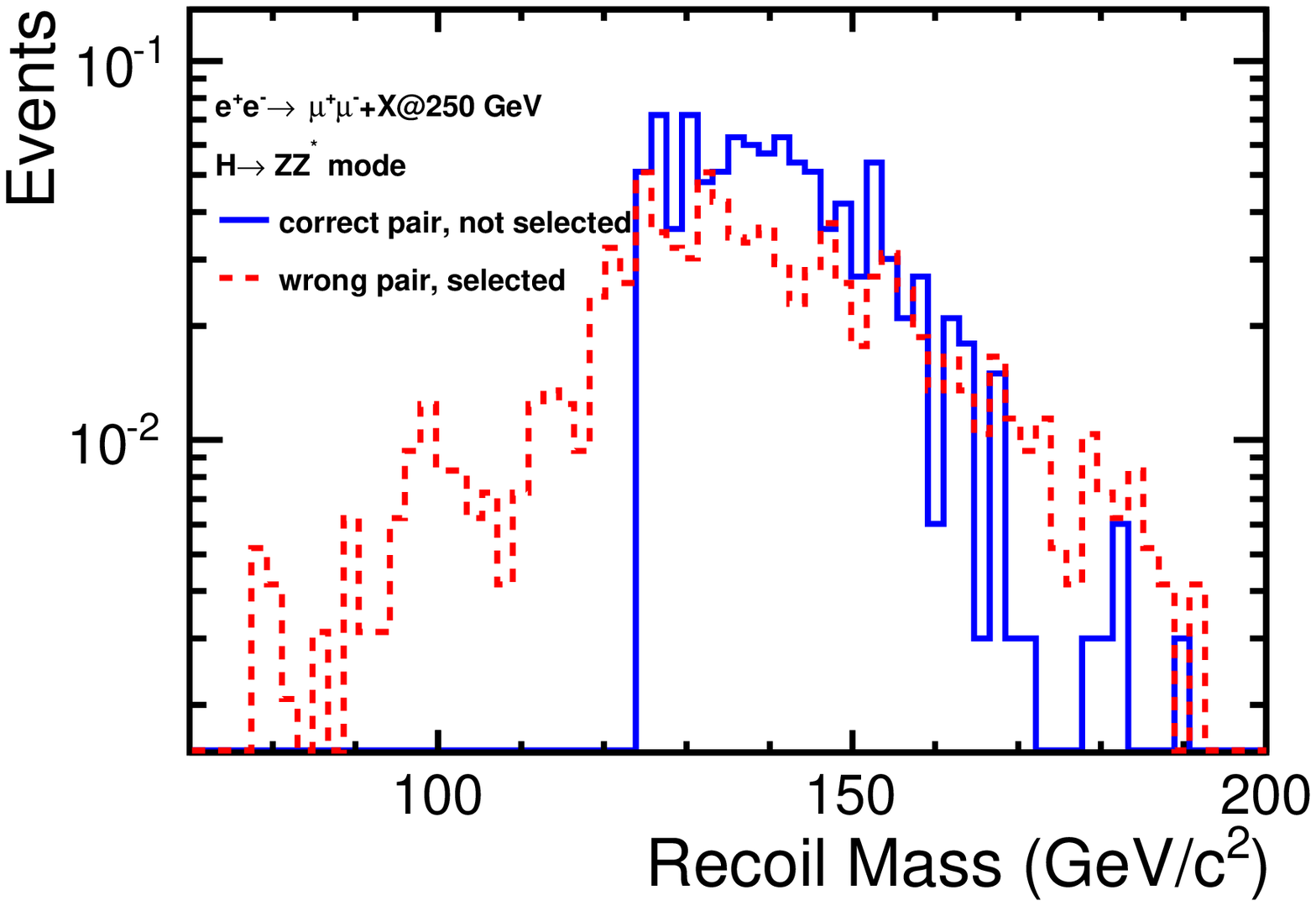}\includegraphics[scale=0.3]{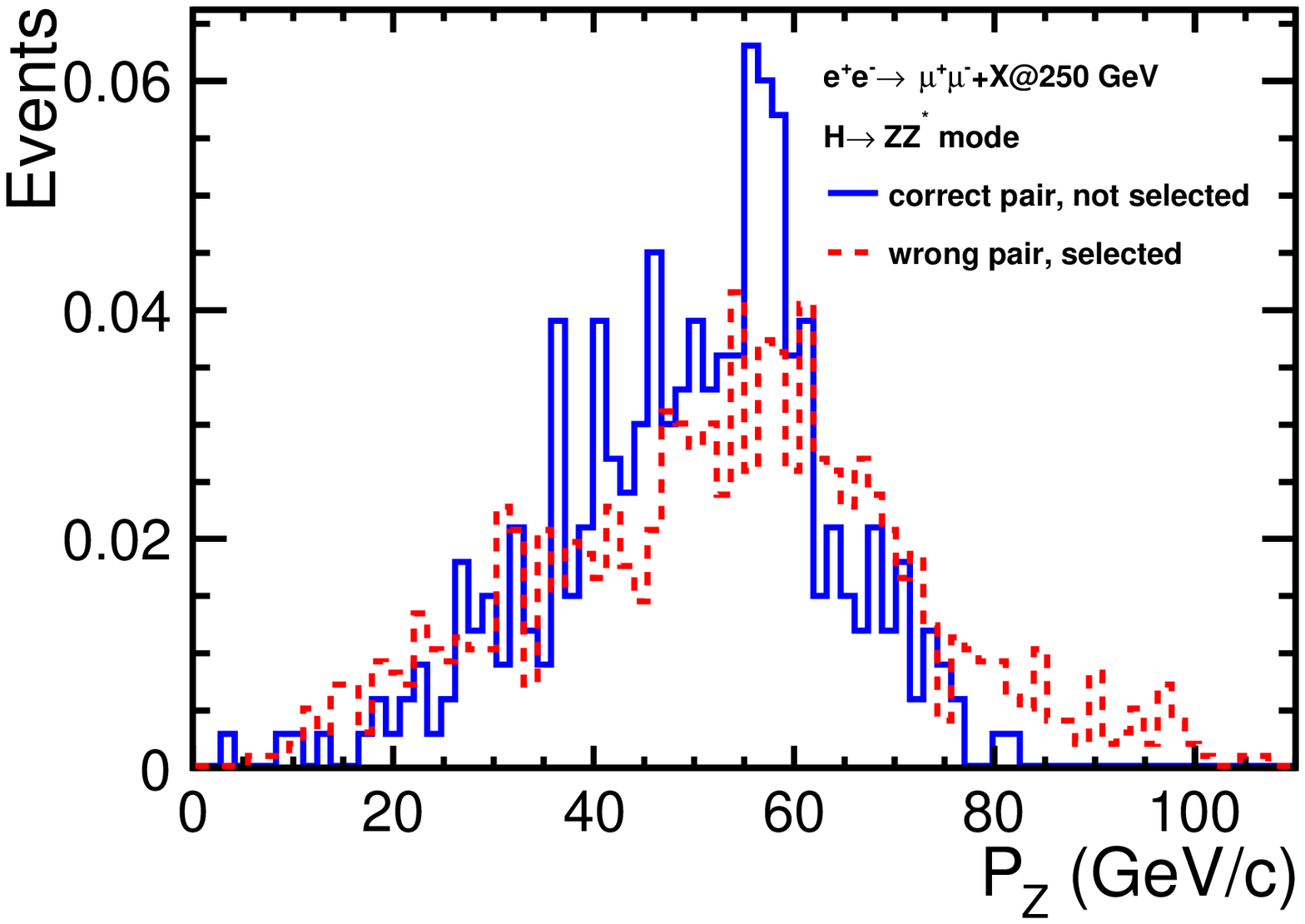} 
\par\end{centering}

\begin{centering}
\includegraphics[scale=0.3]{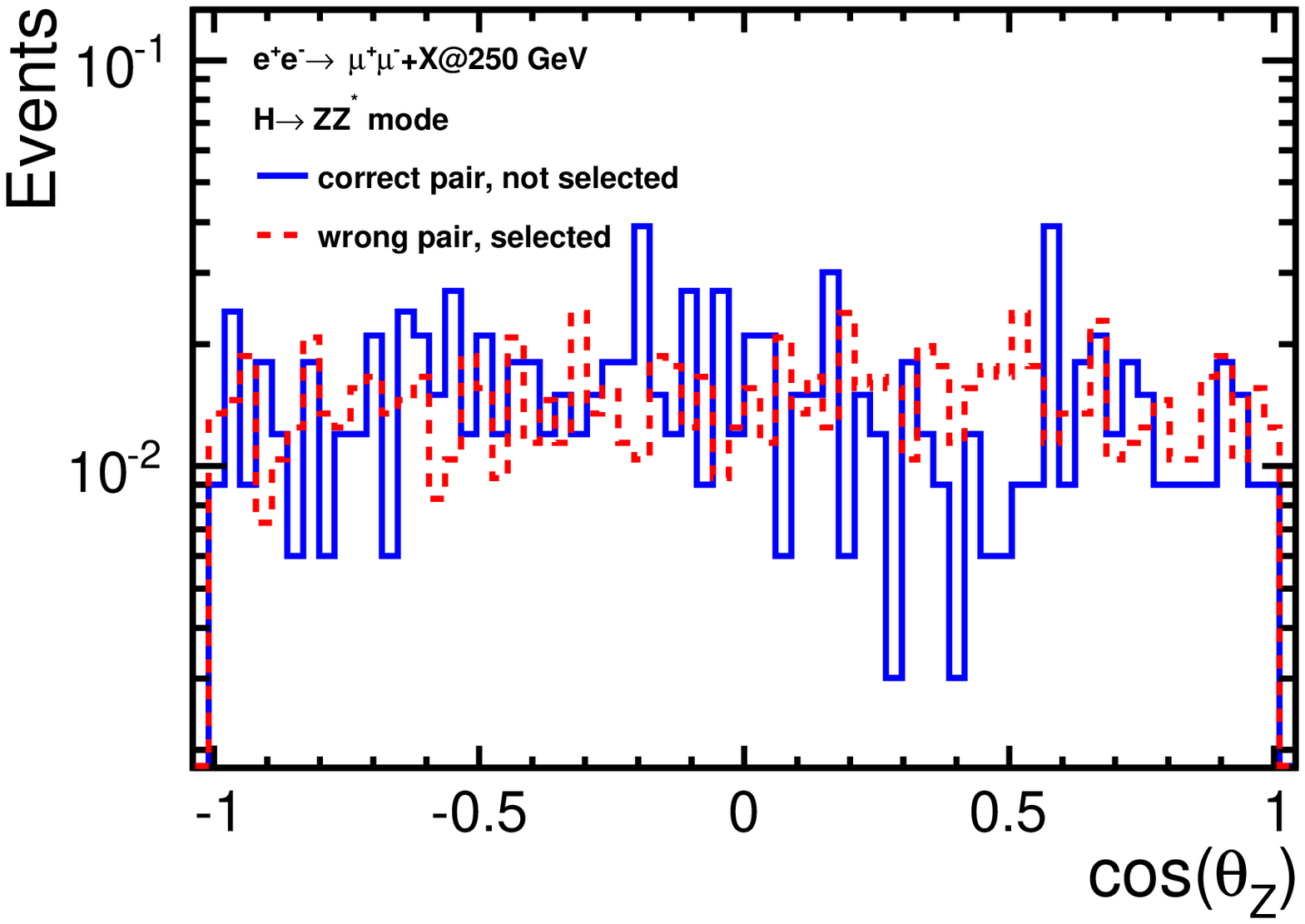}\includegraphics[scale=0.3]{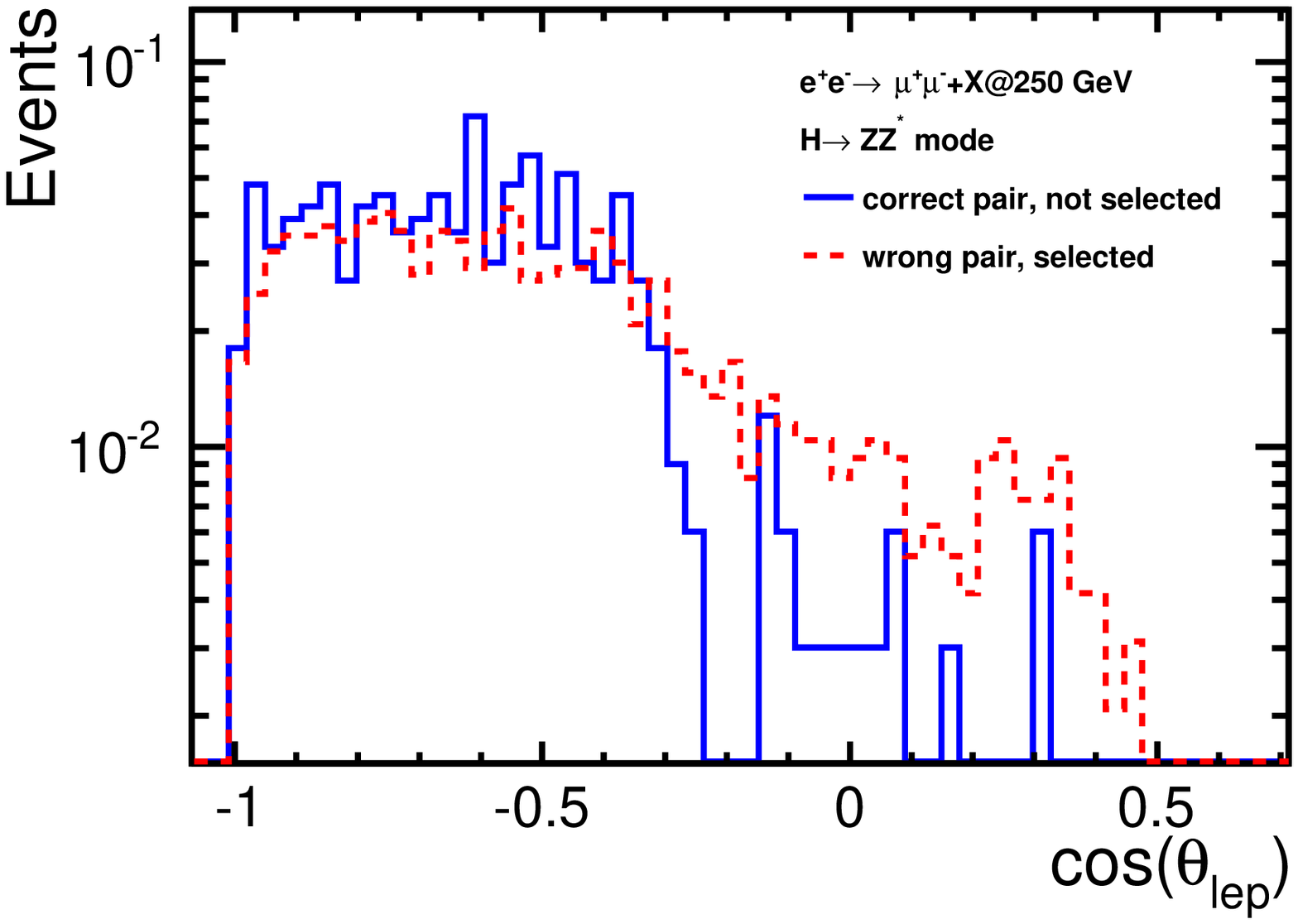}
\par\end{centering}

\protect\caption{The distribution of $M_{\mathrm{l^{+}l^{-}}}$, $M_{\mathrm{rec}}$,
$p_{\mathrm{\mathrm{Z}}}$, $\cos\theta_{\mathrm{Z}}$ and $\cos\theta_{\mathrm{lep}}$
of the ``correct pairs'' which did not get selected, and the ``wrong
pairs'' which got selected, shown for the $\mathrm{H\rightarrow ZZ^{*}}$
mode, $\mathrm{\mathrm{\mu}{}^{+}\mathit{\mathrm{\mu}}^{-}H}$ channel
and $\mathrm{e_{L}^{-}e_{R}^{+}}$ at $\sqrt{s}$=250 GeV.\label{fig:Residual}}
\end{figure}

From Figure\ref{fig:ResidualChi}, it can be seen that while correct
pair events peak sharply at a small $\chi^{2}\left(M_{\mathrm{l^{+}l^{-}}},M_{\mathrm{rec}}\right)$
value, about 1/10 of the peak is occupied by wrong pair events, which
explains the finite amount of pairing mistakes. The fact that this
can not be visibly improved by the MVA method can be understood from
Figure\ref{fig:Residual} which compares the variables $M_{\mathrm{l^{+}l^{-}}}$,
$M_{\mathrm{rec}}$ ,$p_{\mathrm{\mathrm{Z}}}$ , $\cos\theta_{\mathrm{Z}}$,
and $\cos\theta_{\mathrm{lep}}$ of the following two types of events:
(A) A pair consisting of two leptons from the Z boson recoiling against
the Higgs boson, whereas the actually ``selected pair '' contains
at least one lepton from Higgs decay, and (B) A ``selected pair''
consisting of at least one lepton from Higgs decay. With the exception
that the distributions of $p_{\mathrm{\mathrm{Z}}}$ and $M_{\mathrm{rec}}$
are slightly wider for (B), there is no significance difference between
(A) and (B). 

\subsubsection{Other sources of bias\label{sub:BiasTable}}

Following the selection of the isolated lepton pairs, the cuts on
$M_{\mathrm{l^{+}l^{-}}}$, $p_{\mathrm{T}}^{\mathrm{l^{+}l^{-}}}$,
BDT, and $M_{\mathrm{rec}}$ are designed to use only kinematical
information from the selected leptons so as to avoid introducing bias
to the efficiencies of individual Higgs decay modes. On the other
hand, the $\cos\left(\theta_{\mathrm{missing}}\right)$ cut, which
counts the missing momentum from the whole event, in principle uses
information of particles from Higgs decay. The $E_{\mathrm{vis}}$
cut will not introduce additional bias, as it simply categorizes the
events into visible or invisible Higgs decay, as mentioned in Section
\ref{sub:Background-Rejection}. 

\begin{figure}[H]
\begin{centering}
\includegraphics[scale=0.4]{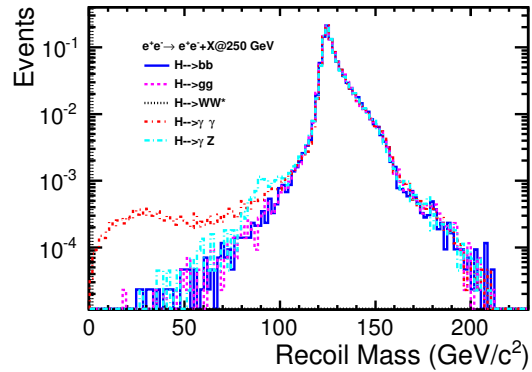}
\par\end{centering}

\centering{}\protect\caption{A comparison of $M_{\mathrm{rec}}$ between the decay modes $\mathrm{H\rightarrow\mathrm{bb}}$,
gg, $\mathrm{WW^{*}}$, $\gamma\gamma$ and $\gamma\mathrm{Z}$ for
the $\mathrm{\mathrm{e}{}^{+}\mathit{\mathrm{e}}^{-}H}$ channel and
$\mathrm{e_{L}^{-}e_{R}^{+}}$ at $\sqrt{s}$ = 250 GeV. \label{MrecMode}}
\end{figure}

Tables \ref{tab:Mode Eff} and \ref{tab:Mode EffZee} show the efficiency
of each decay mode after each cut for the case of $\sqrt{s}$=250
GeV and $\mathrm{e_{L}^{-}e_{R}^{+}}$. The tables for the other channels
are given in Appendix \ref{sec:Efficiency-Table}. The bias is reduced
at higher center-of-mass energies. For example, at $\sqrt{s}$=500
GeV, no bias exists beyond the MC statistical error (< 0.2\%) for
any mode. Based on these results, the bias on the measured $\sigma_{\mathrm{ZH}}$
will be given in Section\ref{sub:Systematic-Error-of}. The following
sources of residual bias can be observed: 
\begin{itemize}
\item The first row ``Lepton Finder'' in the Tables \ref{tab:Mode Eff}
and \ref{tab:Mode EffZee} shows that more lepton pairs are found
for the $\mathrm{H\rightarrow WW^{*}}$, $\mathrm{ZZ^{*}}$, $\mathrm{\tau\tau}$,
and $\mathrm{\gamma Z}$ modes as they contain leptons from Higgs
decay as an extra source of leptons. These efficiencies are slightly
evened out later on by ``Lepton ID'' and cuts on $M_{\mathrm{l^{+}l^{-}}}$
and $M_{\mathrm{rec}}$. On the other hand, the $\mathrm{H\rightarrow gg}$
mode has a slightly lower efficiency of finding isolated leptons due
to the existence of widely spread gluon jets. This effect has already
been minimized by using a $\mathrm{H\rightarrow gg}$ sample to train
the MVA weights in the isolated lepton finder. 
\item The $\mathrm{H\rightarrow\gamma Z}$ mode receives bias from mistaken
lepton ID due to the confusion with the leptons from Higgs decay.
For example, in the $\mathrm{\mu^{+}\mu^{-}X}$ channel, a pair of
electrons decayed from the Z boson from Higgs decay become selected
as an isolated electron pair.
\item The $\mathrm{H\rightarrow\gamma Z}$ mode receives bias from the $\cos\left(\theta_{\mathrm{missing}}\right)$
cut since it contains events with ISR photons going down the beam
pipe but little visible energy other than that of the isolated lepton
pair. The $\cos\left(\theta_{\mathrm{missing}}\right)$ cut is designed
to be very loose so that this bias is very small, while 2-fermion
backgrounds can still be suppressed effectively. 
\item The $\mathrm{H\rightarrow\gamma\gamma}$ mode in the $\mathrm{\mathrm{e}^{+}\mathrm{e}^{-}X}$
channel receives a slight bias from pre-cuts on $M_{\mathrm{rec}}$
due to the FSR/bremsstrahlung process (see Section \ref{sub:FSR-Recovery}).
From Figure \ref{MrecMode}, a bump can be seen in the lower region
of the reconstructed $M_{\mathrm{rec}}$ spectrum ($\lesssim$ 100
GeV) for the $\mathrm{H\rightarrow\gamma\gamma}$ mode; in these events
the relatively energetic photons from Higgs decay are mistakenly recovered
to the isolated leptons. This effect is less significant at higher
center-of-mass energies for which the Higgs decay products are more
boosted. %
\footnote{\begin{itemize}
\item In the $\mathrm{\mu^{+}\mu^{-}X}$ channel, in order to protect the
$\mathrm{H\rightarrow\gamma\gamma}$ mode from this bias, a protection
is set up so that the recovery is undone if the invaraint mass after
the recovery is further away from the Z boson mass than before the
recovery. However because this protection affects the efficiency of
the FSR/bremsstrahlung recovery, hence the rejection of 2-fermion
backgrounds, it cannot be used in the $\mathrm{\mathrm{e}^{+}\mathrm{e}^{-}X}$
channel where 2-fermion backgrounds are dominant. \end{itemize}
}
\end{itemize}

\begin{table}[H]
\begin{centering}
\begin{tabular}{|c|c|c|c|c|c|c|c|c|}
\hline 
$\mathrm{H\rightarrow XX}$ & bb & cc & gg & $\tau\tau$ & $\mathrm{\mathrm{WW^{*}}}$ & $\mathrm{ZZ^{*}}$ & $\gamma\gamma$ & $\gamma\mathrm{Z}$\tabularnewline
\hline 
BR (SM) & 57.8\% & 2.7\% & 8.6\% & 6.4\% & 21.6\% & 2.7\% & 0.23\% & 0.16\%\tabularnewline
\hline 
\hline 
Lepton Finder & 93.70\% & 93.69\% & 93.40\% & 94.02\% & 94.04\% & 94.36\% & 93.75\% & 94.08\%\tabularnewline
\hline 
Lepton ID+Precut & 93.68\% & 93.66\% & 93.37\% & 93.93\% & 93.94\% & 93.71\% & 93.63\% & 93.22\%\tabularnewline
\hline 
$M_{\mathrm{l^{+}l^{-}}}\in${[}73, 120{]} GeV & 89.94\% & 91.74\% & 91.40\% & 91.90\% & 91.82\% & 91.81\% & 91.73\% & 91.47\%\tabularnewline
\hline 
$p_{\mathrm{t}}^{\mathrm{l^{+}l^{-}}}\in${[}10, 70{]} GeV & 89.94\% & 90.08\% & 89.68\% & 90.18\% & 90.04\% & 90.16\% & 89.99\% & 89.71\%\tabularnewline
\hline 
$\mathrm{\left|cos\mathit{\theta_{\mathrm{missing}}}\right|}$< 0.98 & 89.94\% & 90.08\% & 89.68\% & 90.16\% & 90.04\% & 90.16\% & 89.91\% & 89.41\%\tabularnewline
\hline 
MVA & 88.90\% & 89.04\% & 88.63\% & 89.12\% & 88.96\% & 89.11\% & 88.91\% & 88.28\%\tabularnewline
\hline 
$M_{\mathrm{rec}}\in${[}110, 155{]} GeV & 88.25\% & 88.35\% & 87.98\% & 88.43\% & 88.33\% & 88.52\% & 88.21\% & 87.64\%\tabularnewline
\hline 
\end{tabular}
\par\end{centering}

\centering{}\protect\caption{The BR values and the efficiency of the major SM Higgs decay modes,
after each data selection step, shown here for the case of the $\mathrm{\mu^{+}\mu^{-}X}$
channel and $\mathrm{e_{L}^{-}e_{R}^{+}}$ at $\sqrt{s}$=250 GeV.
The statistical uncertainties on these values are below 0.16\%.\label{tab:Mode Eff}}
\end{table}

\begin{table}[H]
\centering{}%
\begin{tabular}{|c|c|c|c|c|c|c|c|c|}
\hline 
$\mathrm{H\rightarrow XX}$ & bb & cc & gg & $\tau\tau$ & $\mathrm{\mathrm{WW^{*}}}$ & $\mathrm{ZZ^{*}}$ & $\gamma\gamma$ & $\gamma\mathrm{Z}$\tabularnewline
\hline 
BR (SM) & 57.8\% & 2.7\% & 8.6\% & 6.4\% & 21.6\% & 2.7\% & 0.23\% & 0.16\%\tabularnewline
\hline 
\hline 
Lepton Finder & 89.12\% & 88.92\% & 88.51\% & 89.50\% & 89.87\% & 90.15\% & 89.83\% & 90.06\%\tabularnewline
\hline 
Lepton ID+Precut & 88.58\% & 88.42\% & 87.99\% & 88.58\% & 88.96\% & 88.37\% & 87.59\% & 87.67\%\tabularnewline
\hline 
$M_{\mathrm{l^{+}l^{-}}}\in${[}73, 120{]} GeV & 86.70\% & 86.42\% & 85.94\% & 86.12\% & 86.19\% & 86.12\% & 85.38\% & 85.64\%\tabularnewline
\hline 
$p_{\mathrm{t}}^{\mathrm{l^{+}l^{-}}}\in${[}10, 70{]} GeV & 84.96\% & 84.76\% & 84.24\% & 84.42\% & 84.47\% & 84.32\% & 83.65\% & 83.77\%\tabularnewline
\hline 
$\mathrm{\left|cos\mathit{\theta_{\mathrm{missing}}}\right|}$< 0.98 & 84.96\% & 84.76\% & 84.24\% & 84.29\% & 84.45\% & 84.18\% & 83.24\% & 83.48\%\tabularnewline
\hline 
MVA & 68.90\% & 68.87\% & 68.52\% & 68.34\% & 68.19\% & 68.31\% & 67.39\% & 67.73\%\tabularnewline
\hline 
$M_{\mathrm{rec}}\in${[}110, 155{]} GeV & 68.61\% & 68.60\% & 68.19\% & 68.04\% & 67.88\% & 68.02\% & 67.08\% & 67.45\%\tabularnewline
\hline 
\end{tabular}\protect\caption{The BR values and the efficiency of the major SM Higgs decay modes,
after each data selection step, shown here for the case of the $\mathrm{\mathrm{e}^{+}\mathrm{e}^{-}X}$
channel and $\mathrm{e_{L}^{-}e_{R}^{+}}$ at $\sqrt{s}$=250 GeV.
The statistical uncertainties on these values are below 0.16\%. \label{tab:Mode EffZee}}
\end{table}

\subsection{Bias on the measured cross section \label{sub:Systematic-Error-of}}

In this section, the potential bias on the measured $\sigma_{\mathrm{ZH}}$
due to residual Higgs decay mode dependence is evaluated from a conservative
perspective. Table \ref{tab:Mode Eff} shows no discrepancy in efficiencies
beyond 1\%, which demonstrates model independence at a level of better
than 0.5\% based on the most conservative scenario B. Note that the
bias is even smaller at higher center-of-mass energies. 

Regarding the most realistic scenario C, the bias is estimated as
follows (using Equations \ref{eq:BR3} and \ref{eq:BR4}). The known
modes are assumed to be $\mathrm{H\rightarrow\mathrm{bb}}$, cc, gg,
$\tau\tau$, $\mathrm{WW^{*}}$, $\mathrm{ZZ^{*}}$, $\gamma\gamma$,
$\gamma\mathrm{Z}$, since they will be measured at the LHC or the
ILC \cite{snowmass,White paper}. Taking into consideration the possibility
of unknown exotic Higgs decay modes, their total branching ratio ($B_{x}$)
is assumed to be 10\%, based on the estimation of the 95\% C.L. upper
limit for branching ratio of BSM decay modes from the HL-LHC \cite{snowmass}.
In fact assigning a large BR of 10\% to unknown modes is a considerably
conservative assumption, because at the ILC the upper limit for BSM
decay will be greatly improved and in general any decay mode with
a few percent branching ratio shall be directly measured. Since the
characteristics of any exotic decay mode is expected to fall within
the wide range of known decay modes being directly investigated, we
obtain $\delta\varepsilon_{\mathrm{max}}$ by assuming that the efficiencies
of the unknown modes will lie in the range of the efficiencies of
known modes; this is, for example, -0.68\% from the $\gamma\mathrm{Z}$
mode in the case of the channel shown in Table \ref{tab:Mode Eff}.
Then for the known modes, each $B_{i}$ is scaled from their SM values
by 90\%, following which $\varepsilon_{0}$ is obtained straightforwardly
from $B_{i}$ and $\epsilon_{i}$. Each $\Delta B_{i}$ is taken conservatively
by fluctuating the BR values by their largest uncertainties predicted
for future measurements at the ILC\cite{White paper} with exceptions
of the $\mathrm{H\rightarrow cc}$ and gg modes which are very difficult
to obtain at the HL-LHC and thus are obtained from the predictions
for the ILC\cite{White paper}. Based on the information in Table
\ref{tab:Mode Eff}, Table \ref{tab:Mode EffWeighed} gives the deviation
in efficiency of each known mode from the average efficiency for the
case of $\mathrm{e_{L}^{-}e_{R}^{+}}$ at $\sqrt{s}$=250 GeV. 

The same analysis is carried out for all channels. Table \ref{tab:ModeBias}
shows for all center-of-mass energies and polarizations in this analysis
the relative bias on $\sigma_{\mathrm{ZH}}$, which is below 0.08\%
for the $\mathrm{\mu{}^{+}\mathit{\mu}^{-}H}$ channel and 0.19\%
for the $\mathrm{\mathrm{e}^{+}\mathrm{e}^{-}X}$ channel. The maximum
contribution to the residual bias comes from either the $\mathrm{H\rightarrow\gamma\gamma}$
mode or the $\mathrm{H\rightarrow\gamma Z}$ mode. 

From the the above and results in Table \ref{tab:ModeBias}, we conclude
that the model independence of $\sigma_{\mathrm{ZH}}$ measurement
at the ILC using Higgsstrahlung events $\mathrm{e^{+}e^{-}\rightarrow ZH\rightarrow\mathrm{\mathit{\mathrm{l^{+}l^{-}}}H}}$
($\mathrm{l}$ = e or $\mu$) is demonstrated to a level well below
even the smallest statistical $\sigma_{\mathrm{ZH}}$ uncertainties
expected from the leptonic recoil measurements in the full H20 run,
by a factor of 5 \cite{Recoil}.

\begin{table}[H]
\begin{centering}
\begin{tabular}{|c|c|c|c|c|c|}
\hline 
$\sqrt{s}$=250 GeV &  & \multicolumn{1}{c}{$\mathrm{\mu^{+}\mu^{-}X}$ } &  & \multicolumn{1}{c}{$\mathrm{e}^{+}\mathrm{e}^{-}X$} & \tabularnewline
\cline{2-6} 
$\mathrm{e_{L}^{-}e_{R}^{+}}$ & Average eff. & 88.32\% &  & 68.40\% & \tabularnewline
\hline 
\hline 
$\mathrm{H\rightarrow XX}$ & BR & efficiency & deviation & efficiency & deviation\tabularnewline
\hline 
bb & 57.8\% & 88.25\% & -0.01\% & 68.61\% & 0.02\%\tabularnewline
\hline 
cc & 2.7\% & 88.35\% & 0.00\% & 68.60\% & 0.02\%\tabularnewline
\hline 
gg & 8.6\% & 87.98\% & -0.03\% & 68.19\% & -0.02\%\tabularnewline
\hline 
$\tau\tau$ & 6.4\% & 88.43\% & 0.01\% & 68.04\% & -0.04\%\tabularnewline
\hline 
$\mathrm{WW^{*}}$ & 21.6\% & 88.33\% & 0.00\% & 67.88\% & -0.05\%\tabularnewline
\hline 
$ZZ^{*}$ & 2.7\% & 88.52\% & 0.02\% & 68.02\% & -0.04\%\tabularnewline
\hline 
$\gamma\gamma$ & 0.23\% & 88.21\% & -0.01\% & 67.08\% & -0.13\%\tabularnewline
\hline 
$\gamma\mathrm{Z}$ & 0.16\% & 87.64\% & -0.07\% & 67.45\% & -0.10\%\tabularnewline
\hline 
\end{tabular}
\par\end{centering}

\protect\caption{The final efficiency and the deviation rom the average efficiency
(weighed by BR) of each major known SM Higgs decay mode, shown for
the case of $\sqrt{s}$=250 GeV and beam polarization $\mathrm{e_{L}^{-}e_{R}^{+}}$.
\label{tab:Mode EffWeighed}}
\end{table}

\begin{table}[H]
\centering{}%
\begin{tabular}{|c|c|c|c|c|c|c|}
\hline 
$\sqrt{s}$ & \multicolumn{1}{c}{250 GeV} &  & \multicolumn{1}{c}{350 GeV} &  & \multicolumn{1}{c}{500 GeV} & \tabularnewline
\hline 
$\mathrm{l^{+}l^{-}}\mathrm{H}$ & $\mathrm{\mu^{+}\mu^{-}X}$  & $\mathrm{e}^{+}\mathrm{e}^{-}X$ & $\mathrm{\mu^{+}\mu^{-}X}$  & $\mathrm{e}^{+}\mathrm{e}^{-}X$ & $\mathrm{\mu^{+}\mu^{-}X}$  & $\mathrm{e}^{+}\mathrm{e}^{-}X$\tabularnewline
\hline 
\hline 
$\mathrm{e_{L}^{-}e_{R}^{+}}$ & 0.08\% & 0.19\% & 0.04\% & 0.11\% & 0.05\% & 0.09\%\tabularnewline
\hline 
$\mathrm{e_{R}^{-}e_{L}^{+}}$ & 0.06\% & 0.13\% & 0.00\% & 0.12\% & 0.02\% & 0.02\%\tabularnewline
\hline 
\end{tabular}\protect\caption{The relative bias on $\sigma_{\mathrm{ZH}}$ evaluated for each center-of-mass
energy and polarization.\label{tab:ModeBias}}
\end{table}

\section{SUMMARY AND CONCLUSIONS\label{sec:SUMMARY-AND-CONCLUSIONS}}

The model independent measurements of the absolute cross section $\sigma_{\mathrm{ZH}}$
at the ILC are essential for providing sensitivity to new physics
beyond the Standard Model. By applying the recoil technique to the
Higgsstrahlung process with the Z boson decaying leptonically as $Z\rightarrow\mathrm{l}^{+}\mathrm{l}^{-}$
($\mathrm{l}$= e or $\mu$), the precision of the measurement of
$\sigma_{\mathrm{ZH}}$ and $M_{\mathrm{H}}$ has been evaluated for
three center of mass energies $\sqrt{s}$ = 250, 350, and 500 GeV,
and two beam polarizations $\left(P\mathrm{e^{-}},P\mathrm{e^{+}}\right)$
=($-$80\%, +30\%) and (+80\%, $-$30\%) in \cite{Recoil}, based
on the full simulation of the ILD. This paper demonstrates in detail
that this analysis is model independent to the sub-percent level.
Methods of signal selection and background rejection are optimized
to not only achieve high $\sigma_{\mathrm{ZH}}$ precisions, but also
to minimize the bias on the measured $\sigma_{\mathrm{ZH}}$ due to
discrepancy in signal efficiencies among Higgs decay modes. Under
conservative assumptions which take into account unknown exotic Higgs
decay modes occupying a BR of 10\%, the maximum relative bias on $\sigma_{\mathrm{ZH}}$
is about 0.08\% for the $\mathrm{\mu^{+}\mu^{-}X}$ channel and about
0.19\% for the $\mathrm{\mathrm{e}^{+}\mathrm{e}^{-}X}$ channel,
which are smaller than even the smallest $\sigma_{\mathrm{ZH}}$ statistical
uncertainties expected from the leptonic recoil measurements in a
full 20 years ILC physics program \cite{Recoil} by a factor of 5.

\section*{Acknowledgements}

The authors would like to thank T. Barklow and colleagues in the ILD
Concept Group for their help in realizing this paper; in particular,
J. Strube, D. Jeans, S. Watanuki, H. Yamamoto, and A. Ishikawa for
their contribution to the Higgs recoil study in general, and J. Strube,
A. Miyamoto, C. Calancha, and M. Berggren for their work in generating
the Monte-Carlo samples. This work has been partially supported by
JSPS Grants-inAid for Science Research No. 22244031 and the JSPS Specially
Promoted Research No. 23000002.

\appendix

\section{Visible energy cut\label{sec:Visible-Energy-Cut}}

This section compares the precisions of $\sigma_{\mathrm{ZH}}$ and
$\Delta M_{\mathrm{H}}$ for $\sqrt{s}$=250 GeV between the cases
where the $E_{\mathrm{vis}}$ cut (see Section \ref{sub:Background-Rejection})
is applied and not applied.

\begin{table}[H]
\begin{centering}
\begin{tabular}{|c|c|c|c|}
\hline 
\multicolumn{1}{|c}{$\sqrt{s}$} & 250 GeV & With $E_{\mathrm{vis}}$ cut & No $E_{\mathrm{vis}}$ cut\tabularnewline
\hline 
\multicolumn{1}{|c}{} &  & $\Delta\sigma_{\mathrm{ZH}}/\sigma_{\mathrm{ZH}}$ & $\Delta\sigma_{\mathrm{ZH}}/\sigma_{\mathrm{ZH}}$\tabularnewline
\hline 
\hline 
$\mathrm{e_{L}^{-}e_{R}^{+}}$ & $\mathrm{\mu^{+}\mu^{-}X}$  & 3.2\% & 3.7\%\tabularnewline
\cline{2-4} 
 & $\mathrm{e^{+}e^{-}X}$ & 4.0\% & 4.5\%\tabularnewline
\cline{2-4} 
 & combined & 2.5\% & 2.9\%\tabularnewline
\hline 
$\mathrm{e_{R}^{-}e_{L}^{+}}$ & $\mathrm{\mu^{+}\mu^{-}X}$  & 3.6\% & 3.8\%\tabularnewline
\cline{2-4} 
 & $\mathrm{e^{+}e^{-}X}$ & 4.7\% & 4.8\%\tabularnewline
\cline{2-4} 
 & combined & 2.9\% & 3.0\%\tabularnewline
\hline 
\end{tabular}
\par\end{centering}

\medskip{}

\begin{centering}
\begin{tabular}{|c|c|c|c|}
\hline 
\multicolumn{1}{|c}{$\sqrt{s}$} & 250 GeV & With $E_{\mathrm{vis}}$ cut & No $E_{\mathrm{vis}}$ cut\tabularnewline
\hline 
\multicolumn{1}{|c}{} &  & $\Delta M_{\mathrm{H}}$ (MeV) & $\Delta M_{\mathrm{H}}$ (MeV)\tabularnewline
\hline 
\hline 
$\mathrm{e_{L}^{-}e_{R}^{+}}$ & $\mathrm{\mu^{+}\mu^{-}X}$  & 39 & 46\tabularnewline
\cline{2-4} 
 & $\mathrm{e^{+}e^{-}X}$ & 121 & 141\tabularnewline
\cline{2-4} 
 & combined & 37 & 44\tabularnewline
\hline 
$\mathrm{e_{R}^{-}e_{L}^{+}}$ & $\mathrm{\mu^{+}\mu^{-}X}$  & 43 & 44\tabularnewline
\cline{2-4} 
 & $\mathrm{e^{+}e^{-}X}$ & 149 & 156\tabularnewline
\cline{2-4} 
 & combined & 41 & 42\tabularnewline
\hline 
\end{tabular}
\par\end{centering}

\protect\caption{The comparison of the precisions between the cases with and without
the $E_{\mathrm{vis}}$ cut. Shown here are the statistical uncertainties
of $\sigma_{\mathrm{ZH}}$ and $M_{\mathrm{H}}$ for $\sqrt{s}$=250
GeV.\label{Resukt250=0000231-2-1-1-1-1}}
\end{table}

\section{Performance of Lepton Pairing\label{sec:Performance-of-Lepton}}

\begin{table}[H]
\centering{}%
\begin{tabular}{|c|c|c|c|c|c|c|c|c|}
\hline 
$\mathrm{H\rightarrow XX}$  & $\mathrm{bb}$  & $\mathrm{cc}$ & $\mathrm{gg}$ & $\mathrm{WW^{*}}$  & $\mathrm{ZZ^{*}}$ & $\mathrm{\tau\tau}$ & $\mathrm{\gamma\gamma}$  & $\mathrm{Z\gamma}$ \tabularnewline
\hline 
C0 & 100\% & 100\% & 100\% & 100\% & 100\% & 100\% & 100\% & 100\%\tabularnewline
\hline 
C1 & 93.70\% & 93.69\% & 93.39\% & 94.01\% & 94.15\% & 94.00\% & 93.70\% & 93.30\%\tabularnewline
\hline 
C2 & 93.69\% & 93.69\% & 93.39\% & 92.86\% & 93.17\% & 93.70\% & 93.70\% & 92.36\%\tabularnewline
\hline 
C3 & 0\% & 0\% & 0\% & 0.831\% & 0.728\% & 0.204\% & 0\% & 0.78\%\tabularnewline
\hline 
C4 & 0\% & 0\% & 0\% & 0\% & 0.342\% & 0.002\% & 0\% & 0.55\%\tabularnewline
\hline 
C5 & 0\% & 0\% & 0\% & 0.315\% & 0.250\% & 0.092\% & 0\% & 0.16\%\tabularnewline
\hline 
C6 & 0\% & 0\% & 0\% & 0.002\% & 0.002\% & 0\% & 0\% & 0.007\%\tabularnewline
\hline 
\end{tabular}\protect\caption{The lepton pairing performance of the $\mathrm{\mu{}^{+}\mathit{\mu}^{-}H}$
channel and $\mathrm{e_{L}^{-}e_{R}^{+}}$ at $\sqrt{s}$=250 GeV.
The categories C1-C6 are defined as in Table \ref{tab:Cate}. The
statistical uncertainties are below 0.1\%. \label{tab:CateZmm250All}}
\end{table}

\begin{table}[H]
\centering{}%
\begin{tabular}{|c|c|c|c|c|c|c|c|c|}
\hline 
$\mathrm{H\rightarrow XX}$  & $\mathrm{bb}$  & $\mathrm{cc}$ & $\mathrm{gg}$ & $\mathrm{WW^{*}}$  & $\mathrm{ZZ^{*}}$ & $\mathrm{\tau\tau}$ & $\mathrm{\gamma\gamma}$  & $\mathrm{Z\gamma}$ \tabularnewline
\hline 
C0 & 100\% & 100\% & 100\% & 100\% & 100\% & 100\% & 100\% & 100\%\tabularnewline
\hline 
C1 & 89.14\% & 88.92\% & 88.51\% & 89.73\% & 87.08\% & 89.43\% & 89.84\% & 88.78\%\tabularnewline
\hline 
C2 & 89.07\% & 88.87\% & 88.49\% & 86.91\% & 85.13\% & 87.96\% & 88.15\% & 86.89\%\tabularnewline
\hline 
C3 & 0\% & 0\% & 0\% & 2.020\% & 1.363\% & 1.005\% & 0\% & 1.39\%\tabularnewline
\hline 
C4 & 0\% & 0\% & 0\% & 0.008\% & 0.548\% & 0.002\% & 0\% & 0.70\%\tabularnewline
\hline 
C5 & 0\% & 0\% & 0\% & 0.801\% & 0.572\% & 0.464\% & 0\% & 0.49\%\tabularnewline
\hline 
C6 & 0\% & 0\% & 0\% & 0.004\% & 0.008\% & 0\% & 0\% & 0.01\%\tabularnewline
\hline 
\end{tabular}\protect\caption{The lepton pairing performance of the $\mathrm{\mathrm{e}{}^{+}\mathit{\mathrm{e}}^{-}H}$
channel and $\mathrm{e_{L}^{-}e_{R}^{+}}$ at $\sqrt{s}$=250 GeV.
The categories C1-C6 are defined as in Table \ref{tab:Cate}. The
statistical uncertaintiesare below 0.1\%. \label{tab:CateZee250All}}
\end{table}

\section{Efficiency Table \label{sec:Efficiency-Table}}

This section shows the efficiency of each major SM decay mode after
each cut mentioned in Section \ref{sec:Event-Selection}, for all
channels except the case of $\sqrt{s}$=250 GeV and $\mathrm{e_{L}^{-}e_{R}^{+}}$,
which is shown in Tables \ref{tab:Mode Eff} and \ref{tab:Mode EffZee}.

\begin{table}[H]
\begin{centering}
\begin{tabular}{|c|c|c|c|c|c|c|c|c|}
\hline 
$\sqrt{s}$=250 GeV & $\mathrm{\mu{}^{+}\mathit{\mu}^{-}H}$  & $\mathrm{e_{R}^{-}e_{L}^{+}}$  & \multicolumn{1}{c}{} & \multicolumn{1}{c}{} & \multicolumn{1}{c}{} & \multicolumn{1}{c}{} & \multicolumn{1}{c}{} & \tabularnewline
\hline 
$\mathrm{H\rightarrow XX}$ & bb & cc & gg & $\tau\tau$ & $\mathrm{\mathrm{WW^{*}}}$ & $\mathrm{ZZ^{*}}$ & $\gamma\gamma$ & $\gamma\mathrm{Z}$\tabularnewline
\hline 
BR (SM) & 57.8\% & 2.7\% & 8.6\% & 6.4\% & 21.6\% & 2.7\% & 0.23\% & 0.16\%\tabularnewline
\hline 
\hline 
Lepton Finder & 93.77\% & 93.67\% & 93.54\% & 93.98\% & 94.25\% & 94.32\% & 93.80\% & 94.19\%\tabularnewline
\hline 
Lepton ID+PreCuts & 93.74\% & 93.65\% & 93.53\% & 93.90\% & 94.13\% & 93.59\% & 93.68\% & 93.33\%\tabularnewline
\hline 
$M_{\mathrm{l^{+}l^{-}}}\in${[}73, 120{]} GeV & 91.73\% & 91.72\% & 91.57\% & 91.93\% & 92.04\% & 91.69\% & 91.78\% & 91.54\%\tabularnewline
\hline 
$p_{\mathrm{t}}^{\mathrm{l^{+}l^{-}}}\in${[}10, 70{]} GeV & 90.03\% & 90.01\% & 89.83\% & 90.25\% & 90.29\% & 90.01\% & 89.97\% & 89.85\%\tabularnewline
\hline 
$\mathrm{\left|cos\mathit{\theta_{\mathrm{missing}}}\right|}$< 0.98 & 90.03\% & 90.01\% & 89.83\% & 90.23\% & 90.28\% & 90.00\% & 89.90\% & 89.53\%\tabularnewline
\hline 
MVA & 83.01\% & 83.10\% & 82.87\% & 83.16\% & 83.18\% & 82.97\% & 82.66\% & 82.63\%\tabularnewline
\hline 
$M_{\mathrm{rec}}\in${[}110, 155{]} GeV & 82.63\% & 82.73\% & 82.47\% & 82.78\% & 82.85\% & 82.61\% & 82.28\% & 82.22\%\tabularnewline
\hline 
\end{tabular}
\par\end{centering}

\medskip{}

\begin{centering}
\begin{tabular}{|c|c|c|c|c|c|c|c|c|}
\hline 
$\sqrt{s}$=250 GeV & $\mathrm{e{}^{+}\mathit{\mathrm{e}}^{-}H}$  & $\mathrm{e_{R}^{-}e_{L}^{+}}$  & \multicolumn{1}{c}{} & \multicolumn{1}{c}{} & \multicolumn{1}{c}{} & \multicolumn{1}{c}{} & \multicolumn{1}{c}{} & \tabularnewline
\hline 
$\mathrm{H\rightarrow XX}$ & bb & cc & gg & $\tau\tau$ & $\mathrm{\mathrm{WW^{*}}}$ & $\mathrm{ZZ^{*}}$ & $\gamma\gamma$ & $\gamma\mathrm{Z}$\tabularnewline
\hline 
BR (SM) & 57.8\% & 2.7\% & 8.6\% & 6.4\% & 21.6\% & 2.7\% & 0.23\% & 0.16\%\tabularnewline
\hline 
\hline 
Lepton Finder & 88.60\% & 88.95\% & 88.52\% & 89.32\% & 89.56\% & 89.77\% & 89.60\% & 89.64\%\tabularnewline
\hline 
Lepton ID+PreCuts & 88.12\% & 88.35\% & 87.97\% & 88.43\% & 88.67\% & 87.93\% & 87.38\% & 87.19\%\tabularnewline
\hline 
$M_{\mathrm{l^{+}l^{-}}}\in${[}73, 120{]} GeV & 86.03\% & 86.29\% & 85.96\% & 85.88\% & 85.90\% & 85.57\% & 85.15\% & 85.10\%\tabularnewline
\hline 
$p_{\mathrm{t}}^{\mathrm{l^{+}l^{-}}}\in${[}10, 70{]} GeV & 84.36\% & 84.60\% & 84.26\% & 84.19\% & 84.12\% & 83.82\% & 83.40\% & 83.34\%\tabularnewline
\hline 
$\mathrm{\left|cos\mathit{\theta_{\mathrm{missing}}}\right|}$< 0.98 & 84.36\% & 84.60\% & 84.26\% & 84.06\% & 84.11\% & 83.71\% & 82.99\% & 83.04\%\tabularnewline
\hline 
MVA & 67.00\% & 67.17\% & 66.94\% & 66.66\% & 66.56\% & 66.41\% & 65.95\% & 65.75\%\tabularnewline
\hline 
$M_{\mathrm{rec}}\in${[}110, 155{]} GeV & 66.70\% & 66.92\% & 66.66\% & 66.39\% & 66.28\% & 66.13\% & 65.66\% & 65.49\%\tabularnewline
\hline 
\end{tabular}
\par\end{centering}

\centering{}\protect\caption{The BR values and the efficiency of the major SM Higgs decay modes,
after each data selection step, shown here for the case of the $\mathrm{\mu^{+}\mu^{-}X}$
(top) and $\mathrm{\mathrm{e}^{+}\mathrm{e}^{-}X}$ (bottom) channels
and $\mathrm{e_{R}^{-}e_{L}^{+}}$ at $\sqrt{s}$=250 GeV. The statistical
uncertainties on these values are below 0.17\%.\label{tab:Mode Eff-1-2}}
\end{table}

\begin{table}[H]
\begin{centering}
\begin{tabular}{|c|c|c|c|c|c|c|c|c|}
\hline 
$\sqrt{s}$=350 GeV & $\mathrm{\mu{}^{+}\mathit{\mu}^{-}H}$  & $\mathrm{e_{L}^{-}e_{R}^{+}}$  & \multicolumn{1}{c}{} & \multicolumn{1}{c}{} & \multicolumn{1}{c}{} & \multicolumn{1}{c}{} & \multicolumn{1}{c}{} & \tabularnewline
\hline 
$\mathrm{H\rightarrow XX}$ & bb & cc & gg & $\tau\tau$ & $\mathrm{\mathrm{WW^{*}}}$ & $\mathrm{ZZ^{*}}$ & $\gamma\gamma$ & $\gamma\mathrm{Z}$\tabularnewline
\hline 
BR (SM) & 57.8\% & 2.7\% & 8.6\% & 6.4\% & 21.6\% & 2.7\% & 0.23\% & 0.16\%\tabularnewline
\hline 
\hline 
Lepton Finder & 93.65\% & 93.79 & 93.49\% & 94.05\% & 94.13\% & 94.20\% & 94.00\% & 93.97\%\tabularnewline
\hline 
Lepton ID+Precut & 93.60\% & 93.72 & 93.40\% & 93.94\% & 93.96\% & 93.53\% & 93.84\% & 93.09\%\tabularnewline
\hline 
$M_{\mathrm{l^{+}l^{-}}}\in${[}73, 120{]} GeV & 91.49\% & 91.71 & 91.32\% & 91.78\% & 91.73\% & 91.46\% & 91.64\% & 91.00\%\tabularnewline
\hline 
$p_{\mathrm{t}}^{\mathrm{l^{+}l^{-}}}\in${[}10, 140{]} GeV & 91.07\% & 91.29 & 90.94\% & 91.36\% & 91.32\% & 91.03\% & 91.24\% & 90.55\%\tabularnewline
\hline 
$\mathrm{\left|cos\mathit{\theta_{\mathrm{missing}}}\right|}$< 0.98 & 91.07\% & 91.29 & 90.94\% & 91.36\% & 91.31\% & 91.02\% & 91.18\% & 90.35\%\tabularnewline
\hline 
MVA & 66.70\% & 66.79 & 66.63\% & 66.66\% & 66.57\% & 66.63\% & 66.83\% & 66.54\%\tabularnewline
\hline 
$M_{\mathrm{rec}}\in${[}100, 200{]} GeV & 64.85\% & 64.90 & 64.81\% & 64.80\% & 64.72\% & 64.82\% & 64.89\% & 64.61\%\tabularnewline
\hline 
\end{tabular}
\par\end{centering}

\medskip{}

\begin{centering}
\begin{tabular}{|c|c|c|c|c|c|c|c|c|}
\hline 
$\sqrt{s}$=350 GeV & $\mathrm{\mathrm{e}{}^{+}\mathrm{e}^{-}H}$  & $\mathrm{e_{L}^{-}e_{R}^{+}}$  & \multicolumn{1}{c}{} & \multicolumn{1}{c}{} & \multicolumn{1}{c}{} & \multicolumn{1}{c}{} & \multicolumn{1}{c}{} & \tabularnewline
\hline 
$\mathrm{H\rightarrow XX}$ & bb & cc & gg & $\tau\tau$ & $\mathrm{\mathrm{WW^{*}}}$ & $\mathrm{ZZ^{*}}$ & $\gamma\gamma$ & $\gamma\mathrm{Z}$\tabularnewline
\hline 
BR (SM) & 57.8\% & 2.7\% & 8.6\% & 6.4\% & 21.6\% & 2.7\% & 0.23\% & 0.16\%\tabularnewline
\hline 
\hline 
Lepton Finder & 78.35\% & 78.22\% & 77.96\% & 80.08\% & 81.22\% & 80.19\% & 79.40\% & 80.15\%\tabularnewline
\hline 
Lepton ID+Precut & 71.92\% & 71.91\% & 71.49\% & 73.51\% & 74.81\% & 73.01\% & 72.22\% & 72.33\%\tabularnewline
\hline 
$M_{\mathrm{l^{+}l^{-}}}\in${[}73, 120{]} GeV & 65.18\% & 65.12\% & 64.93\% & 65.81\% & 66.43\% & 65.70\% & 65.33\% & 65.41\%\tabularnewline
\hline 
$p_{\mathrm{t}}^{\mathrm{l^{+}l^{-}}}\in${[}10, 140{]} GeV & 64.35\% & 64.25\% & 64.10\% & 65.00\% & 65.54\% & 64.86\% & 64.51\% & 64.57\%\tabularnewline
\hline 
$\mathrm{\left|cos\mathit{\theta_{\mathrm{missing}}}\right|}$< 0.98 & 64.35\% & 64.25\% & 64.10\% & 64.98\% & 65.54\% & 64.86\% & 64.45\% & 64.40\%\tabularnewline
\hline 
MVA & 37.70\% & 37.51\% & 37.38\% & 37.33\% & 37.33\% & 37.29\% & 37.64\% & 37.11\%\tabularnewline
\hline 
$M_{\mathrm{rec}}\in${[}100, 200{]} GeV & 37.03\% & 36.89\% & 36.72\% & 36.70\% & 36.68\% & 36.66\% & 37.05\% & 36.51\%\tabularnewline
\hline 
\end{tabular}
\par\end{centering}

\centering{}\protect\caption{The BR values and the efficiency of the major SM Higgs decay modes,
after each data selection step, shown here for the case of the $\mathrm{\mu^{+}\mu^{-}X}$
(top) and $\mathrm{\mathrm{e}^{+}\mathrm{e}^{-}X}$ (bottom) channels
and $\mathrm{e_{L}^{-}e_{R}^{+}}$ at $\sqrt{s}$=350 GeV. The statistical
uncertainties on these values are below 0.17\%.\label{tab:Mode Eff-1}}
\end{table}

\begin{table}[H]
\begin{centering}
\begin{tabular}{|c|c|c|c|c|c|c|c|c|}
\hline 
$\sqrt{s}$=350 GeV & $\mathrm{\mu{}^{+}\mathit{\mu}^{-}H}$  & $\mathrm{e_{R}^{-}e_{L}^{+}}$  & \multicolumn{1}{c}{} & \multicolumn{1}{c}{} & \multicolumn{1}{c}{} & \multicolumn{1}{c}{} & \multicolumn{1}{c}{} & \tabularnewline
\hline 
$\mathrm{H\rightarrow XX}$ & bb & cc & gg & $\tau\tau$ & $\mathrm{\mathrm{WW^{*}}}$ & $\mathrm{ZZ^{*}}$ & $\gamma\gamma$ & $\gamma\mathrm{Z}$\tabularnewline
\hline 
BR (SM) & 57.8\% & 2.7\% & 8.6\% & 6.4\% & 21.6\% & 2.7\% & 0.23\% & 0.16\%\tabularnewline
\hline 
\hline 
Lepton Finder & 93.72\% & 93.66\% & 93.48\% & 94.02\% & 93.99\% & 94.21\% & 94.08\% & 94.29\%\tabularnewline
\hline 
Lepton ID+Precut & 93.66\% & 93.58\% & 93.40\% & 93.91\% & 93.82\% & 93.53\% & 93.94\% & 93.46\%\tabularnewline
\hline 
$M_{\mathrm{l^{+}l^{-}}}\in${[}73, 120{]} GeV & 91.43\% & 91.45\% & 91.22\% & 91.66\% & 91.48\% & 91.33\% & 91.79\% & 91.30\%\tabularnewline
\hline 
$p_{\mathrm{t}}^{\mathrm{l^{+}l^{-}}}\in${[}10, 140{]} GeV & 91.00\% & 91.03\% & 90.81\% & 91.23\% & 91.06\% & 90.91\% & 91.37\% & 90.88\%\tabularnewline
\hline 
$\mathrm{\left|cos\mathit{\theta_{\mathrm{missing}}}\right|}$< 0.98 & 91.00\% & 91.03\% & 90.81\% & 91.21\% & 91.06\% & 90.89\% & 91.32\% & 90.69\%\tabularnewline
\hline 
MVA & 76.48\% & 76.62\% & 76.45\% & 76.42\% & 76.43\% & 76.33\% & 76.56\% & 76.29\%\tabularnewline
\hline 
$M_{\mathrm{rec}}\in${[}100, 200{]} GeV & 72.94\% & 72.99\% & 72.89\% & 72.84\% & 72.93\% & 72.83\% & 72.81\% & 72.95\%\tabularnewline
\hline 
\end{tabular}
\par\end{centering}

\medskip{}

\begin{centering}
\begin{tabular}{|c|c|c|c|c|c|c|c|c|}
\hline 
$\sqrt{s}$=350 GeV & $\mathrm{\mathrm{e}{}^{+}\mathrm{e}^{-}H}$  & $\mathrm{e_{R}^{-}e_{L}^{+}}$  & \multicolumn{1}{c}{} & \multicolumn{1}{c}{} & \multicolumn{1}{c}{} & \multicolumn{1}{c}{} & \multicolumn{1}{c}{} & \tabularnewline
\hline 
$\mathrm{H\rightarrow XX}$ & bb & cc & gg & $\tau\tau$ & $\mathrm{\mathrm{WW^{*}}}$ & $\mathrm{ZZ^{*}}$ & $\gamma\gamma$ & $\gamma\mathrm{Z}$\tabularnewline
\hline 
BR (SM) & 57.8\% & 2.7\% & 8.6\% & 6.4\% & 21.6\% & 2.7\% & 0.23\% & 0.16\%\tabularnewline
\hline 
\hline 
Lepton Finder & 81.30\% & 81.12\% & 80.08\% & 82.56\% & 83.45\% & 82.78\% & 82.11\% & 82.80\%\tabularnewline
\hline 
Lepton ID+Precut & 75.69\% & 75.57\% & 75.46\% & 76.94\% & 77.74\% & 76.57\% & 75.67\% & 75.98\%\tabularnewline
\hline 
$M_{\mathrm{l^{+}l^{-}}}\in${[}73, 120{]} GeV & 70.00\% & 69.92 & 69.97\% & 70.57\% & 70.78\% & 70.50\% & 69.70\% & 69.98\%\tabularnewline
\hline 
$p_{\mathrm{t}}^{\mathrm{l^{+}l^{-}}}\in${[}10, 140{]} GeV & 69.08\% & 69.02\% & 69.02\% & 69.63\% & 69.86\% & 69.61\% & 68.83\% & 68.99\%\tabularnewline
\hline 
$\mathrm{\left|cos\mathit{\theta_{\mathrm{missing}}}\right|}$< 0.98 & 69.08\% & 69.02\% & 69.02\% & 69.61\% & 69.86\% & 69.60\% & 68.78\% & 68.85\%\tabularnewline
\hline 
MVA & 39.47\% & 39.33\% & 39.32\% & 39.10\% & 38.85\% & 39.26\% & 38.85\% & 38.19\%\tabularnewline
\hline 
$M_{\mathrm{rec}}\in${[}100, 200{]} GeV & 38.89\% & 38.76\% & 38.73\% & 38.55\% & 38.25\% & 38.69\% & 38.31\% & 37.67\%\tabularnewline
\hline 
\end{tabular}
\par\end{centering}

\centering{}\protect\caption{The BR values and the efficiency of the major SM Higgs decay modes,
after each data selection step, shown here for the case of the $\mathrm{\mu^{+}\mu^{-}X}$
(top) and $\mathrm{\mathrm{e}^{+}\mathrm{e}^{-}X}$ (bottom) channels
and $\mathrm{e_{R}^{-}e_{L}^{+}}$ at $\sqrt{s}$=350 GeV. The statistical
uncertainties on these values are below 0.17\%.\label{tab:Mode Eff-1-3}}
\end{table}

\begin{table}[H]
\begin{centering}
\begin{tabular}{|c|c|c|c|c|c|c|c|c|}
\hline 
$\sqrt{s}$=500 GeV & $\mathrm{\mu{}^{+}\mathit{\mu}^{-}H}$  & $\mathrm{e_{L}^{-}e_{R}^{+}}$  & \multicolumn{1}{c}{} & \multicolumn{1}{c}{} & \multicolumn{1}{c}{} & \multicolumn{1}{c}{} & \multicolumn{1}{c}{} & \tabularnewline
\hline 
$\mathrm{H\rightarrow XX}$ & bb & cc & gg & $\tau\tau$ & $\mathrm{\mathrm{WW^{*}}}$ & $\mathrm{ZZ^{*}}$ & $\gamma\gamma$ & $\gamma\mathrm{Z}$\tabularnewline
\hline 
BR (SM) & 57.8\% & 2.7\% & 8.6\% & 6.4\% & 21.6\% & 2.7\% & 0.23\% & 0.16\%\tabularnewline
\hline 
\hline 
Lepton Finder & 93.29\% & 93.32\% & 93.29\% & 93.77\% & 93.74\% & 93.95\% & 93.63\% & 93.79\%\tabularnewline
\hline 
Lepton ID+Precut & 84.69\% & 84.75\% & 84.68\% & 84.85\% & 84.79\% & 84.69\% & 84.80\% & 84.49\%\tabularnewline
\hline 
$M_{\mathrm{l^{+}l^{-}}}\in${[}73, 120{]} GeV & 83.18\% & 83.18\% & 83.17\%\% & 83.32\% & 83.23\% & 83.15\% & 83.26\% & 82.96\%\tabularnewline
\hline 
$p_{\mathrm{t}}^{\mathrm{l^{+}l^{-}}}\in${[}10, 230{]} GeV & 83.14\% & 83.14\% & 83.12\% & 83.28\% & 83.18\% & 83.10\% & 83.20\% & 82.93\%\tabularnewline
\hline 
$\mathrm{\left|cos\mathit{\theta_{\mathrm{missing}}}\right|}$< 0.98 & 83.14\% & 83.14\% & 83.12\% & 83.28\% & 83.18\% & 83.09\% & 83.19\% & 82.81\%\tabularnewline
\hline 
MVA & 66.61\% & 66.56\% & 66.25\% & 66.58\% & 66.63\% & 66.40\% & 66.62\% & 66.28\%\tabularnewline
\hline 
$M_{\mathrm{rec}}\in${[}100, 250{]} GeV & 62.51\% & 62.39\% & 62.24\% & 62.38\% & 62.48\% & 62.32\% & 62.50\% & 62.18\%\tabularnewline
\hline 
\end{tabular}
\par\end{centering}

\medskip{}

\begin{centering}
\begin{tabular}{|c|c|c|c|c|c|c|c|c|}
\hline 
$\sqrt{s}$=500 GeV & $\mathrm{\mathrm{e}{}^{+}\mathrm{e}^{-}H}$  & $\mathrm{e_{L}^{-}e_{R}^{+}}$  & \multicolumn{1}{c}{} & \multicolumn{1}{c}{} & \multicolumn{1}{c}{} & \multicolumn{1}{c}{} & \multicolumn{1}{c}{} & \tabularnewline
\hline 
$\mathrm{H\rightarrow XX}$ & bb & cc & gg & $\tau\tau$ & $\mathrm{\mathrm{WW^{*}}}$ & $\mathrm{ZZ^{*}}$ & $\gamma\gamma$ & $\gamma\mathrm{Z}$\tabularnewline
\hline 
BR (SM) & 57.8\% & 2.7\% & 8.6\% & 6.4\% & 21.6\% & 2.7\% & 0.23\% & 0.16\%\tabularnewline
\hline 
\hline 
Lepton Finder & 41.29\% & 41.40\% & 41.52\% & 45.74\% & 48.75\% & 46.84\% & 43.37\% & 46.19\%\tabularnewline
\hline 
Lepton ID+Precut & 29.88\% & 30.07\% & 30.02\% & 32.34\% & 33.79\% & 32.17\% & 30.99\% & 31.27\%\tabularnewline
\hline 
$M_{\mathrm{l^{+}l^{-}}}\in${[}73, 120{]} GeV & 28.27\% & 28.50\% & 28.47\% & 29.71\% & 30.41\% & 29.75\% & 28.91\% & 29.09\%\tabularnewline
\hline 
$p_{\mathrm{t}}^{\mathrm{l^{+}l^{-}}}\in${[}10, 230{]} GeV & 28.21\% & 28.45\% & 28.40\% & 29.63\% & 30.33\% & 29.67\% & 28.85\% & 29.01\%\tabularnewline
\hline 
$\mathrm{\left|cos\mathit{\theta_{\mathrm{missing}}}\right|}$< 0.98 & 28.21\% & 28.45\% & 28.40\% & 29.63\% & 30.33\% & 29.66\% & 28.83\% & 28.98\%\tabularnewline
\hline 
MVA & 20.56\% & 20.92\% & 20.75\% & 20.89\% & 20.89\% & 20.93\% & 20.77\% & 20.79\%\tabularnewline
\hline 
$M_{\mathrm{rec}}\in${[}100, 250{]} GeV & 19.41\% & 19.66\% & 19.49\% & 19.67\% & 19.68\% & 19.63\% & 19.59\% & 19.53\%\tabularnewline
\hline 
\end{tabular}
\par\end{centering}

\begin{centering}

\par\end{centering}

\centering{}\protect\caption{The BR values and the efficiency of the major SM Higgs decay modes,
after each data selection step, shown here for the case of the $\mathrm{\mu^{+}\mu^{-}X}$
(top) and $\mathrm{\mathrm{e}^{+}\mathrm{e}^{-}X}$ (bottom) channels
and $\mathrm{e_{L}^{-}e_{R}^{+}}$ at $\sqrt{s}$=500 GeV. The statistical
uncertainties on these values are below 0.17\%.\label{tab:Mode Eff-1-4}}
\end{table}

\begin{table}[H]
\begin{centering}
\begin{tabular}{|c|c|c|c|c|c|c|c|c|}
\hline 
$\sqrt{s}$=500 GeV & $\mathrm{\mu{}^{+}\mathit{\mu}^{-}H}$  & $\mathrm{e_{R}^{-}e_{L}^{+}}$  & \multicolumn{1}{c}{} & \multicolumn{1}{c}{} & \multicolumn{1}{c}{} & \multicolumn{1}{c}{} & \multicolumn{1}{c}{} & \tabularnewline
\hline 
$\mathrm{H\rightarrow XX}$ & bb & cc & gg & $\tau\tau$ & $\mathrm{\mathrm{WW^{*}}}$ & $\mathrm{ZZ^{*}}$ & $\gamma\gamma$ & $\gamma\mathrm{Z}$\tabularnewline
\hline 
BR (SM) & 57.8\% & 2.7\% & 8.6\% & 6.4\% & 21.6\% & 2.7\% & 0.23\% & 0.16\%\tabularnewline
\hline 
\hline 
Lepton Finder & 93.26\% & 93.22\% & 93.16\% & 93.68\% & 93.60\% & 93.02\% & 93.66\% & 93.88\tabularnewline
\hline 
Lepton ID+PreCuts & 84.55\% & 84.55\% & 84.46\% & 84.77\% & 84.59\% & 84.49\% & 84.76\% & 84.58\%\tabularnewline
\hline 
$M_{\mathrm{l^{+}l^{-}}}\in${[}73, 120{]} GeV & 82.96\% & 82.90\% & 82.83\% & 83.18\% & 82.98\% & 82.91\% & 83.13\% & 83.02\%\tabularnewline
\hline 
$p_{\mathrm{t}}^{\mathrm{l^{+}l^{-}}}\in${[}10, 230{]} GeV & 92.91\% & 82.86\% & 82.78\% & 83.14\% & 82.94\% & 82.87\% & 83.08\% & 82.98\%\tabularnewline
\hline 
$\mathrm{\left|cos\mathit{\theta_{\mathrm{missing}}}\right|}$< 0.98 & 82.91\% & 82.86\% & 82.78\% & 83.13\% & 82.94\% & 82.86\% & 83.05\% & 82.88\%\tabularnewline
\hline 
MVA & 64.95\% & 64.88\% & 64.97\% & 65.02\% & 64.96\% & 64.74\% & 64.92\% & 64.60\%\tabularnewline
\hline 
$M_{\mathrm{rec}}\in${[}100, 250{]} GeV & 60.94\% & 60.89\% & 60.93\% & 60.97\% & 60.97\% & 60.80\% & 60.95\% & 60.85\%\tabularnewline
\hline 
\end{tabular}
\par\end{centering}

\medskip{}

\begin{centering}
\begin{tabular}{|c|c|c|c|c|c|c|c|c|}
\hline 
$\sqrt{s}$=500 GeV & $\mathrm{\mathrm{e}{}^{+}\mathrm{e}^{-}H}$  & $\mathrm{e_{R}^{-}e_{L}^{+}}$  & \multicolumn{1}{c}{} & \multicolumn{1}{c}{} & \multicolumn{1}{c}{} & \multicolumn{1}{c}{} & \multicolumn{1}{c}{} & \tabularnewline
\hline 
$\mathrm{H\rightarrow XX}$ & bb & cc & gg & $\tau\tau$ & $\mathrm{\mathrm{WW^{*}}}$ & $\mathrm{ZZ^{*}}$ & $\gamma\gamma$ & $\gamma\mathrm{Z}$\tabularnewline
\hline 
BR (SM) & 57.8\% & 2.7\% & 8.6\% & 6.4\% & 21.6\% & 2.7\% & 0.23\% & 0.16\%\tabularnewline
\hline 
\hline 
Lepton Finder & 48.91\% & 48.96\% & 48.77\% & 52.69\% & 55.18\% & 53.84\% & 50.93\% & 53.17\%\tabularnewline
\hline 
Lepton ID+PreCuts & 36.95\% & 37.14\% & 36.98\% & 38.87\% & 40.33\% & 38.98\% & 38.02\% & 38.13\%\tabularnewline
\hline 
$M_{\mathrm{l^{+}l^{-}}}\in${[}73, 120{]} GeV & 35.24\% & 35.43\% & 35.26\% & 36.26\% & 37.18\% & 36.62\% & 35.91\% & 35.99\%\tabularnewline
\hline 
$p_{\mathrm{t}}^{\mathrm{l^{+}l^{-}}}\in${[}10, 230{]} GeV & 35.15\% & 35.35\% & 35.19\% & 36.17\% & 37.09\% & 36.53\% & 35.83\% & 35.91\%\tabularnewline
\hline 
$\mathrm{\left|cos\mathit{\theta_{\mathrm{missing}}}\right|}$< 0.98 & 35.15\% & 35.35\% & 35.19\% & 36.17\% & 37.09\% & 36.52\% & 35.80\% & 35.88\%\tabularnewline
\hline 
MVA & 16.76\% & 16.91\% & 16.81\% & 16.82\% & 16.77\% & 16.97\% & 16.78\% & 16.64\%\tabularnewline
\hline 
$M_{\mathrm{rec}}\in${[}100, 250{]} GeV & 16.41\% & 16.52\% & 16.44\% & 16.44\% & 16.39\% & 16.56\% & 16.38\% & 16.28\%\tabularnewline
\hline 
\end{tabular}
\par\end{centering}

\centering{}\protect\caption{The BR values and the efficiency of the major SM Higgs decay modes,
after each data selection step, shown here for the case of the $\mathrm{\mu^{+}\mu^{-}X}$
(top) and $\mathrm{\mathrm{e}^{+}\mathrm{e}^{-}X}$ (bottom) channels
and $\mathrm{e_{R}^{-}e_{L}^{+}}$ at $\sqrt{s}$=500 GeV. The statistical
uncertainties on these values are below 0.17\%.\label{tab:Mode Eff-1-1}}
\end{table}

\end{document}